\documentclass[12pt]{iopart}
\usepackage{graphicx}
\usepackage{mathptmx}
\usepackage{flushend}
\usepackage[numbers,sort&compress]{natbib}
\usepackage[colorlinks,citecolor=blue,urlcolor=blue,linkcolor=blue]{hyperref}
\expandafter\let\csname equation*\endcsname\relax
\expandafter\let\csname endequation*\endcsname\relax
\usepackage{amsmath,amssymb}
\usepackage{multirow}
\usepackage{xspace}
\usepackage[caption=false]{subfig}

\def\MET{\ensuremath{E_\text{T}^\text{miss}}\xspace}
\def\e{\ensuremath{\epsilon}}
\def\d{\ensuremath{\partial}}

\def\gonehww{\ensuremath{g^{(1)}_{_{SWW}}}}
\def\gtwohww{\ensuremath{g^{(2)}_{_{SWW}}}}

\def\gthww{\ensuremath{\tilde g_{_{SWW}}}}

\begin{document}

\title[Multi-lepton signatures of additional BSM scalars at the LHC]{Multi-lepton signatures of additional scalar bosons beyond the Standard Model at the LHC}

\author{Stefan von Buddenbrock,$^1$ Alan S. Cornell,$^1$ Abdualazem Fadol Mohammed,$^1$ Mukesh Kumar,$^1$ Bruce Mellado,$^{1,2}$ and Xifeng Ruan$^1$}

\address{$^1$School of Physics and Institute for Collider Particle Physics, University of the Witwatersrand, Johannesburg, Wits 2050, South Africa}
\address{$^2$iThemba LABS, National Research Foundation, PO Box 722, Somerset West 7129, South Africa}

\eads{\mailto{stef.von.b@cern.ch}, \mailto{alan.cornell@wits.ac.za}, \mailto{amohammed@aims.ac.tz}, \mailto{mukesh.kumar@cern.ch}, \mailto{bruce.mellado@wits.ac.za}, \mailto{xifeng.ruan@cern.ch}}

\begin{abstract}
Following a prediction made in Refs.~\cite{vonBuddenbrock:2015ema,Kumar:2016vut,vonBuddenbrock:2016rmr}, this paper focuses on multi-lepton signatures arising from two new hypothetical scalar bosons, $H$ and $S$, at the Large Hadron Collider (LHC).
These two new bosons are an extension to the Standard Model (SM) and interact with the SM Higgs boson, $h$.
We consider two production modes for $H$, one being gluon fusion and the other being in association with top quarks.
The $H \to S h$ decay mode is considered, where leptonic final states are studied.
The CP properties of $S$ are characterised by considering effective couplings derived from dimension six operators through $SWW$ vertices.
The nature of the $S$ boson is considered in two separate contexts.
Firstly in a simplified model, it is considered to have Higgs-like couplings.
Secondly, we consider a heavy neutrino model and its interactions with the $Z, W$ and $S$ bosons.
The predictions of the models are compared both to ATLAS and CMS results at $\sqrt{s} = 8$ and $13$~TeV, where appropriate.
The data is interpreted using a simplified model where all the signal comes from $H \to S h$, assuming $S$ to be Higgs-like, $m_H=270$~GeV and $m_S=150$~GeV.
The combined result yields gives a best fit value for the parameter $\beta_g$ (the strength of the Yukawa coupling of $H$ to top quarks), $\beta_g^2=1.38\pm 0.22$.
A number of regions of the phase space are suggested to the experiments for further exploration.
\end{abstract}

\submitto{\jpg}


\maketitle

\section{Introduction}
\label{sec:intro}

In the aftermath of the observation of a new boson at the Large Hadron Collider (LHC) that is consistent with the Standard Model (SM) Higgs boson $h$ reported by the ATLAS and CMS collaborations~\cite{Aad:2012tfa,Chatrchyan:2012xdj,Aad:2013xqa,Chatrchyan:2012jja,Aad:2015zhl,Khachatryan:2016vau}, a new challenge is faced.
The current goal of the experimental collaborations is to confirm its properties via measurements of its couplings, decay width, and differential distributions (of observables such as transverse momentum $p_\text{T}^h$, rapidity $y^h$ etc.).
In addition to this, the expectation is that physics beyond the SM (BSM) might be identified with larger datasets from the LHC at the current centre of mass energy $\sqrt{s} = 13$~TeV and possibly even in the $8$~TeV data.
A plethora of BSM models are being considered in the literature, including additional scalar/vector bosons, fermions or exotic BSM objects.
In fact, the existence of non-zero masses for the neutrinos is clearly an interesting BSM scenario that is expected to be studied both at present and future colliders~\cite{Atre:2009rg,Degrande:2016aje,Alva:2014gxa,Golling:2016gvc}.

Of particular interest here is the BSM scenario considered in Refs.~\cite{vonBuddenbrock:2015ema,Kumar:2016vut,vonBuddenbrock:2016rmr,Kumar:2016wzt,Mthembu:2017vix,vonBuddenbrock:2017bqf,Fang:2017tmh,vonBuddenbrock:2017jqp}.
Here, the additional scalars $H$ and $S$ were introduced to explain a number of features in the LHC Run 1 data.
From the limited datasets in Run 1, the following features served as motivation for the model: distortions of the Higgs boson transverse momentum spectrum  $p_\text{T}^h$, which in some cases showed an enhancement in the range $20~\text{GeV}<p_\text{T}^h<100~\text{GeV}$ often accompanied by elevated associated jet activity, elevated rates of leptons in association with $b$-tagged jets, most notably visible in the search for the associated production of the Higgs boson with top quarks in multi-lepton final states, and the search results for double Higgs boson and weak boson production through a heavy resonance, a combination of which indicates mild excesses around resonance masses of $270$~GeV~\cite{vonBuddenbrock:2017jqp}.
A simplified model was introduced and tested against both the ATLAS and the CMS datasets, a full description of which exists in Ref.~\cite{vonBuddenbrock:2015ema}.
This model suggests the production and decay modes of these scalars which could have significant signals at the LHC, which can be tested with newer and statistically more precise datasets.

In this work, the mass of the BSM scalar $H$ is fixed at $m_H = 270$~GeV, based largely on the study done in Ref.~\cite{vonBuddenbrock:2015ema}.
The mass of $S$ is studied in the range $m_h\lesssim m_S\lesssim m_H-m_h$, where $m_h$ is the mass of the SM Higgs boson, although it was found that a mass of $m_S=150$~GeV fits the data best in most cases, and so only this mass point is used for our results.
For the most part the features of the data that triggered the investigation reported in Ref.~\cite{vonBuddenbrock:2015ema} remain or are magnified in results with Run 2 data reported so far, excluding $tth$ production (see Ref.~\cite{Mellado_HDAYS2017} for a recent review).
In this article we study some relevant multi-lepton signatures of these scalars in different scenarios.
Two production modes of $H$ are considered: gluon fusion ($gg$F), the assumed dominant production mode ($g g \to H$), and top associated $H$ production ($p p \to t H,\bar{t}H$ and $p p \to t\bar t H$).
Here we consider that the $H\to Sh$ branching ratio (BR) is 100\%,\footnote{In the language of Ref.~\cite{vonBuddenbrock:2016rmr}, we are in the limit where $a_1\to0$.} and that the BR of $S$ to dark matter is 0\%.
The $S$ boson is modelled in two different ways. Firstly, $S$ is considered to have globally re-scaled Higgs-like couplings, such that its BRs are the same as for a Higgs boson with a mass $m_S$. Secondly, a model with heavy neutrinos ($N_i$) is introduced which can interact with $S$ in non-standard ways. We study leptonic final states via the $S \to N_i \nu_l$ decay modes.

In Ref.~\cite{vonBuddenbrock:2016rmr}, emphasis was made on rare multi-lepton final states.
These would include the production of four leptons from the production of four $W$s, and the production of three same sign leptons from the production of six $W$s.
Here we concentrate on other multi-lepton signatures.
In addition, a study is reported here where the tensor structure of the $SWW$ coupling is modified to evaluate possible model dependencies of the kinematics of four leptons due to four $W$ production.
Comparisons between the prediction made previously and the data are performed in this work, with the aim of constraining the fit parameter $\beta_g^2$, which shall be discussed below.

The article is organised as follows.
\autoref{sec:formalism} gives a brief introduction to the simplified models used in this study.
Details about simulation and analysis of the signals are then shown in \autoref{sec:analysis} together with comparisons with available data. The results are discussed and concluded in \autoref{sec:conc}.

\section{Formalism}
\label{sec:formalism}

In this section we provide the interaction vertices that are used for this study.
The production and decay modes of $H$ make use of a simplified model approach.
$S$, on the other hand, is considered to be produced dominantly through the decay of $H$.
The nature of the $S$ decay modes should include multiple leptons, but we still have to assume how it decays in order to reduce the number of free parameters it introduces to the model.
For the $S$ decays, two models are considered.
First, the $S$ is considered to have Higgs-like decay modes, and therefore we have referred to it as the ``Higgs-like $S$ model''.
The second model considered has interactions between $N_i, Z, W$ and $S$.
We have referred to this model as the ``heavy neutrino model''.

As in Refs.~\cite{vonBuddenbrock:2015ema,Kumar:2016vut,vonBuddenbrock:2016rmr,Kumar:2016wzt,Mthembu:2017vix,vonBuddenbrock:2017bqf,Fang:2017tmh,vonBuddenbrock:2017jqp}, the mass ranges for the new scalars are as follows:
\begin{itemize}
\item Light Higgs boson ($h$): $m_h = 125$~GeV (the SM Higgs boson),
\item Heavy Higgs boson ($H$): $2 m_h < m_H < 2 m_t$, where $m_t$ is the top-quark mass, and
\item Additional scalar ($S$):\footnote{Note that this range is not strictly adhered to, since the original idea was to keep the $H\to Sh$ decay mode on-shell. In this study, off-shell decays deserve special consideration.} $m_h \lesssim m_S \lesssim m_H - m_h$.
\end{itemize}
The masses of the heavy neutrinos ($N_i$), however, are not as well constrained and therefore have been considered in the range $m_{N_i} = 100 - 1000$~GeV.

\subsection{Simplified modelling of $H$ and $S$}

\begin{figure}
  \centering
  \subfloat[]{\includegraphics[width=0.4\textwidth]{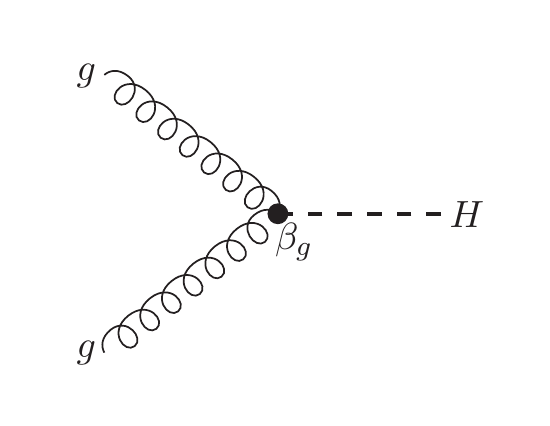}\label{fig:ggh}}
  \subfloat[]{\includegraphics[width=0.4\textwidth]{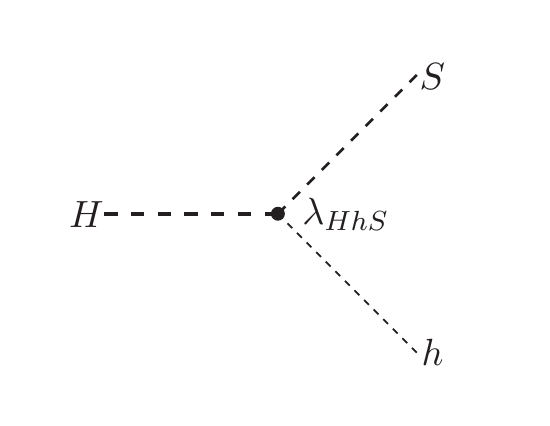}\label{fig:hhs}}
  \caption{Feynman diagrams contributing to (a) the dominant production mode of $H$ ($gg$F), and (b) the dominant decay mode ($H \to Sh$).
  The coupling parameters $\beta_g$ and $\lambda_{HhS}$ are defined in \autoref{vh}
  and \autoref{lsh}.}
\label{fig:feynH}
\end{figure}

\label{efft}
Following a simplified model approach, and after electroweak symmetry breaking (EWSB), the required vertices for this study are,
\begin{align}
\mathcal{L}_{\text{G}} =& -\frac{1}{4}~\beta_{g} \kappa_{_{Hgg}}^{\text{SM}}~G_{\mu\nu}G^{\mu\nu}H
+\beta_{_V}\kappa_{_{hVV}}^{\text{SM}}~V_{\mu}V^{\mu}H,  \label{vh} \\
\mathcal{L}_{\text{Y}} =& -\frac{1}{\sqrt{2}}~\Big[y_{_{ttH}}\bar{t} t H + y_{_{bbH}} \bar{b} b H\Big], \label{H_yuk}
\end{align}
where $\beta_g = y_{ttH}/y_{ttH}^\text{SM}$ is a scale factor with respect to the SM-like Yukawa top coupling for $H$, and therefore multiplies the effective $ggH$ coupling; in Ref.~\cite{vonBuddenbrock:2015ema}, for Run~1 fit results, $\beta_g$ was constrained to be $1.5 \pm 0.6$.
A similar factor $\beta_V$ is used for $VVH$ couplings, where $V = W$ and $Z$ bosons.
$\kappa_{Hgg}^{\rm SM}$ is an effective SM-like $ggH$ coupling (a reduction of the top quark loop) and $v$ is the standard vacuum expectation value (VEV).
It should be noted that in this simplified model, the $gg$F production of $H$ occurs through the effective vertex in \autoref{vh}, which is equivalent to $gg$F through a top quark loop.
The top Yukawa coupling in \autoref{H_yuk} is necessary for $ttH$ production.

These interactions are independent of the SM Lagrangian, ${\cal{L}_{\text{SM}}}$, and the $H$ sector is given by,
\begin{equation}
\mathcal{L}_{H} =\, \frac{1}{2} \partial_\mu H \partial^\mu H - \frac{1}{2} m_H^2 H H
 + \mathcal{L}_{\text{G}} + \mathcal{L}_\text{Y}.\label{lag}
\end{equation}
It is crucial for the reader to understand that the simplified model described above only consists of the vertices needed for the production of $H$ in this study.
For a more comprehensive discussion of how this simplified model can be expanded and embedded into a two Higgs doublet model, see Ref.~\cite{vonBuddenbrock:2016rmr}.

Similarly, the Lagrangian for the real singlet scalar $S$ can be written as,
\begin{equation}
{\cal L}_{S} = {\cal L}_\text{K} + {\cal L}_{SVV^\prime} + {\cal L}_{Sf\bar f} + {\cal L}_{hHS},
\end{equation}
where
\begin{equation}
{\cal L}_\text{K} =\, \frac{1}{2} \partial_\mu S \partial^\mu S - \frac{1}{2} m_S^2 S S, \label{lsk}
\end{equation}
\begin{align}
{\cal L}_{SVV^\prime} =&\, \frac{1}{4} \kappa_{_{Sgg}} \frac{\alpha_{s}}{12 \pi v} S G^{a\mu\nu}G_{\mu\nu}^a
+ \frac{1}{4} \kappa_{_{S\gamma\gamma}} \frac{\alpha}{\pi v} S F^{\mu\nu}F_{\mu\nu} \notag \\
& + \frac{1}{4} \kappa_{_{SZZ}} \frac{\alpha}{\pi v} S Z^{\mu\nu}Z_{\mu\nu}
  + \frac{1}{4} \kappa_{_{SZ\gamma}} \frac{\alpha}{\pi v} S Z^{\mu\nu}F_{\mu\nu} \notag \\
& + \frac{1}{4} \kappa_{_{SWW}} \frac{2 \alpha}{\pi s_w^2 v} S W^{+\mu\nu}W^{-}_{\mu\nu}, \label{lsvv}
\end{align}
\begin{equation}
{\cal L}_{Sf\bar f} =\, - \sum_f \kappa_{_{Sf}} \frac{m_f}{v} S \bar f f, \label{lsf}
\end{equation}
\begin{align}
{\cal L}_{HhS} =\, -\frac{1}{2}~v\Big[\lambda_{_{hhS}} hhS + \lambda_{_{hSS}} hSS +
\lambda_{_{HHS}} HHS \notag \\ & + \lambda_{_{HSS}} HSS + \lambda_{_{HhS}} HhS\Big]. \label{lsh}
\end{align}
The notation of $V$ here differs slightly from what is seen above, where in this case $V \equiv g, \gamma, Z, W$ and $W^\pm_{\mu\nu} = D_\mu W^\pm_\nu - D_\nu W^\pm_\mu $, $D_\mu W^\pm_\nu = \left[ \partial_\mu \pm i e A_\mu \right] W^\pm_\nu$.
The parameters $\alpha_s$ and $\alpha$ are the strong and weak couplings, respectively.
Other possible self interaction terms for $S$ are omitted here since they are not of any phenomenological interest in this study.
The total effective Lagrangian is therefore,
\begin{equation}
\mathcal{L}_{\text{tot}} = \mathcal{L}_\text{SM} + \mathcal{L}_H + \mathcal{L}_S. \label{lagt}
\end{equation}

As mentioned above, the dominant production mechanism for $H$ is assumed to be $gg$F, and the dominant decay mode is $H\to Sh$. The Feynman diagrams for this production and decay chain can be seen in \autoref{fig:feynH}.

\subsection{Modelling the decays of $S$}

Of particular interest in this article is the ability of the BSM scenario we consider to produce multiple leptons. Since the process $pp\to H\to Sh$ is considered, one should be able to produce multi-leptonic signatures through the decay of $S$, while the Higgs boson decays dominantly to $b$-quarks.
In an attempt to model this we have considered two different models for the decay of $S$.

\subsubsection{The Higgs-like $S$ model}

Following the logic of Ref.~\cite{vonBuddenbrock:2016rmr}, the decays of $S$ can be made simple through the assumption that it has globally re-scaled Higgs-like couplings.
These couplings are assumed to be suppressed by some unspecified BSM physics, such that the relative coupling hierarchy to the SM particles is Higgs-like.
$S$ is therefore not produced directly with a large cross section, however, its BRs should be the same as those for a SM-like Higgs boson with a higher mass.
This greatly reduces the number of free parameters in the model.
If this is the case, then the $S$ should decay more dominantly to $W$ and $Z$ boson pairs as its mass gets closer to $\sim 2\,m_W$.
Therefore, the $S$ becomes a source of multiple leptons through the decays of the gauge bosons.

\subsubsection{The heavy neutrino model}

\begin{figure}[t]
  \centering
    \subfloat[]{\includegraphics[width=0.4\textwidth]{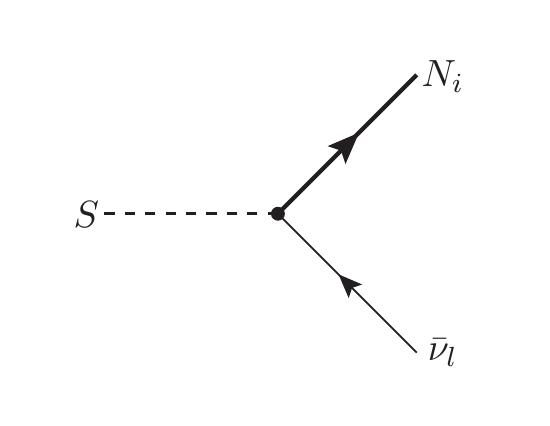}\label{fig:snn}}
    \subfloat[]{\includegraphics[width=0.4\textwidth]{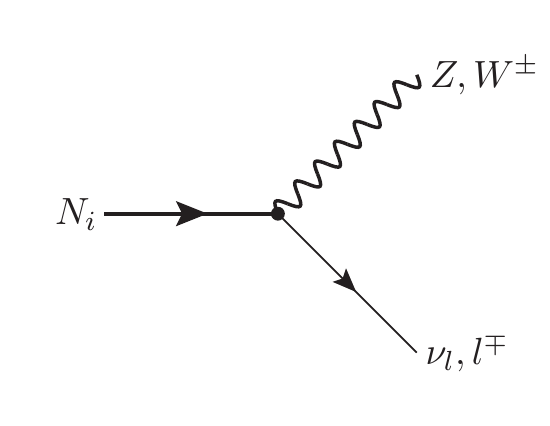}\label{fig:nln}}
  \caption{The vertices that build the decay chain of $S$ in the heavy neutrino model. (a) The decay $S \to N_i\bar\nu_l$, and (b) the decay $N_i \to Z\nu_l$, $N_i \to W^\pm l^\mp$.}
\label{fig:feynHN}
\end{figure}

\label{hnm}
As mentioned above, we have also taken an opportunity to study the possibilities of heavy neutrinos at the LHC.
Thus, we introduce the interactions discussed in Refs.~\cite{Atre:2009rg,Degrande:2016aje,Alva:2014gxa,Golling:2016gvc}, where heavy mass neutrino eigenstates $N_i$ couple to the EW bosons $Z, W$ and $S$ via mixing with left-handed (LH) neutrinos $\nu_L$.

Following the notation used in Ref.~\cite{Atre:2009rg}, the flavour state $\nu_l$ in the mass basis is
\begin{equation}
\nu_l = \sum_{m=1}^{3} U_{lm} \nu_m + \sum_{m^\prime=1}^{n} V_{lm^\prime} N^c_{m^\prime},
\end{equation}
where $U_{lm}$ and $V_{lm^\prime}$ are the observed light neutrino mixing and active-heavy mixing matrices, respectively.
Considering only one heavy mass eigenstate $N$, for simplicity, the interaction Lagrangian with the electroweak bosons is,
\begin{align}
 {\cal L}_{\rm int}
 =
 &- \dfrac{g}{\sqrt{2}} W^{+}_\mu \sum_{l=e}^{\tau} \sum_{m=1}^{3}
 \bar{\nu_m} U^{*}_{lm}\gamma^\mu P_L l^{-} \notag - \dfrac{g}{\sqrt{2}} W^{+}_\mu \sum_{l=e}^{\tau}
 \bar{N^c} V^{*}_{lN}\gamma^\mu P_L l^{-} \notag \\
 &- \dfrac{g}{2\,\cos\theta_W} Z_\mu \sum_{l=e}^{\tau} \sum_{m=1}^{3}
 \bar{\nu_m} U^{*}_{lm}\gamma^\mu P_L \nu_l \notag - \dfrac{g}{2\,\cos\theta_W} Z_\mu \sum_{l=e}^{\tau}
 \bar{N^c} V^{*}_{lN}\gamma^\mu P_L \nu_l \notag \\
 &- \dfrac{gm_N}{2M_W} S \sum_{l=e}^{\tau}
 \bar{N^c} V^{*}_{lN}\gamma^\mu P_L \nu_l + \text{h.c.}
 \label{laghn}
\end{align}
The precise values of $V_{lN}$ are model dependent, and are constrained by neutrino-less double beta decays, oscillation and collider experiments, and tests of lepton universality, as discussed in Refs.~\cite{Atre:2009rg,Antusch:2014woa,deGouvea:2015euy}. The production of multiple leptons due to heavy neutrinos at the LHC has been studied in Refs.~\cite{Das:2015toa,Das:2016hof}.

In the heavy neutrino model $S$ is still a source of leptons, due to the interaction between the heavy neutrinos and the gauge bosons.
The Feynman diagrams showing the decay chain of $S$ in the heavy neutrino model can be seen in \autoref{fig:feynHN}.

\section{Simulation and analysis}
\label{sec:analysis}

Monte Carlo (MC) events were generated and analysed in order to study the cross sections and kinematic distributions of the BSM scenario we have discussed in this article.
Firstly, model files were built based on the Lagrangians in \autoref{lag}, \autoref{lagt} and \autoref{laghn} using the \textsc{FeynRules} package~\cite{Alloul:2013naa}.
The various processes were simulated using \textsc{MG5\_aMC@NLO}~\cite{Alwall:2014hca}.
These hard scatter events were passed to \textsc{Pythia8}~\cite{Sjostrand:2014zea} for the resonant decays (including $t\to Wb$), the parton shower (PS) and hadronisation, and thereafter through \textsc{Delphes3}~\cite{deFavereau:2013fsa} to estimate the detector response for both ATLAS and CMS.
In this study, $H$ was assumed to decay only to $Sh$.
Furthermore, the decays of $S$ and $h$ were left open to explore all possible combinations of leptonic decays.
Jets were clustered using \textsc{FastJet}~\cite{Cacciari:2011ma} with the anti-$k_T$ algorithm~\cite{Cacciari:2008gp} using the distance parameter, $R = 0.4$.
For $b$-tagged jets $p_\text{T} > 25$\,GeV within $|\eta| < 2.4$ were used.
Only signal processes were generated for study; all contributing backgrounds were extracted from the relevant ATLAS and CMS papers, as referenced in the text.

We have considered two production mechanisms of $H$ to be studied:
\begin{itemize}
\item Direct production through $gg$F: $g g \to H \to S h$.
\item Top associated production: $p p \to t H + \bar t H,~t\bar t H$, and further $H \to S h$.\footnote{The $H$ decays to pairs of vector bosons has been neglected, since the BRs are small and the effect on the final fit result would be negligible.}
\end{itemize}

\subsection{Direct production of $H$}
\label{ppHhs}

As mentioned above, we assume that the $H$ is produced dominantly through $gg$F, and has a 100\% BR to $Sh$.
Thereafter, $h$ is allowed to decay to all possible modes according to the available phase space.
$S$, on the other hand, decays according to the two scenarios discussed in \autoref{sec:formalism}.
Firstly, through SM-like Higgs boson decay modes, $S$ decays into all possible SM final states.
Secondly, $S$ decays through $N_i$ to leptons and missing energy (\MET) in the heavy neutrino model (as shown in \autoref{fig:feynHN}).

By considering $S$ to be a Higgs-like scalar boson, the number of free parameters is drastically reduced.
For this reason, we consider the Higgs-like $S$ model for comparisons to data and fitting the masses of $H$ and $S$.
This choice also enhances the number of leptons via the decay of $S$, since for a Higgs-like scalar boson with a mass near $2\,m_W$, the BR for $S \to WW$ becomes dominant.
Thereafter, the $W$ bosons in the final state can decay leptonically to provide a unique signature for the process.
In contrast with this, the heavy neutrino model produces $N_i$, which naturally give leptonic signatures either through charged or neutral leptons in association with $W$ and $Z$ bosons.
We also note that for $m_{N_i} > m_S$ the $N_i$ become off-shell, and this affects the features of kinematic observables, as opposed to the case where $m_{N_i} < m_S$.

\subsubsection{Di-lepton final states}
\label{sec:2l}

For di-leptonic final states, there are two natural signatures which arise for event selections based on the charges of the leptons ($\ell$).
That is, opposite-sign (OS) and same-sign (SS) di-leptons.
In this article we have only considered $e^\pm$ and $\mu^\pm$ as leptons, barring possibilities from $\tau^\pm$ leptons.
In the OS selection we have required $e^\pm e^\mp,~\mu^\pm \mu^\mp$ or $e^\pm \mu^\mp$ pairs, and for SS pairs we have required $e^\pm e^\pm,~\mu^\pm \mu^\pm$ or $e^\pm \mu^\pm$.
The event selection based on kinematic cuts is the same for both cases, and is as follows:
\begin{itemize}
\item Exactly two OS or SS leptons,
\item Absolute pseudo-rapidities: $|\eta^{\mu^\pm}| < 2.4$, $|\eta^{e^\pm}| < 2.47$, excluding $1.37 < |\eta^{e^\pm}| < 1.52$ to align with detector acceptances,
\item Transverse momentum of leading (sub-leading) leptons: $p_\text{T}^\ell > 25 (20)$~GeV,
\item $|\eta^j| < 4.5$, transverse momentum of jets: $p_\text{T}^j > 25$~GeV, and
\item Total transverse missing energy: $\MET > 20$~GeV.
\end{itemize}

Based on the above requirements, the basic kinematics are shown in \autoref{fig:dl1}, \autoref{fig:dl2} and \autoref{fig:dl3} for the mass point of $m_H=270$~GeV, $m_S=145$~GeV and $m_N=1$~TeV.
An SM prediction of Higgs boson production through $gg$F is also shown for the OS di-leptonic events, whereas for the SS selection the SM acceptance was far too low to generate a reasonable amount of statistics.
Even if the SM can produce a few SS di-leptonic events, they are constructed from fake leptons and should be ignored.

A number of the features seen in \autoref{fig:dl1}, \autoref{fig:dl2} and \autoref{fig:dl3} deserve discussion.
As opposed to the production of the SM Higgs boson via $gg$F, the direct production of di-leptons with $H\rightarrow Sh$ displays a significantly larger jet multiplicity.
Di-leptons are studied at ATLAS and CMS within the context of $h$-related measurements, and the measurement of the non-resonant $WW$ production by applying restrictions on the number of hadronic jets and $b$-tagged jets.
In this light it would be interesting to study the di-lepton invariant mass distributions in events with more than one hadronic jet, and with the application of a veto on $b$-tagged jets.
OS di-leptons display the feature that about a third of the events contain at least one $b$-tagged jet, where top-related backgrounds contribute strongly (see \autoref{sec:2lcomp}).
The production mechanism of $S$ suggested here is distinct from the potential direct production via $gg$F, due to the significantly enhanced jet activity and the associated production of $b$-tagged jets.

\subsubsection{Comparison with ATLAS and CMS {\rm OS} di-lepton invariant mass spectra}
\label{sec:2lcomp}

The BSM scenario discussed in this article is of particular interest with regards to several ATLAS and CMS SM measurements of di-lepton invariant mass spectra, since the BSM OS di-leptonic signature has some sensitivity to them. For this reason, various 8~TeV ATLAS and CMS results were selected for study.
The BSM signal was generated and reduced according to the event selections prescribed in the experimental publications, and then added to the SM prediction. The best fit event yield was calculated by varying the BSM normalisation and fitting the SM+BSM prediction to the data, using a likelihood fit in the \textsc{RooFit} framework. A list of the measurements that were considered can be seen in \autoref{tab:dlfits}. The SM MCs are scaled to the data in the region of the di-lepton mass, $m_{ll}>110$~GeV, where the signal is expected to be negligible (see \autoref{fig:dl3}).
The data with $m_{ll}<100$~GeV is compared to the re-scaled SM MC plus the BSM signal by means of a $\chi^2$ function, which is defined in \ref{app1}.
The mass of the heavy scalar is set to $m_H=270$~GeV and the mass of $S$ is varied.

As discussed in \autoref{sec:formalism}, the cross section of the BSM signal is proportional to $\beta_g^2$, which is a free parameter.
Therefore, from the fits to the invariant mass spectra, the value of $\beta_g^2$ can be calculated along with its standard deviation.
In addition to this, a test statistic can be used to determine the significance of the BSM signal's improvement with respect to the SM-only hypothesis.
The test statistic that was considered is $\Delta\chi^2\equiv\chi^2_\text{SM} -\chi^2_\text{SM+BSM}$.
The significance of the fit can be estimated as the square root of this quantity.
The combined best fit is obtained for $m_S=150\pm5$~GeV.
The results of the individual fits are compared with the prediction for the direct production of $H\rightarrow Sh$.
The cross section is estimated by using a combined value of $\beta_g^2=1.38\pm0.32$.
This number is obtained from the different Higgs boson signal strengths reported so far (see Ref.~\cite{Mellado_HDAYS2017}) and the analysis of the multi-lepton data in association with $b$-tagged jets performed here (see \autoref{sec:tth}).
The results are reported in \autoref{tab:dlfits}, and distributions of the resulting BSM spectrum overlaying the data are shown in \autoref{Fig:12} and \autoref{Fig:13}.

Rescaling the SM MC to the data in the region of $m_{ll}>110$~GeV has the advantage of reducing the systematic uncertainties on the normalisation of the SM background.
The analysis of the results reported in Ref.~\cite{Aaboud:2017ujq} is most sensitive to the assumed systematic uncertainty, while the other measurements are dominated by statistical uncertainty, and are therefore largely insensitive to the assumed systematic uncertainty.
A systematic error of 2\% is assumed for the analysis of Ref.~\cite{Aaboud:2017ujq}, correlated over all bins and without any constraints.
The systematic uncertainty is not incorporated in the likelihood fit; it was found that the far more conservative approach was to use the combined systematic and statistical uncertainty in the calculation of the $\chi^2$.
This is because the systematic terms in the likelihood are heavily constrained by the tail of the distribution, and so the effect of systematics becomes negligible in the likelihood fit.
Therefore, all the best fit parameters in \autoref{tab:dlfits} were calculated with a likelihood and the significance of those fits can be deduced through the $\Delta\chi^2$ test statistic.

A combined fit was performed with a common value of $\beta_g^2$.
This tests the ability of the simplified model to describe the data as a whole.
The best fit obtained here corresponds to $\beta_g^2=1.22\pm0.38$.
The value of $\beta_g^2$ obtained in this section is in good agreement with the combined value discussed above.
It should be noted that this best fit value does not account for the systematics in the SM distributions, as discussed above.

\begin{table}
\centering
\begin{tabular}{lccccc}
\hline
\textbf{Measurement} & \textbf{Ref.} & \begin{tabular}[c]{@{}c@{}}\textbf{BSM event yield} \\ ($\beta_g^2=1.38\pm0.32$)\end{tabular} & \begin{tabular}[c]{@{}c@{}}\textbf{BSM event yield} \\ (best fit)\end{tabular} & \textbf{Best fit $\beta_g^2$} & $\Delta\chi^2$ \\ \hline
\begin{tabular}[c]{@{}l@{}}ATLAS, 20.2 fb$^{-1}$ \\ $e^{\pm}\mu^{\mp}$ \\ $N_{b\text{-jet}}\geq 1$\end{tabular} & \cite{Aaboud:2017ujq}
& $112\pm26$ & $397\pm93$ & $4.89\pm1.15$ & $12.1$ \\ \hline
\begin{tabular}[c]{@{}l@{}}ATLAS, 20.3 fb$^{-1}$ \\ $e^{\pm}\mu^{\mp}$ \\ $N_\text{jet}= 0$\end{tabular} & \cite{Aad:2016wpd}
& $28\pm6$ & $48\pm46$ & $2.37\pm2.27$ & $0.43$ \\ \hline
\begin{tabular}[c]{@{}l@{}}ATLAS, 20.3 fb$^{-1}$ \\ $e^+ e^-,~\mu^+\mu^-$\\ $N_\text{jet}= 0$\end{tabular} & \cite{Aad:2016wpd}
& $16\pm4$ & $82\pm20$ & $7.07\pm1.73$ & $7.31$ \\ \hline
\begin{tabular}[c]{@{}l@{}}ATLAS, 20.3 fb$^{-1}$ \\ $e^{\pm}\mu^{\mp}$ \\ $N_\text{jet}= 1$\end{tabular} & \cite{Aaboud:2016mrt}
& $70\pm16$ & $20\pm36$ & $0.39\pm0.71$ & $0.16$ \\ \hline
\begin{tabular}[c]{@{}l@{}}CMS, 19.4 fb$^{-1}$ \\ $e^+ e^-,~\mu^+\mu^-,~e^{\pm}\mu^{\mp}$ \\  $N_\text{jet} = 0$\end{tabular} & \cite{Khachatryan:2015sga}
& $46\pm11$ & $136\pm58$ & $4.08\pm1.74$ & $3.31$ \\ \hline
\begin{tabular}[c]{@{}l@{}}CMS, 19.4 fb$^{-1}$ \\ $e^+e^-,~\mu^+\mu^-,~e^{\pm}\mu^{\mp}$ \\  $N_\text{jet} = 1$\end{tabular} & \cite{Khachatryan:2015sga}
& $111\pm26$ & $46\pm43$ & $0.57\pm0.53$ & $0.58$ \\ \hline
\begin{tabular}[c]{@{}l@{}}CMS, 5.3 fb$^{-1}$ \\ $e^+e^-,~\mu^+\mu^-,~e^{\pm}\mu^{\mp}$ \\  $N_\text{jet}\geq 2,~N_{b\text{-jet}}\geq 2$\end{tabular} & \cite{Chatrchyan:2013faa}
& $25\pm6$ & $17\pm58$ & $0.94\pm3.20$ & $-0.04$ \\ \hline
\end{tabular}
\caption{BSM fit results for the Run 1 di-lepton invariant mass spectra reported by ATLAS and CMS at $\sqrt{s}=8$~TeV.
The event yields reflect the number of BSM events to fit the data in excess of the SM prediction, first using the previously calculated value of $\beta_g^2$ and then at the best fit point for the measurement in question.
The value of $\beta_g^2$ corresponding to each measurement's fit is reported along with the test statistic $\Delta\chi^2$, the square root of which gives an estimate of the fit significance (incorporating systematics).
The negative value of $\Delta\chi^2$ for the last measurement arises through the difference in fitting with a likelihood and then calculating the test statistic with a $\chi^2$.
The mass of the heavy scalar is fixed at $m_H=270$~GeV and the mass of $S$ is fixed at its best fit point, $m_S=150$~GeV.
}
\label{tab:dlfits}
\end{table}

A few remarks are in order here. A simplified assumption is made here whereby the BSM contribution arises solely from the direct production of $H\rightarrow Sh$.
A non-negligible contribution from $H\rightarrow hh$ is expected, as detailed in Refs.~\cite{vonBuddenbrock:2015ema,Kumar:2016vut,vonBuddenbrock:2016rmr}.
This contribution can impact the di-lepton invariant mass distributions.
The analysis of the complete set of results with the data taken in 2016 is not yet available.
As a result, the $hh$ decay is not taken into account here.
The contribution from OS di-leptons in the associated production of $H$ (see \autoref{sec:tth}) is neglected here.
The contribution from the decay $H\rightarrow WW\rightarrow \ell\nu\ell\nu$ is also neglected.
As the mass of $H$ becomes less than $m_h+m_S$, the di-boson decay includes off-shell effects.
This is expected to have a non-trivial impact on the expected yields of the di-leptons and hadronic jets, including $b$-tagged jets.
It can be anticipated that this effect will improve the self-consistency for the interpretation of the di-lepton data with the simplified model.
This effect is not studied in this work and will be taken into account when more data is available.

It is relevant to note that the discrepancy in the di-lepton invariant mass reported in Ref.~\cite{Aaboud:2017ujq} appears to be correlated with the di-lepton azimuthal angle difference reported in Ref~\cite{Aad:2014mfk} in the context of the $t\overline{t}$ spin correlation analysis with the same dataset.
This indicates that the apparent discrepancy between the data and the MC in the di-lepton azimuthal angle difference is correlated with low di-lepton invariant mass events.
The CMS experiment reported the same analysis with the Run 1 dataset in Ref.~\cite{Khachatryan:2016xws}, where the di-lepton azimuthal angle difference is very similar to that reported by the ATLAS experiment in Ref.~\cite{Aad:2014mfk}.
Ref.~\cite{ATLAS:2018gcr} reports results using Run 2 data with the same final state but with exactly one $b$-tagged jet (see Figure 7d in the publication) where the top quark related backgrounds need to be normalised to the data in the transverse mass range $m_\text{T}>200$~GeV (following a similar procedure performed here).
The size of the discrepancy in Ref.~\cite{ATLAS:2018gcr} is compatible with what the simplified model used here predicts (with $m_H=270$~GeV and $m_S=150$~GeV), after taking the event selection into account.

\subsubsection{Tri-lepton final states}
\label{sec:3l}
Another important final state to check is the production of three leptons with a total charge of $\pm1$, since the prediction of  the model is not excluded by the existing data.
We have studied tri-lepton final states with the following event selection criteria:
\begin{itemize}
\item Exactly three leptons, two of them must be an OS and same-flavour (SF) lepton pair ($eee, \mu\mu\mu,
e\mu\mu$ or $ee\mu$),
\item $|\eta^{\mu^\pm}| < 2.5$, $|\eta^{e^\pm}| < 1.37$ or $1.52 < |\eta^{e^\pm}| < 2.47$ to align with detector acceptances,
\item $p_\text{T}^\ell > 15$~GeV,
\item $p_\text{T}^j > 25$~GeV, $|\eta^j| < 4.5$.
\end{itemize}

Kinematic distributions for key variables in the tri-lepton final state can be seen in \autoref{fig:tl}.
The differences between the SM and BSM predictions are not as stark as in the di-lepton case.
Arguably the best discriminating variables in this case are the transverse mass variables of the $W$ and $WZ$ systems (defined in \ref{app:directprod}).

\subsubsection{4-lepton final states}
\label{sec:4l}

The production of four $W$s from a resonance is a unique signature leading to the production of four isolated charged leptons with \MET. The LHC experiments have not reported on this signature to date, though due to low backgrounds one can expect an excellent signal to background ratio. In this channel, fake leptons are also under control and could result in a unique signature of the BSM scenario considered in this article.
A kinematic study on this final state with the $H$ mass dependence has been done in the Ref.~\cite{vonBuddenbrock:2016rmr}. In this section we use this channel to study CP-properties of $S$. The mass of the intermediate boson $S$ is constrained with the di-lepton data as shown in \autoref{sec:2lcomp}. It is also important to characterise $S$ with respect to CP properties, and for that we follow the approach used in Refs.~\cite{Biswal:2012mp,Alloul:2013naa,Kumar:2015kca}.

The most general Lagrangian which can account for possible $SWW$ couplings relevant for the phenomenology of $S$ at the LHC are the three-point and four-point interactions involving at least one $S$ boson field. It can be written as,\footnote{Note that we modify the convention for the $SWW$ vertex (given in \autoref{lsvv})
to be consistent with the studies in Refs.~\cite{Biswal:2012mp,Alloul:2013naa,Kumar:2015kca}.}
\begin{align}
 {\cal L}^{(3)}_{_{SWW}}
 =
 - g \bigg[ & \dfrac{\gonehww}{2m_W} W^{\mu\nu} W^\dag_{\mu\nu} S
 + \dfrac{\gtwohww}{m_W} ( W^\nu \d^\mu W^\dag_{\mu\nu} S + {\rm h.c} )
 \notag\\
 &\quad+  \dfrac{\gthww}{2m_W} W^{\mu\nu} \widetilde W^\dag_{\mu\nu} S \bigg].
 \label{lag3}
\end{align}
Here $g^{(i)}_{(\cdots)}$ ($i = 1,2$) and $\tilde g_{(\cdots)}$ are the real coefficients corresponding to the respective CP-even and CP-odd couplings of the $SWW$ anomalous vertices, $W_{\mu\nu} = \d_\mu W_\nu - \d_\nu W_\mu$ and $\widetilde W_{\mu\nu} = \frac{1}{2} \e_{\mu\nu\rho\sigma} W^{\rho\sigma}$.
The Lorentz structures of \autoref{lag3} can be derived from the $SU(2)_\text{L}\otimes U(1)_Y$ gauge invariant dimension six operators given in Ref.~\cite{Alloul:2013naa}.

The complete Lagrangian we have worked with is as follows,
\begin{equation}
 {\cal L}
 =
 {\cal L}_{\rm SM} + {\cal L}^{(3)}_{_{SWW}}.
 \label{lagsww}
\end{equation}
The most general effective vertices take the form,
\begin{align}
 &
 \Gamma_{SW^-W^+}
 = g m_W
   \bigg[\bigg\{1 + \dfrac{\gonehww}{m^2_W} p_2 \cdot p_3
                + \dfrac{\gtwohww}{m^2_W} (p^2_2 + p^2_3)
         \bigg\} \eta^{\mu_2 \mu_3}
 \notag\\
 & \qquad\,\qquad\quad\quad
        - \dfrac{\gonehww}{m^2_W} p_2^{\mu_3} p_3^{\mu_2}
        - \dfrac{\gtwohww}{m^2_W} (p_2^{\mu_2} p_2^{\mu_3} + p_3^{\mu_2} p_3^{\mu_3})
 \notag\\
 & \qquad\,\qquad\quad\quad
        - i \dfrac{\gthww}{m^2_W} \e_{\mu_2\mu_3\mu\nu} p_2^\mu p_3^\nu
   \bigg].
\label{cplag}
\end{align}

In \autoref{fig:fl}, a few kinematic distributions can be seen for the process $gg \to H \to Sh$, where $h \to WW$ and using the above Lagrangian $S \to WW$.
The effect of CP-even and CP-odd $SWW$ couplings are studied by setting \gonehww, \gtwohww and \gthww = $\pm 1$, where we consider one BSM coupling at a time, keeping the others to be SM-like.
From the kinematic distributions in \autoref{fig:fl}, the only apparent discriminating variable sensitive to the CP nature of the $S$ boson can be seen in $\Delta\phi_{\ell_1\ell_2}$; the rest are almost insensitive to the CP structure of the  couplings. The largest deviation is observed when the parameter $g^{(1)}_{SWW}=+1$. This moderately affects observables, such as the missing transverse energy, the azimuthal angle between leptons and the invariant mass of the four-lepton system.


\subsection{Production of $H$ in association with top quarks}
\label{sec:tth}

\begin{figure}[t]
  \centering
  \subfloat[]{\includegraphics[width=0.4\textwidth]{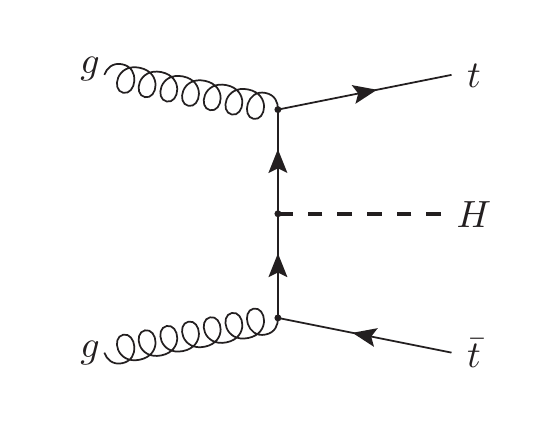}\label{fig:tth}}
  \subfloat[]{\includegraphics[width=0.4\textwidth]{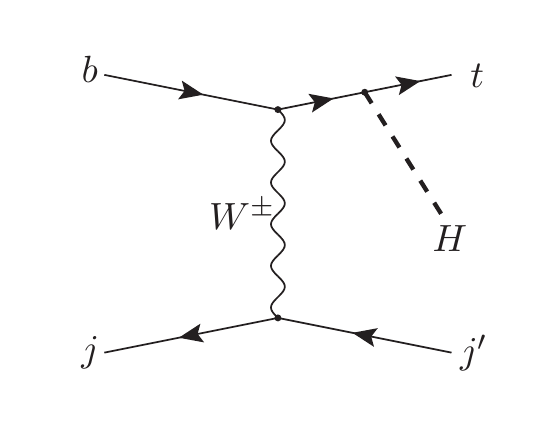}\label{fig:th}}
  \caption{The leading order Feynman diagrams for (a) $ttH$ and (b) $tH$ production.}
\label{fig:feyn2}
\end{figure}

While the assumed dominant production mechanism for $H$ is $gg$F, it should be noted that the same theoretical vertices required for $gg$F can be used to construct processes for $H$ production in association with top-quarks ($ttH$) that have a non-negligible cross section.\footnote{Since $H$ production in $gg$F
mode is possible through one-loop only, where the dominant contribution is due to top-quark Yukawa vertices, $y_{ttH}$.}
This has tantalising prospects, since both Run~1 and Run~2 searches for top associated Higgs boson production ($tth$) have shown deviations from the SM in leptonic channels.
In Ref.~\cite{vonBuddenbrock:2017bqf} it is shown that the signal strength of $tth$ using $\mu_{tth} = \sigma_{tth}^{\rm obs}/\sigma_{tth}^{\rm SM}$, by combining the experimental search results in leptonic final state yields a value of $\mu_{tth} = 1.92 \pm 0.38$.
It should be noted that since this combination was made, updated $tth$ results reported by ATLAS~\cite{Aaboud:2017jvq} and CMS~\cite{Sirunyan:2018shy} in the search for $tth$ with leptonic channels have been released.
The deviations in the new data are not as strong as those used to construct the combination of $\mu_{tth}$ shown above, but they still show a systematic enhancement of the $tth$ cross section.\footnote{The ATLAS result has the added caveat of a veto on events with 3 or more $b$-tagged jets, which is a significant signal for the model studied in this paper. It is therefore biased towards the SM and would not provide much sensitivity towards constraining any BSM physics that predicts more than 2 $b$-tagged jets in multi-lepton final states.}
We have therefore considered the production of $ttH$ in order to understand whether it can shed light on the $tth$ deviations.
Here again we have looked at the $H \to Sh$ decay mode only, such that the leptonic signatures searched for in several existing experimental $tth$ search channels are mimicked.

Typically in $tth$ searches, signal contributions can come from the associated production of a Higgs boson with one ($th$) or two ($tth$) top quarks.
The same is true for $H$, and hence $H$ production with single-top and double-top are labelled as $tH$ and $ttH$ here onwards.\footnote{Note here that we consider a 5 flavour proton, so the initial state for $tH$ involves a $b$ quark coming from the proton.}
The dominant Feynman diagrams for $ttH$ and $tH$ are shown in \autoref{fig:feyn2}.
Note that in the SM, due to negative interference between the Yukawa couplings ($h$-$t$-$\bar{t}$ vertex) and the Higgs boson couplings with vector bosons ($h$-$V$-$V$ vertex), the cross section of $th$ is significantly smaller than that of $tth$ production.
But this is not true for $tH$ production, since the $HVV$ couplings ($\beta_V$) are assumed to be small. By following the logic suggested in Ref.~\cite{Farina:2012xp}, it turns out~\cite{vonBuddenbrock:2015ema}:
\begin{equation}
\sigma_{tH} \simeq \sigma_{ttH}.
\end{equation}
This suggests that both $tH$ and $ttH$ production cross sections are non-negligible in the BSM scenario considered in this article.
To obtain $\sigma_{ttH} (m_H)$, the corresponding cross section for $\sigma_{tth} (m_h = m_H)$ can be taken from the CERN yellow book~\cite{deFlorian:2016spz}, and scaled by the square of the parameter $\beta_g$ accordingly.

In this article we have studied the comparison of the production mechanisms $ttH$ and $tH$ with the CMS Run~2 search for a single top quark in association with a Higgs boson~\cite{CMS:2017uzk}. While the search is branded as a single top search, it is sensitive both to single and double top production in association with $H$. In this reference CMS provides an unbiased sample with a simple event selection that suits our study well:
\begin{itemize}
  \item No lepton pair with $m_{\ell\ell} < 12$~GeV,
  \item $N_{b\text{-jet}}\geq1$,
  \item $N_{\text{jet}}\geq1$ (not including $b$-tagged jets).
\end{itemize}
Only one mass point has been considered in this study, since the kinematic distributions do not have a significant dependence on the masses of $H$ and $S$.

\begin{figure}
        \centering
        \subfloat[]{\includegraphics[width=0.49\textwidth]{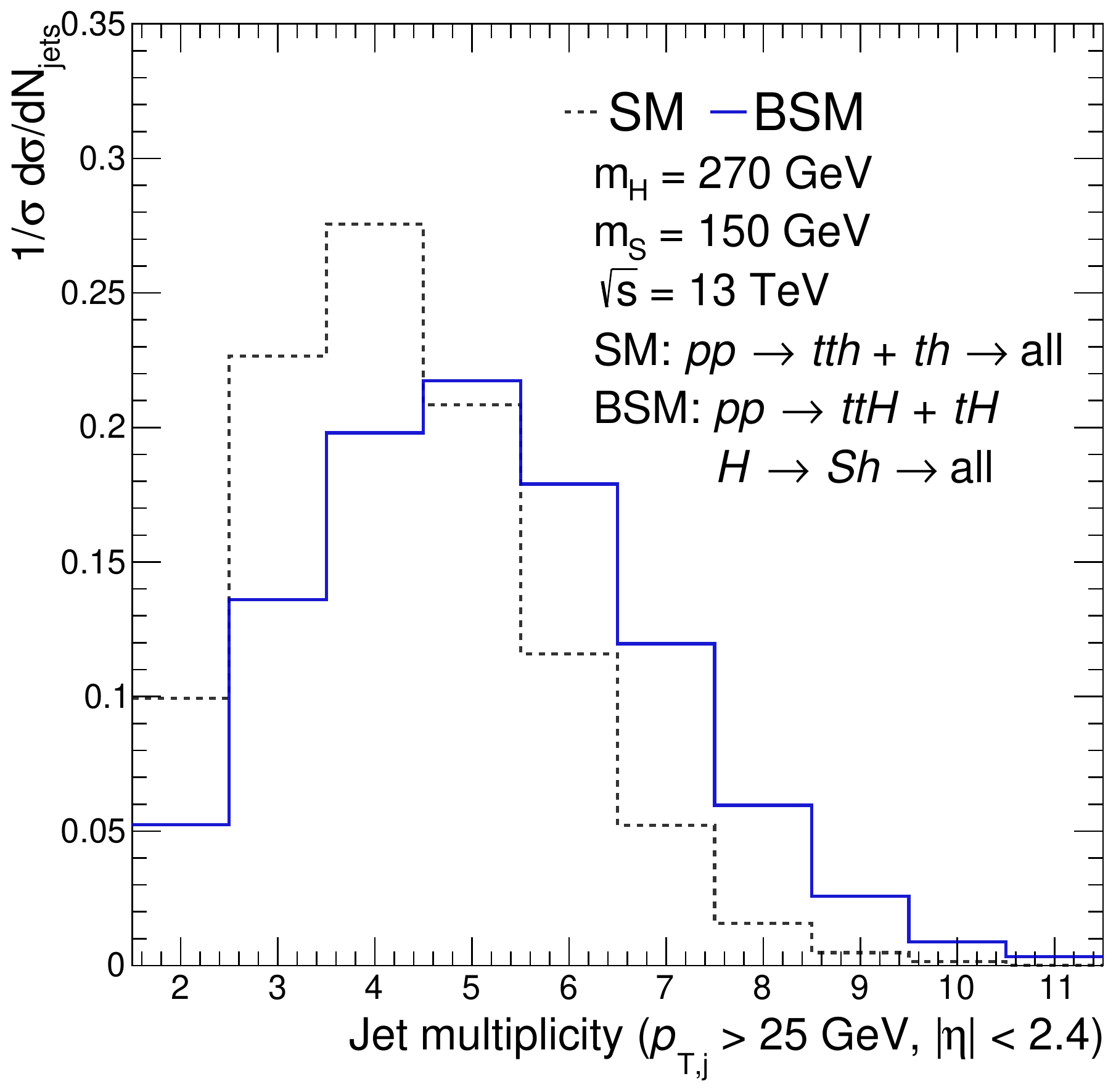}}
        \subfloat[]{\includegraphics[width=0.49\textwidth]{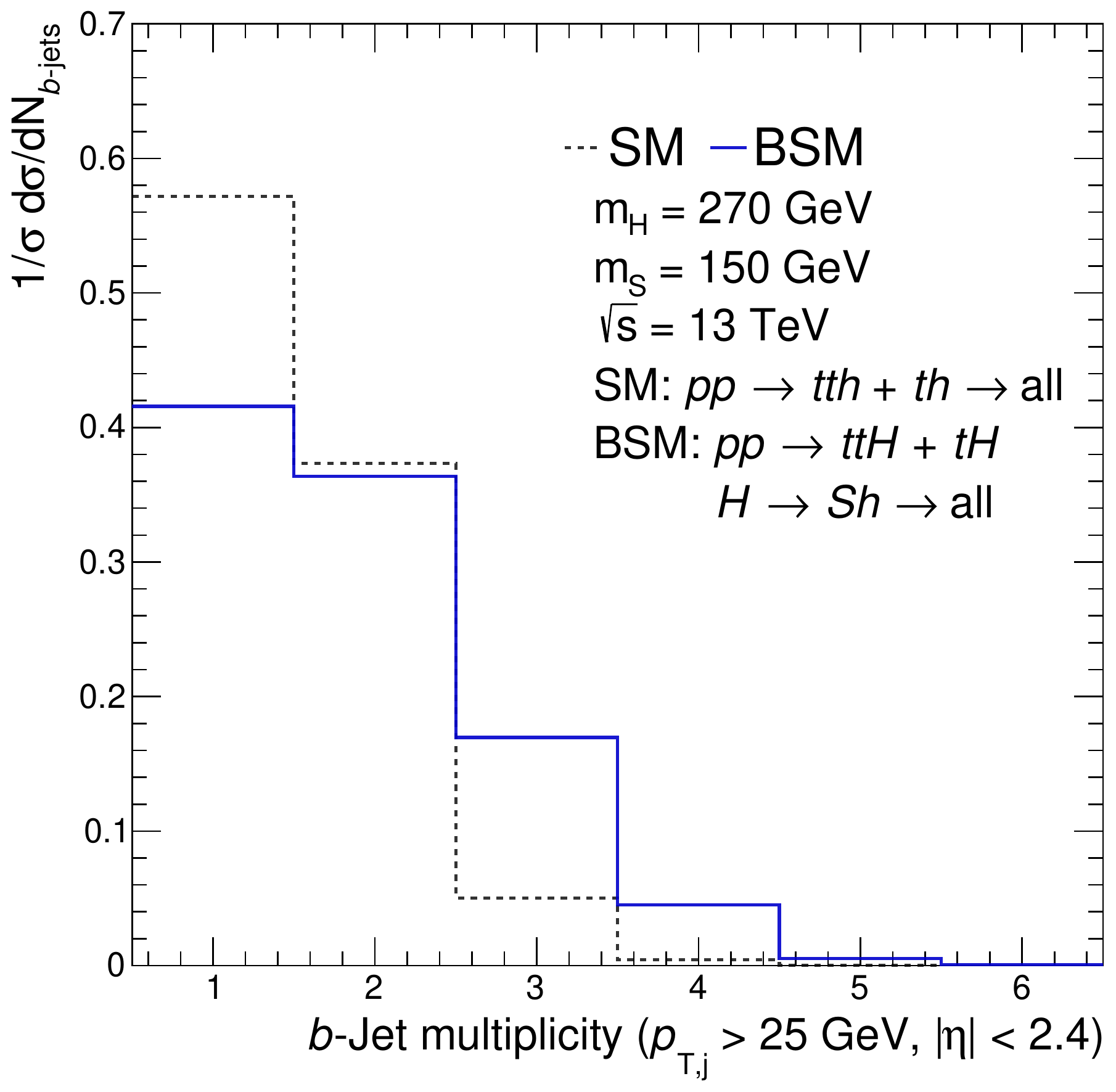}} \\
        \subfloat[]{\includegraphics[width=0.49\textwidth]{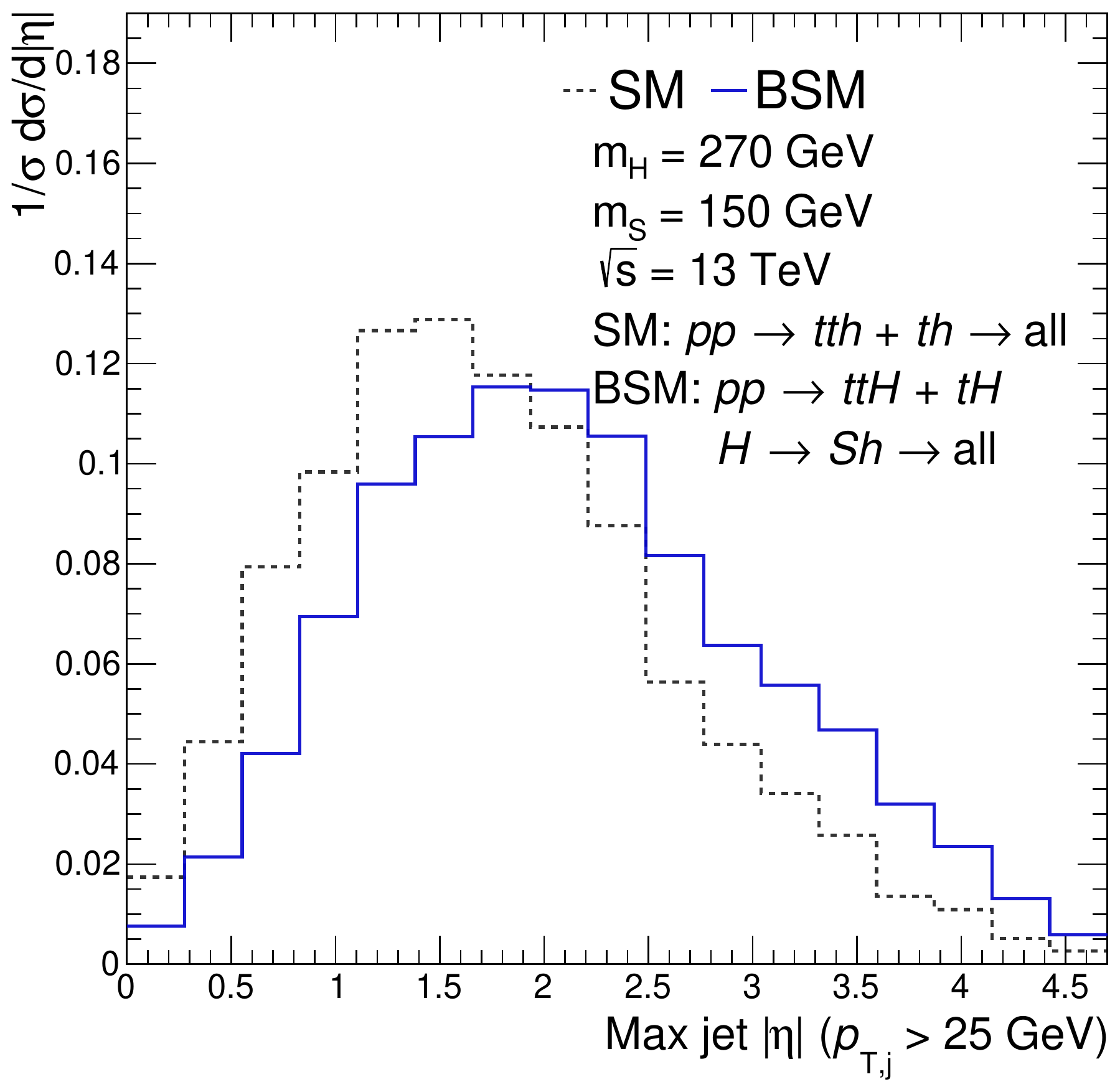}}
        \subfloat[]{\includegraphics[width=0.49\textwidth]{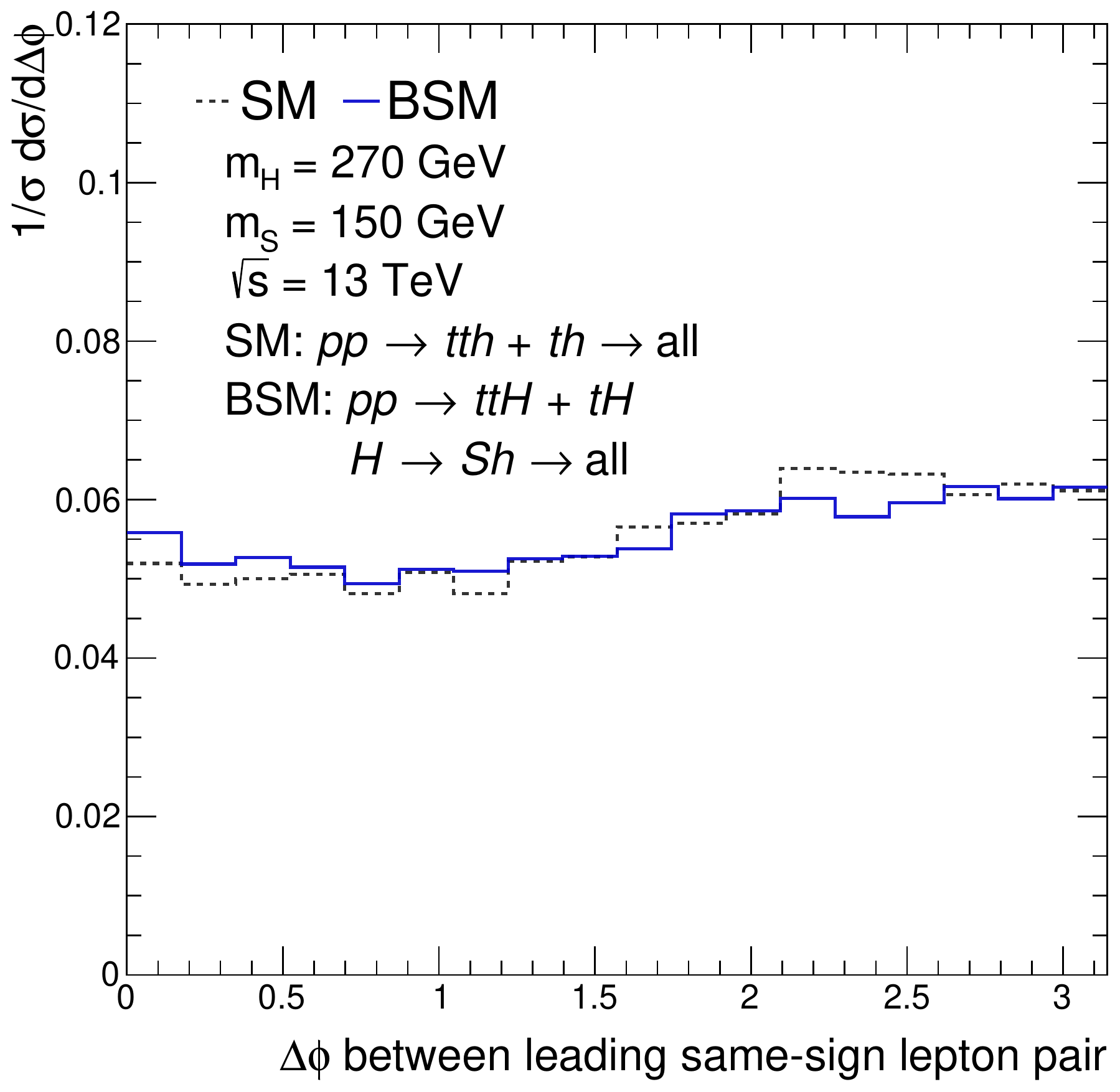}}
        \caption{\label{Fig:2} Comparisons between the SM and BSM predictions of a Higgs boson in association with one or two top quarks. The BSM process considered is $pp\to ttH,~tH$ where $H\to Sh$. The distributions have been normalised so that their shapes can be compared. The BSM mass point considered is $m_H=270$~GeV and $m_S=150$~GeV, as in \autoref{sec:2lcomp}.}
\end{figure}

The plots in \autoref{Fig:2} show comparisons between the SM and BSM prediction for different observables after the application of the event selection of the CMS Run~2 result.
A number of differences are worth discussing.
The significant contribution of $tH$ production tends to lower the hadronic jet activity.
On the flip side, the decay $H\rightarrow Sh$ enhances the hadronic jet activity.
As a result of these two competing effects the jet multiplicity is enhanced somewhat.
The fraction of events with at least three $b$-tagged jets is enhanced significantly.
The fraction of events with at least three $b$-tagged jets increases from about 6\% in the SM case to over 20\% in the scenario considered here.
This is mostly due to the fact that $b$-tagged jets can come from the decay of both the Higgs boson and the top quarks.
The production of SS di-leptons with at least three jets is about 24\%, whereas for three and more leptons it is about 19\%.
This effect opens new corners of the phase space for detailed exploration by the experiments.
In particular, the production of SS di-leptons and three leptons in association with at least three $b$-tagged jets is a rare process in the SM.

The presence of a significant contribution from the $tH$ production forces the leading hadronic jet to be more forward, as seen in \autoref{Fig:2}.
The distribution of the azimuthal angle difference between the leading SS pair shows that in both cases the leptons appear almost uncorrelated as they originate from the decays of different particles.

Beyond the event pre-selection, the CMS search further categorises events into three mutually exclusive categories.
Firstly, events with exactly two SS leptons are grouped together, and are subdivided into $e\mu$ and $\mu\mu$ categories.
The third category contains events with exactly three leptons.

\subsubsection{Same-sign 2-leptons}
\label{sec:ss2lep}
The SS 2 lepton categories select exactly 2 leptons of the same charge with the following criteria:
\begin{itemize}
  \item $\ell\ell=e\mu$~\text{or}~$\mu\mu$,
  \item Leading lepton $p_\text{T} >$ 25~GeV,
  \item Sub-leading lepton $p_\text{T} >$ 15~GeV.
\end{itemize}

These categories are of chief importance in this work, since their signal to background ratios and selection acceptances are far greater than the tri-lepton category.

\subsubsection{Tri-leptons}
\label{sec:trilep}

The tri-lepton category has a far smaller selection acceptance and, therefore, is not as sensitive in this study, however its contribution is still considered.
The selection criteria are as follows:
\begin{itemize}
  \item Exactly 3 leptons,
  \item Leading lepton $p_\text{T} >$ 25~GeV,
  \item Second and third lepton $p_\text{T} >$ 15~GeV,
  \item No lepton pair with $\left|m_{\ell\ell}-m_Z\right|<15$~GeV.
\end{itemize}

\subsubsection{Comparison with CMS data}
\label{sec:compare}

Using a MC prediction of $ttH$ and $tH$ production, we are able to constrain the parameter $\beta_g$ by making fits to the CMS data~\cite{CMS:2017uzk}.
Recall that $\beta_g$ is a re-scaling of the Yukawa coupling, and therefore the cross section for both $ttH$ and $tH$ productions are proportional to $\beta_g^2$.

The CMS search mentioned above is performed using a multivariate technique making use of boosted decision trees, and so a direct comparison to their final results is difficult. However, their paper also contains a few key distributions based on the event pre-selection and categorisation mentioned above. Again, only one mass point was considered for the BSM prediction: $m_H=270$~GeV and $m_S=150$~GeV. Here we have considered only the Higgs-like $S$ model for fits to the data. The SM background and signal predictions were taken directly from the CMS publication.

The three discriminating variables given in the CMS publication are the largest absolute pseudo-rapidity of any jet in the event, the azimuthal separation between the leading SS lepton pair, and the jet multiplicity.
The BSM predictions were scaled to the appropriate SM-like cross sections, and then fit to the CMS data by varying the normalisation (i.e. $\beta_g^2$). By inspection of \autoref{Fig:2}, it can be seen that the number of $b$-tagged jets is an important discriminant between the SM and BSM predictions of the search. However, since CMS did not show any distributions for the $b$-tagged jet multiplicity, arguably the next best discriminant is the jet multiplicity, regardless of the flavour tagging. Therefore, we considered the jet multiplicity to be the distribution to which the BSM event yield can be fit. The total number of observed events in excess of the SM prediction for $N_\text{jet}\geq3$ (since the 2 jet bin is saturated by the SM and offers an insignificant contribution to the bulk of the BSM signal) was calculated and used to provide an estimate of the value of $\beta_g^2$, along with an error that incorporates both statistical and systematic uncertainties from the SM prediction and the data.

The results of this process can be seen in \autoref{tab:fit}.
An error weighted mean is calculated for the three sets of spectra, where the combined value of $\beta_g^2=1.69\pm0.54$ is obtained.
This result is consistent with the value of $\beta_g^2=1.22\pm0.38$ obtained in \autoref{sec:2lcomp} using OS di-leptons.
Plots showing the SM and BSM contributions to the fit can be seen in \autoref{Fig:tth_dilepton} for the di-lepton categories and \autoref{Fig:tth_trilepton} for the tri-lepton category.

 \begin{table}
   \small
  \centering
  \begin{tabular}{ccc}
    \hline
    {\bf{Channel}}	& \textbf{Best fit BSM event yield} & $\beta_g^2$ \\ \hline
    $e\mu$	& $37.04\pm12.10$ & $3.03 \pm 0.99$ \\
    $\mu\mu$ & $37.22\pm17.52$ & $4.25 \pm 2.00$ \\
    Tri-lepton & $6.00\pm5.52$ & $0.75 \pm 0.69$ \\ \hline
    {\bf{Combined}}	& & $1.69 \pm 0.54$ \\ \hline
  \end{tabular}
  \caption{The number of BSM candidate events and the corresponding values of $\beta_g^2$ for each channel in the CMS Run~2 search in Ref.~\cite{CMS:2017uzk}.
  The combined result is calculated as the error weighted mean of the individual values calculated for each channel.}
  \label{tab:fit}
\end{table}

This study suggests that the evaluation of the production of multiple leptons in association with $b$-tagged jets in the framework of the search for $tth$ alone appears insufficient. While the extracted signal strength $\mu_{tth}$ seems elevated, this measurement on its own does not seem to fully characterise potential discrepancies between the data and the SM in these final states. As illustrated here, the recent measurement from CMS yields $\mu_{tth}=1.5\pm0.5$, which does not seem to be a compelling deviation from unity on its own. However, when one evaluates the rate of leptons with $b$-tagged jets (as detailed here) the discrepancy with the SM increases, as seen from the combined result in \autoref{tab:fit}.
Here it is suggested that the experiments report on the production of di-leptons and tri-leptons in association with more than two $b$-tagged jets. In the context discussed here it is very important to understand the potential correlation between the elevated rate of SS di-leptons and tri-leptons observed with 1 or 2 $b$-tagged jets, compared to $N_{b\text{-jet}}\geq 3$. This study goes beyond the measurement of $\mu_{tth}$.
\section{Summary and conclusions}
\label{sec:conc}

The BSM scenario presented in this article introduces the scalars $H$ and $S$ in order to explain a number of features of the Run 1 data, which persist in the Run 2 data. Here we elaborated on the phenomenology discussed in Refs.~\cite{vonBuddenbrock:2015ema,Kumar:2016vut,vonBuddenbrock:2016rmr} by discussing various multi-lepton final states. This includes the production of SS and OS di-leptons and three leptons. We have considered Higgs-like decays of $S$ as well as the leptonic decays via heavy neutrinos. We study the impact of the CP structure of the $SWW$ couplings on the kinematics of leptons in the decay $H\rightarrow Sh \rightarrow 4W \rightarrow 4l$; this is found to be small. The production of SS di-leptons and tri-leptons from the production of $H$ in association with single and double top quarks is also studied.

Available data is compared against predictions from a simplified scenario.
This includes the production of OS di-leptons with \MET in conjunction with jets and $b$-tagged jets, and the production of SS di-leptons and tri-leptons in association with $b$-tagged jets.
In general, the inclusion of the simplified BSM scenario (where $H$ decays only to $Sh$) improves the description of the data compared to the SM-only hypothesis; the extent of which is demonstrated by the results in \autoref{tab:dlfits} and \autoref{tab:fit}.

The available OS di-lepton invariant mass spectra are fit to the spectra predicted from the direct production of $H\rightarrow Sh$ assuming $m_H=270$~GeV.
The mass of $m_S$ is scanned, where the best fit is obtained with $m_S=150\pm5$~GeV.
The hadronic jet activity in events with SS di-leptons and tri-leptons in association with $b$-tagged jets reported by CMS is studied in the context of the production of $H\rightarrow Sh$ in association with one and two top quarks.
In both cases the BSM signal yields with $m_H=270$~GeV and $m_S=150$~GeV are fit by varying $\beta_g^2$, where $\beta_g=0$ corresponds to the absence of a BSM signal.
For the OS di-lepton invariant mass spectra, it is found that $\beta_g^2=1.22\pm0.38$, whereas in the multi-leptons in association with $b$-tagged jets, we find that $\beta_g^2=1.69\pm0.54$.
These compare well with $\beta_g^2=1.38\pm0.32$ obtained from available measurements of various Higgs boson signal strengths, excluding the production of $tth$.
Using an error weighted mean, the combined result yields $\beta_g^2=1.38\pm 0.22$.
It should be noted that this value does not take into account the full effects of correlations among systematic uncertainties; a formal combination of the results is beyond the scope of this paper.

The recently reported results with di-leptons in association with a $b$-tagged jet in Ref.~\cite{ATLAS:2018gcr} are compatible with the simplified model used here (with $m_H=270$~GeV and $m_S=150$~GeV) after taking the event selection into account.
The same applies to the recent results with di-leptons reported by CMS~\cite{Sirunyan:2018lcp} and ATLAS~\cite{ATLAS:2018rgl}.
These results are not added to the determination of $\beta_g$.
The results reported here are also compatible with previous studies on the BSM scenario we have considered, an encouraging insight into potential BSM physics which may already be visible in the existing ATLAS and CMS data.
A comprehensive study of these and other effects present in the data that are related to the model described here is beyond the scope of this article.

As a final word, the authors point out to corners of the phase space with leptons and hadronic jets that have not been explored at the LHC, where the simplified model considered here predicts an anomalous production.
It is suggested that the experiments explore the production of OS di-leptons with missing transverse energy in association with two or more hadronic jets and vetoing $b$-tagged jets.
This region of the phase space is not explored in the measurements of production cross section of the non-resonant $WW$ production, which to date are constrained to the measurement of the cross section with a full jet veto and in association with exactly one jet.

The rate of SS di-leptons and tri-leptons in association with $b$-tagged jets used in the search for $tth$ remains elevated in Run 2.
The measurement of the signal strength of $tth$ does not seem to capture the full extent of the discrepancy between the data and the SM. In the context of the model discussed here, this would imply that an anomalously large rate of SS di-leptons and tri-leptons in association with at least three $b$-tagged jets is expected. So far the experiments have not reported results in this corner of the phase space. It is important to study the possible connection between the elevated rate of SS di-leptons and tri-leptons with 1 or 2 $b$-tagged jets observed so far, compared with the rate in association $N_{b\text{-jet}}\geq 3$. It is also suggested exploring the production of OS di-leptons in association with $N_{b\text{-jet}}\geq 3$.

\ack
\textsc{NITheP} is acknowledged by SvB and MK for funding this work. The support from the Department of Science and Technology via the SA-CERN consortium is acknowledged. The support from the National Research Foundation and the Research Office of the University of the Witwatersrand is also acknowledged. AF acknowledges the funding support of the Research office of the University of the Witwatersrand and the African Institute for Mathematical Sciences in Tanzania (AIMS-Tanzania). The authors want to thank Kehinde Tomiwa for his assistance.

\appendix
\section{Method for best fit value of $\beta_g^2$}
\label{app1}
For fits of the BSM signal to data and evaluating test statistics, a $\chi^2$ approach was used.
A global $\chi^2$ was constructed by adding a $\chi^2$ value for each bin and in each observable per channel.
For each bin $i$, the global $\chi^2$ therefore takes the particular form of Pearson's test statistic,
\begin{equation}
\chi^2 = \sum_i \frac{\left( N_i^{\rm Data} - N_i^{\rm SM} - \beta_g^2 N_i^{\rm BSM} \right)^2}{\left( \Delta N_i^{\rm Data} \right)^2 + \left( \Delta N_i^{\rm SM} \right)^2}, \label{chi2}
\end{equation}
where $N_i$ is the event yield for bin $i$, and $\Delta N_i$ is the associated statistical and systematic uncertainty combination.
The denominator shows the result of adding the uncertainty in the data and theory in quadrature.
The BSM uncertainty is not included since it is dominated by the SM and statistical uncertainty.
As mentioned in \autoref{sec:formalism}, the BSM signal is typically proportional to $\beta_g^2$, and therefore the BSM event yield is proportional to it as well.
The best fit value of $\beta_g^2$ is obtained by minimising \autoref{chi2} while leaving $\beta_g^2$ free.
A one-$\sigma$ uncertainty on this best fit value is taken as the envelope around which \autoref{chi2} can vary by one unit away from the best fit value of $\beta_g^2$.

\section{Direct production kinematic distributions}
\label{app:directprod}
\begin{figure}
        \centering
        \subfloat[]{\includegraphics[width=0.49\textwidth]{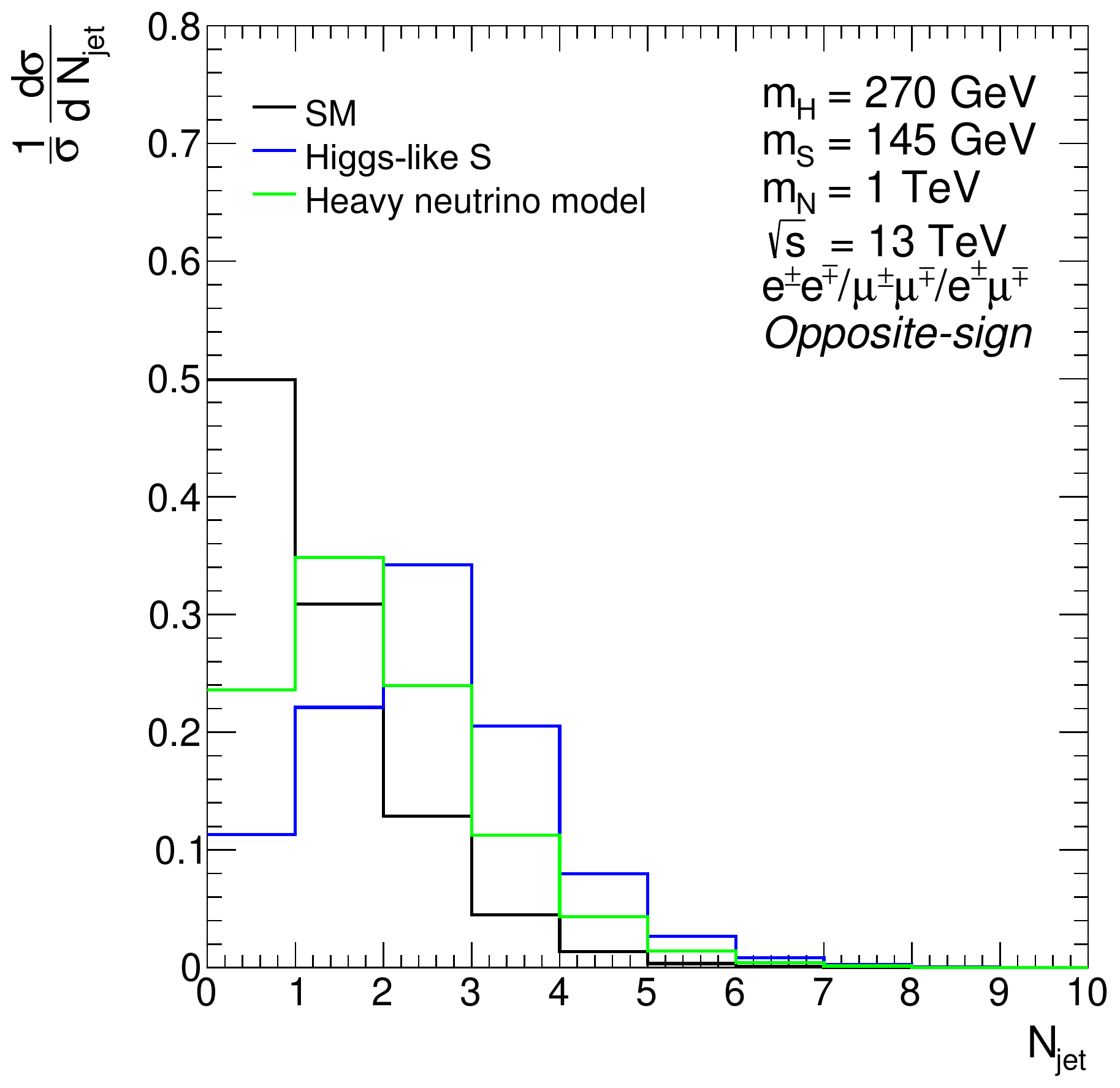}}
        \subfloat[]{\includegraphics[width=0.49\textwidth]{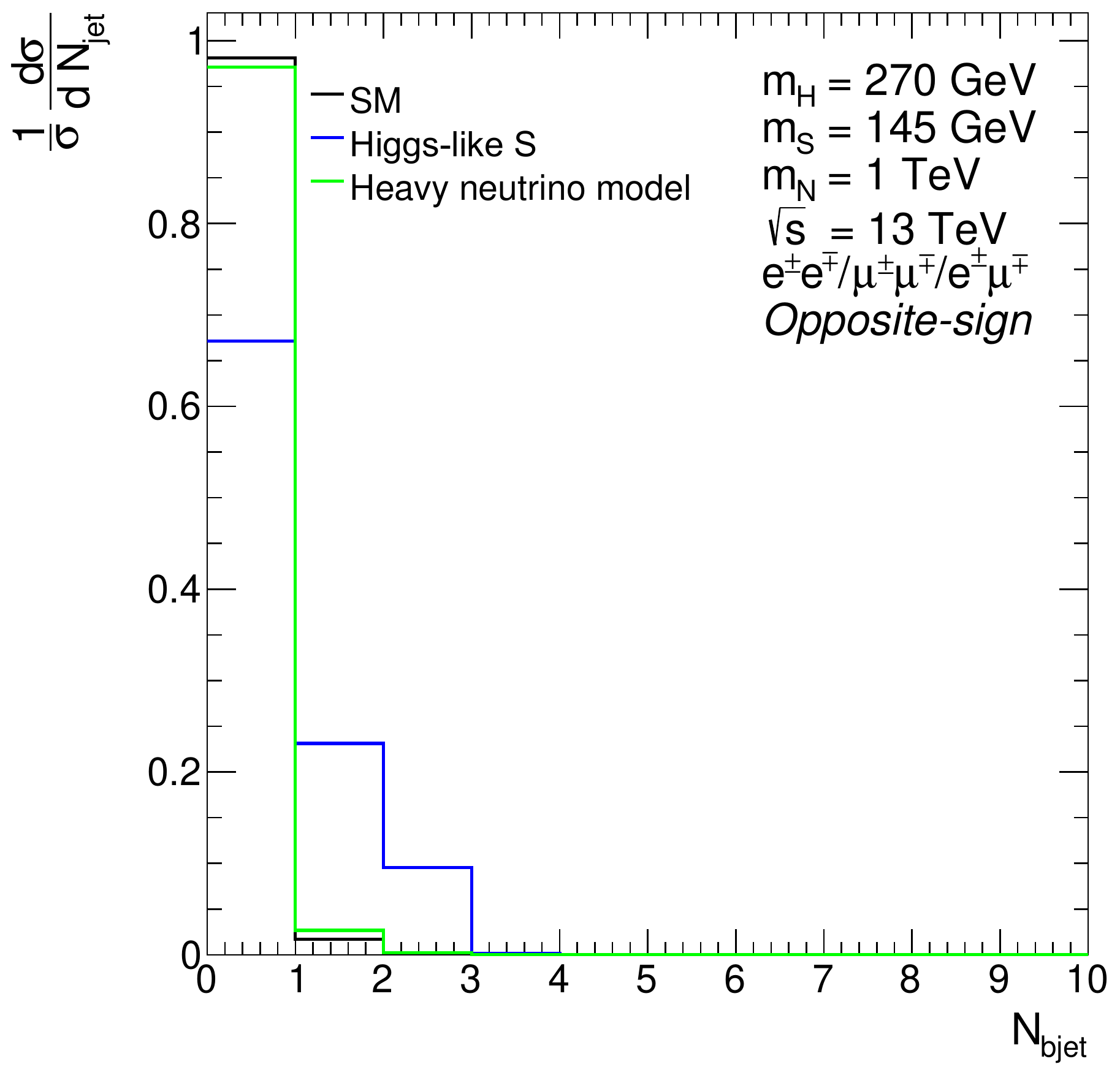}} \\
        \subfloat[]{\includegraphics[width=0.49\textwidth]{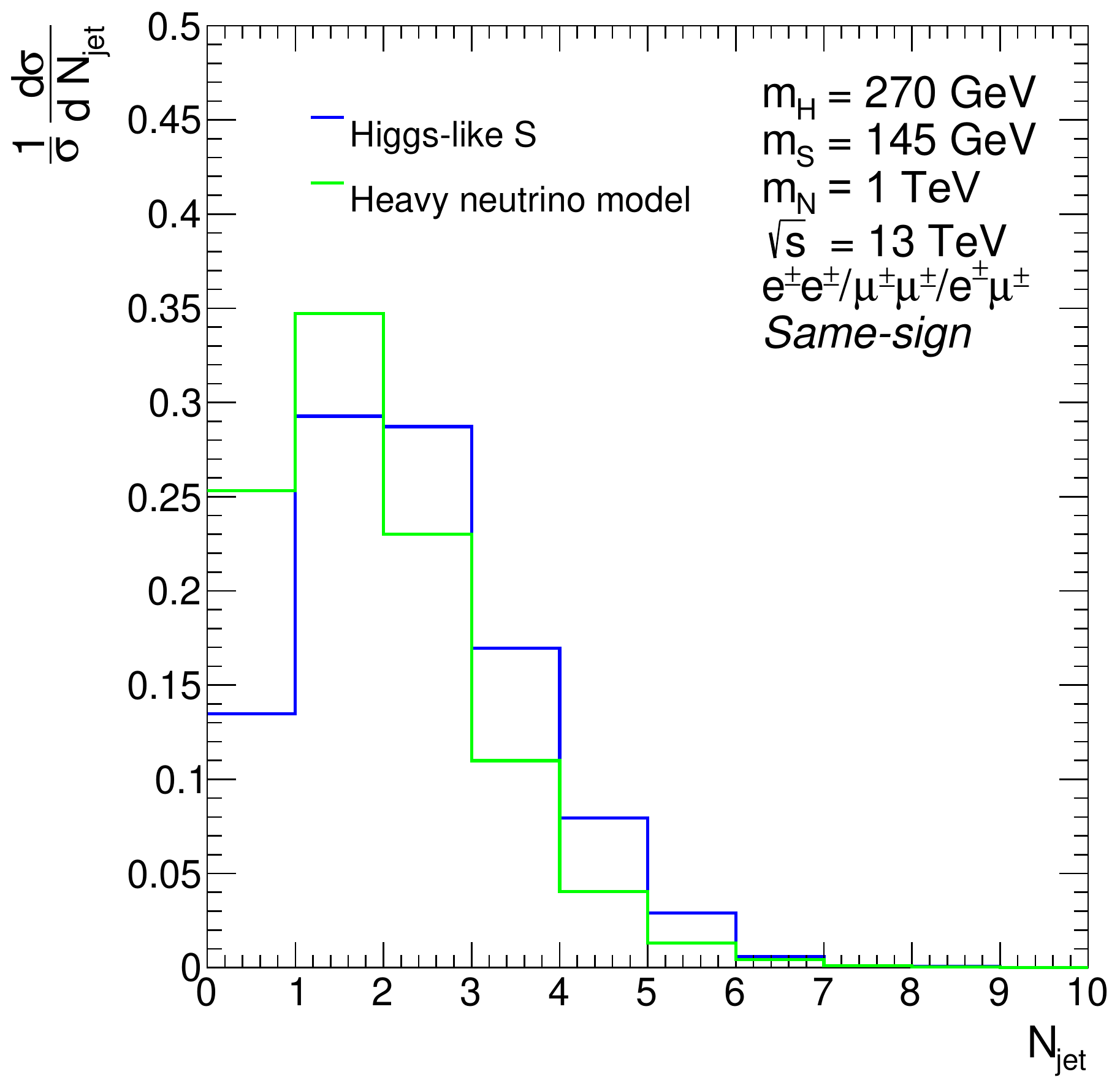}}
        \caption{\label{fig:dl1} Normalised distributions of jet and $b$-tagged jet multiplicities for the event pre-selection in case of
        the OS (in (a) and (b)) and SS (in (c)) channels, where the process is $gg  \to H \to Sh$.
        The SM prediction comes from the process $gg\to h$.
        The comparison with the SM has been omitted for the SS case, due to a lack of statistics.
        The $b$-tagged jet multiplicity for the SS channel has also been omitted, since all events with two SS leptons have no $b$-tagged jets.
        }
\end{figure}

\begin{figure}
        \centering
        \subfloat[]{\includegraphics[width=0.49\textwidth]{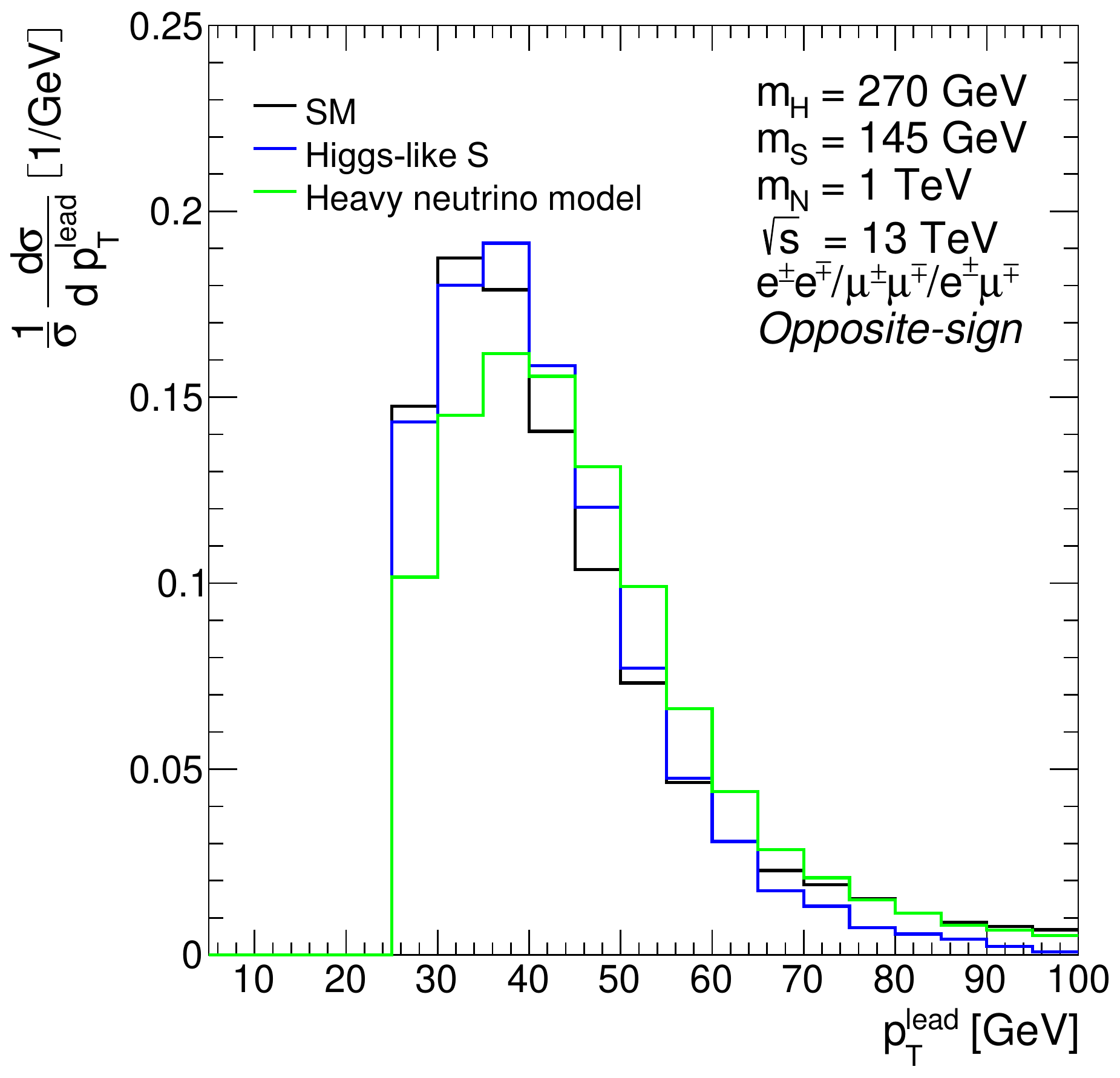}}
        \subfloat[]{\includegraphics[width=0.49\textwidth]{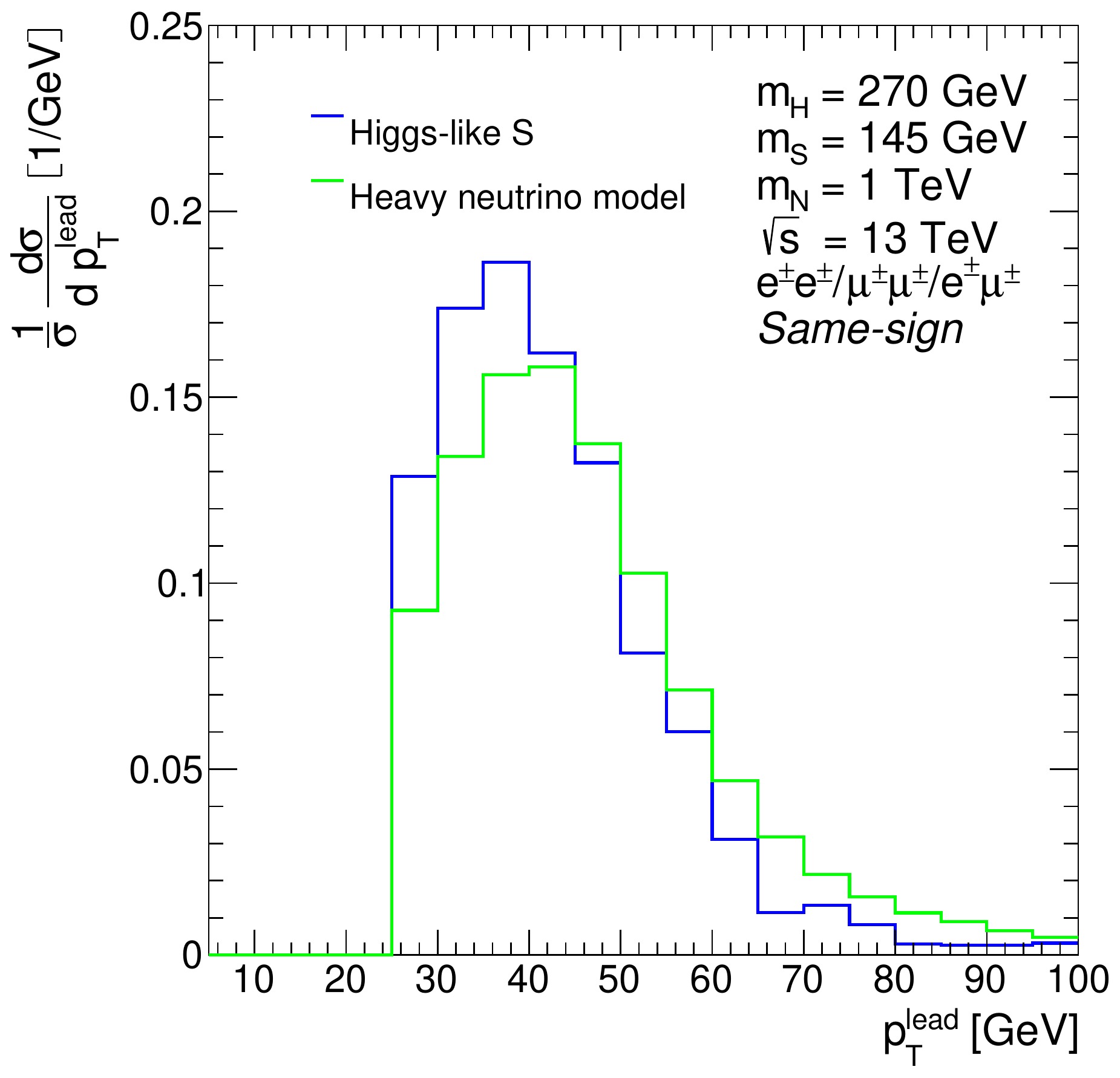}} \\
        \subfloat[]{\includegraphics[width=0.49\textwidth]{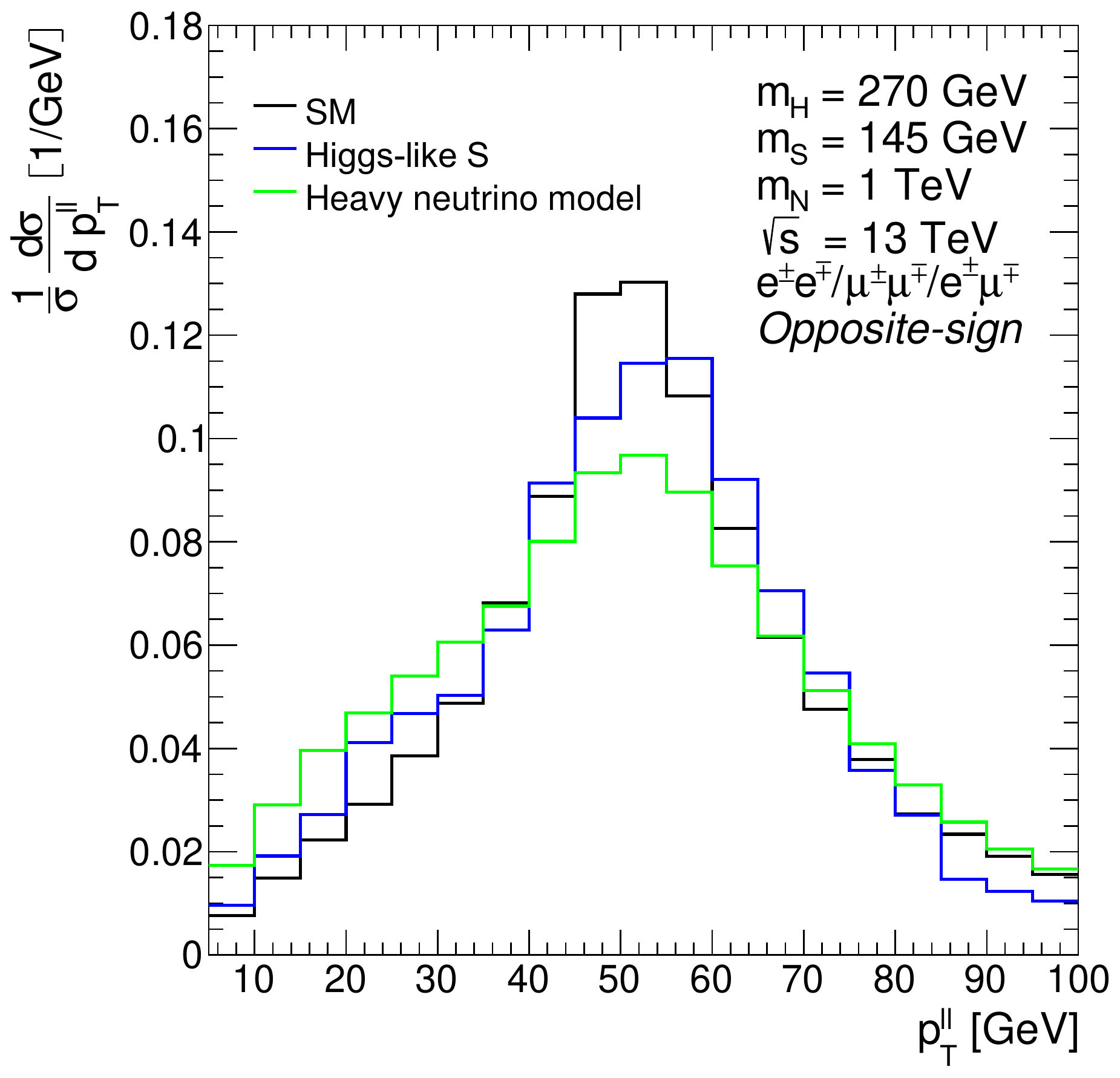}}
        \subfloat[]{\includegraphics[width=0.49\textwidth]{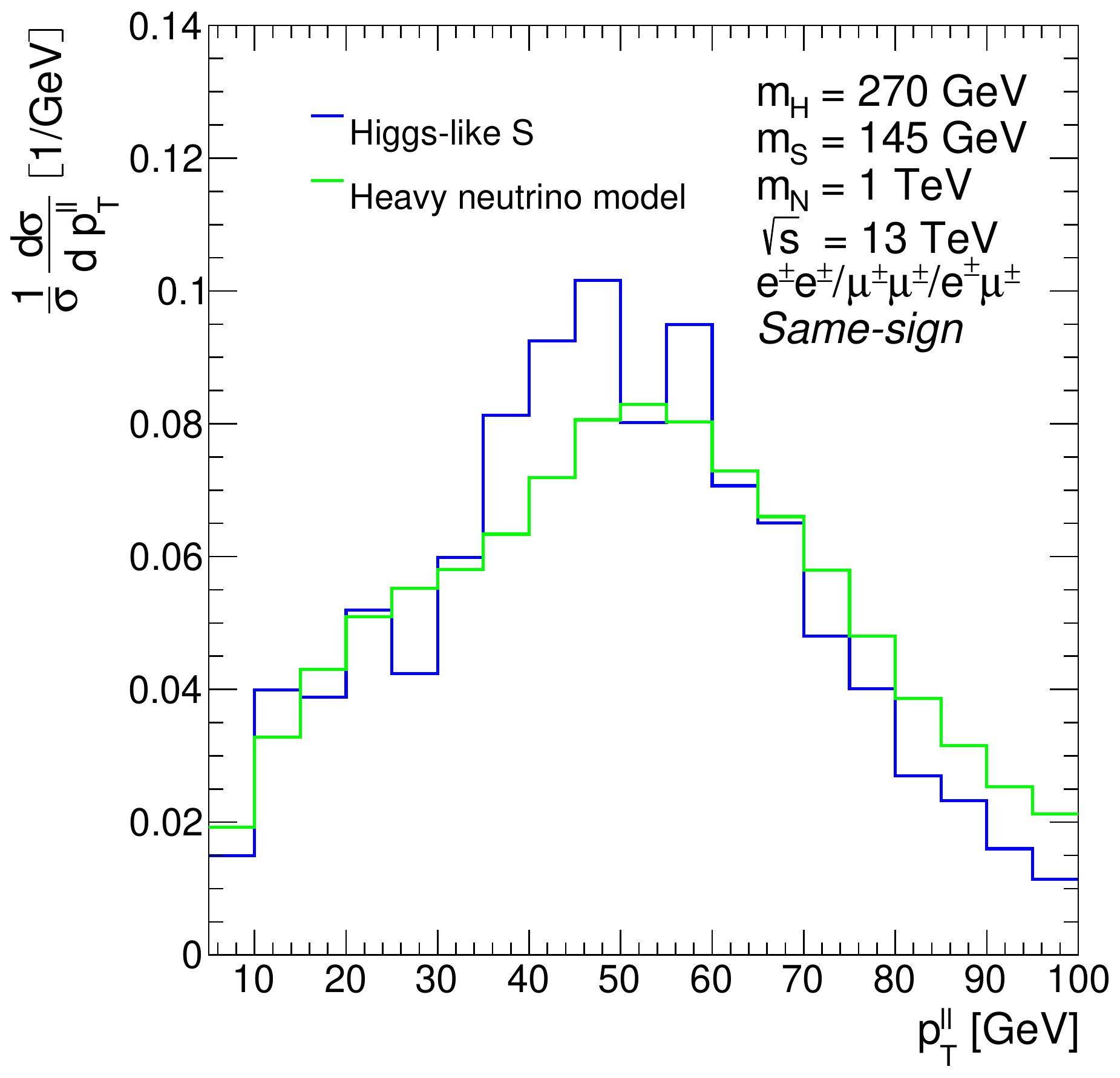}}
        \caption{\label{fig:dl2} Normalised distributions of the leading-lepton and di-lepton $p_\text{T}$ for the event pre-selection in case of the OS (left) and SS (right) channels, where the process is $gg  \to H \to Sh$.
        The SM prediction comes from the process $gg\to h$.
        The comparison with the SM has been omitted for the SS case, due to a lack of statistics.
        In the case of the Higgs-like $S$ model, the SS channel observables also suffer from a lack of statistics, although the difference between the predictions is still evident.
        }
\end{figure}

\begin{figure}
        \centering
        \vspace{-20pt}
        \subfloat[]{\includegraphics[width=0.41\textwidth]{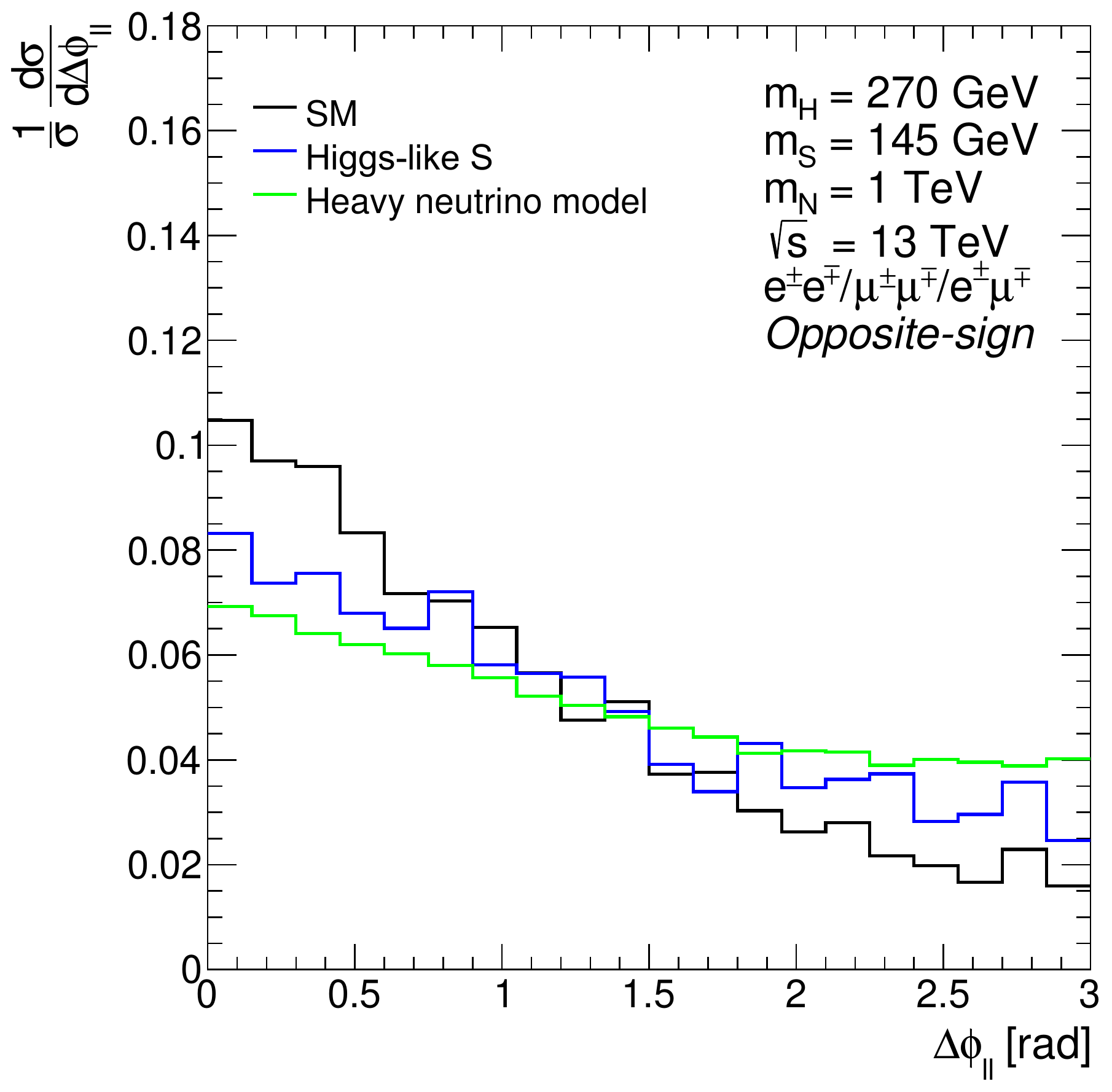}}
        \subfloat[]{\includegraphics[width=0.41\textwidth]{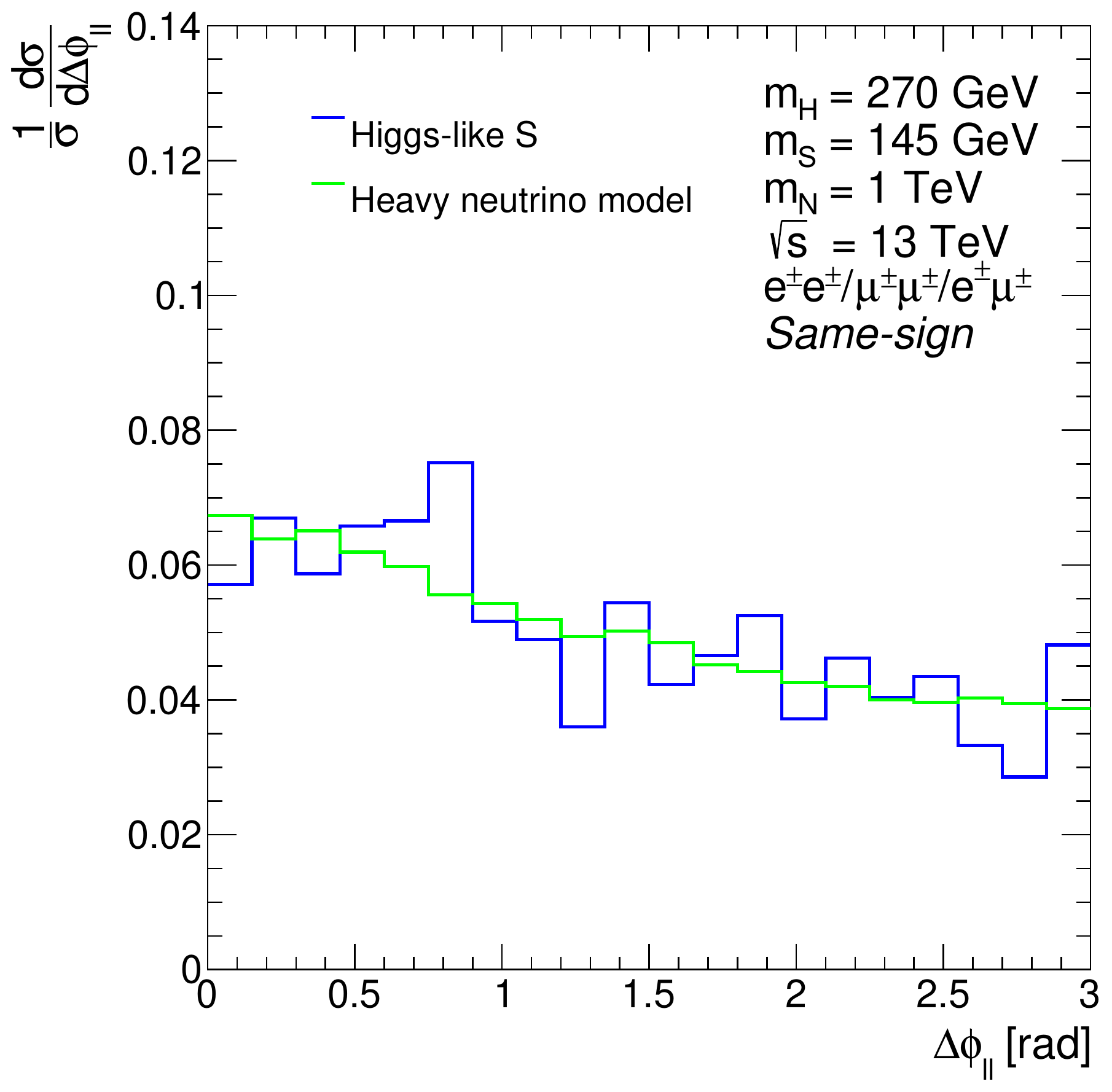}}\vspace{-10pt} \\
        \subfloat[]{\includegraphics[width=0.41\textwidth]{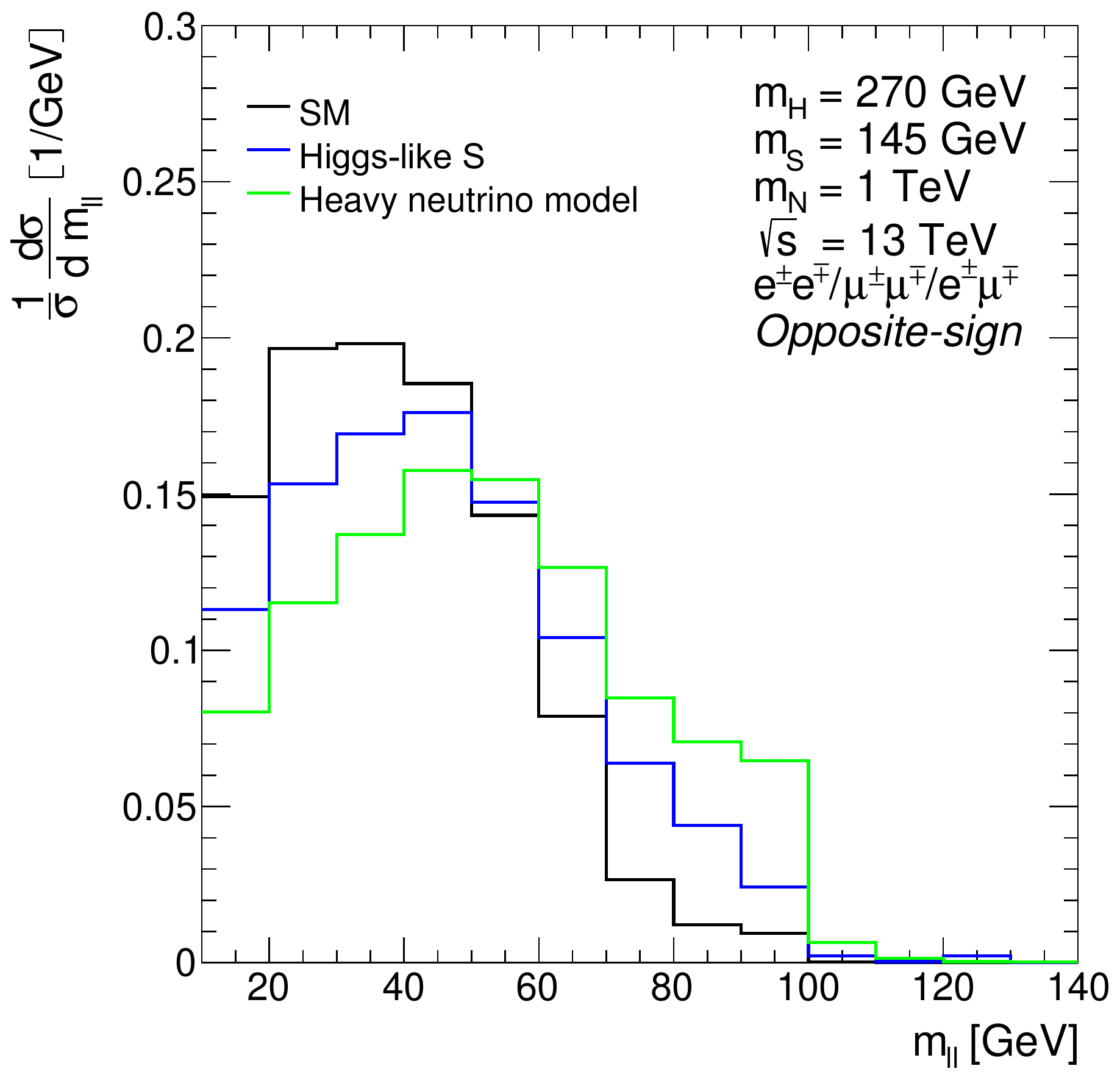}}
        \subfloat[]{\includegraphics[width=0.41\textwidth]{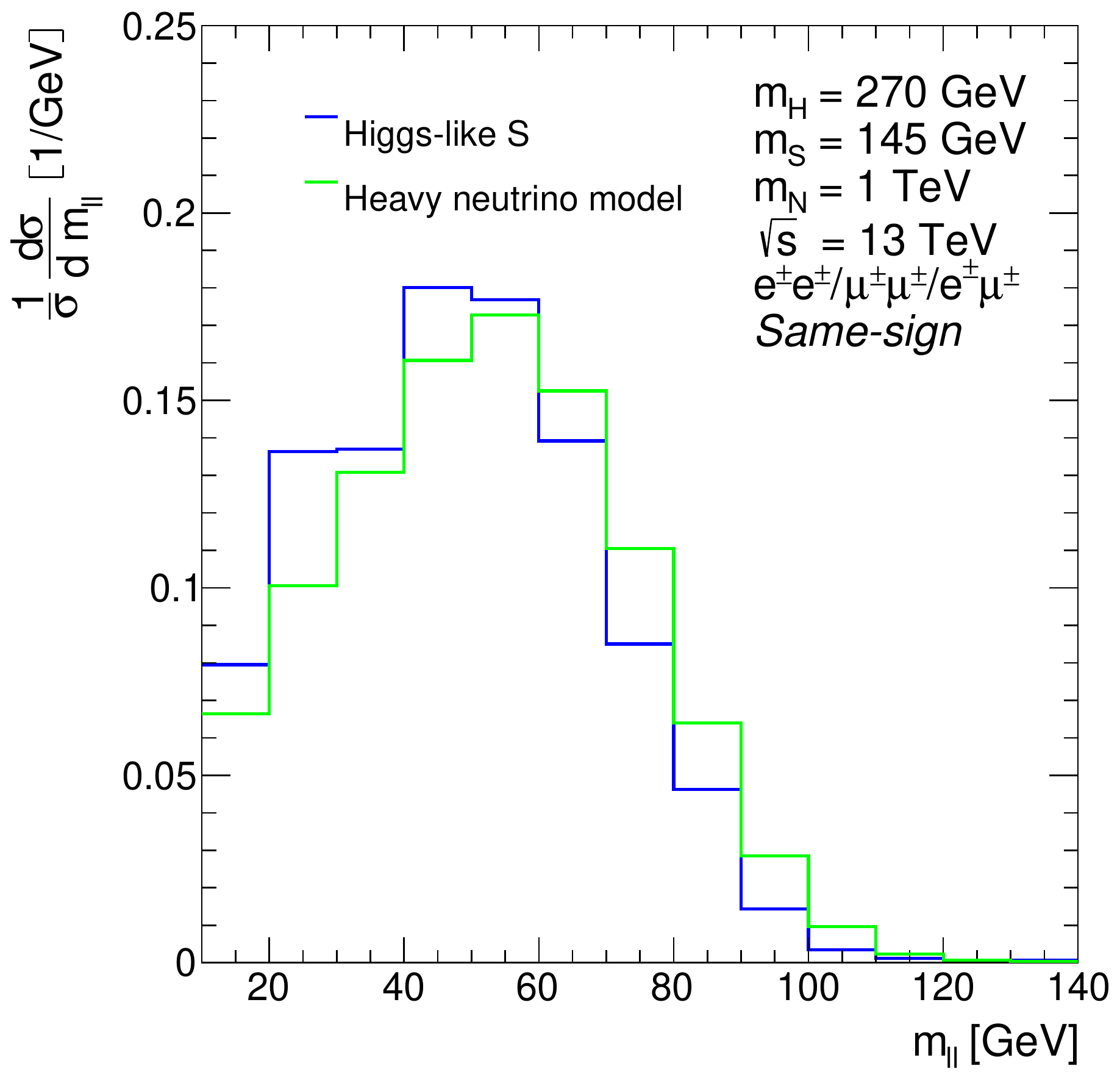}}\vspace{-10pt} \\
        \subfloat[]{\includegraphics[width=0.41\textwidth]{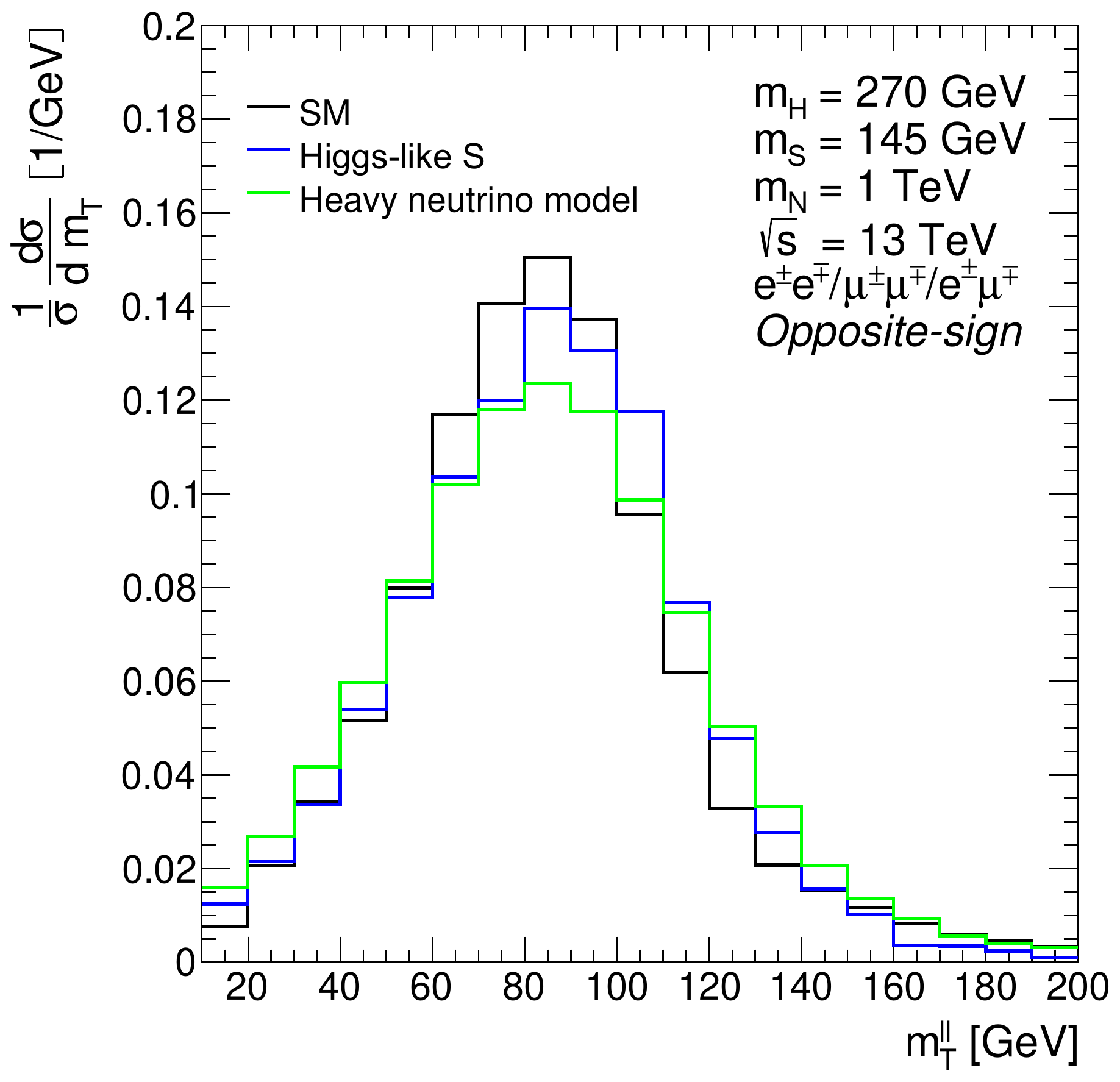}}
        \subfloat[]{\includegraphics[width=0.41\textwidth]{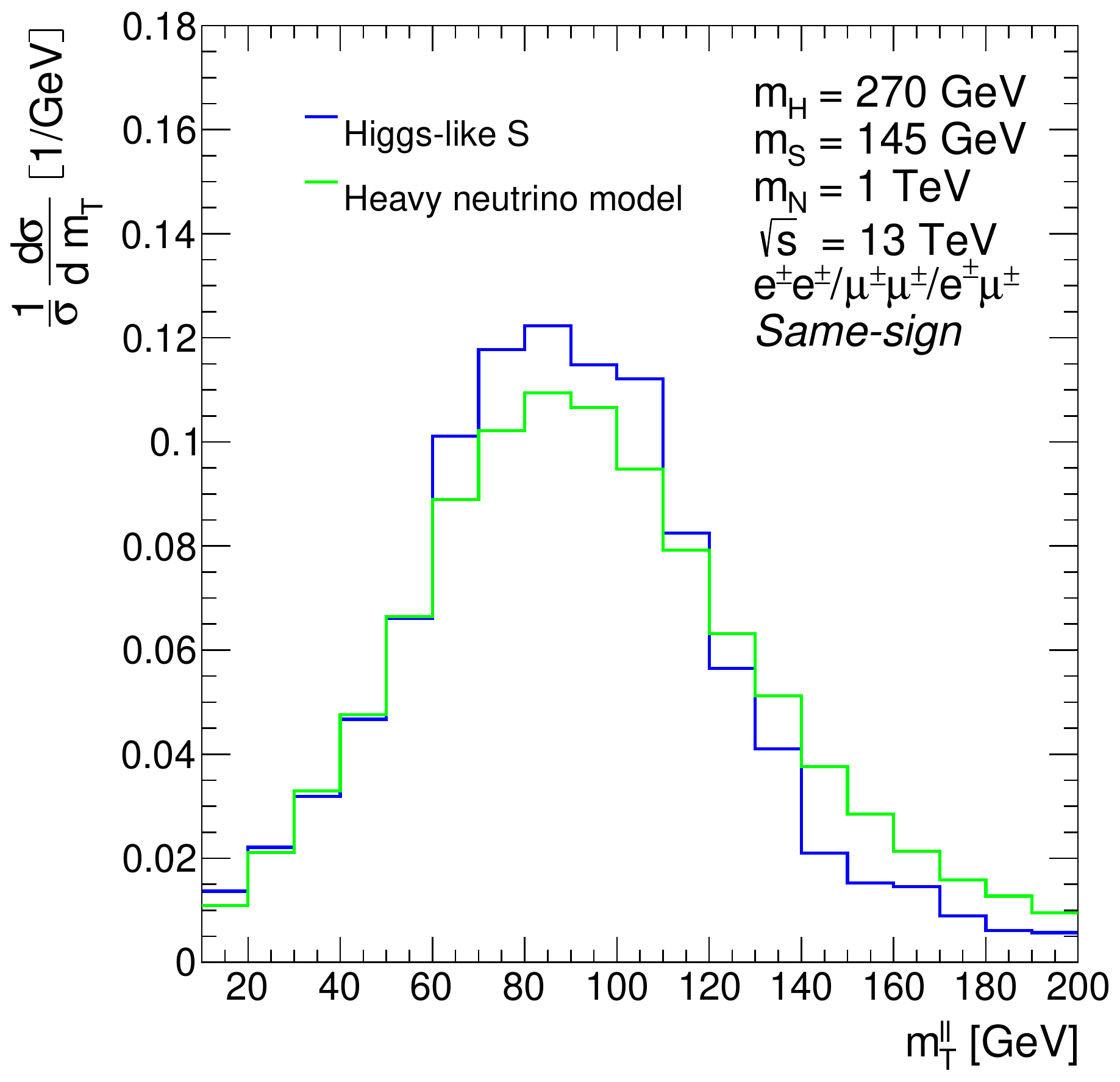}}
        \caption{\label{fig:dl3} Normalised distributions of $\Delta\phi_{\ell\ell}$, the invariant mass of the di-lepton system and the invariant transverse mass of di-lepton system for the event pre-selection in the OS (left) and SS (right) channels, where the process is $gg  \to H \to Sh$.
        The SM prediction comes from the process $gg\to h$.
        The comparison with the SM has been omitted for the SS case, due to a lack of statistics.
        In the case of the Higgs-like $S$ model, the SS channel observables also suffer from a lack of statistics, most notably in the $\Delta\phi_{\ell\ell}$ distribution, although there is very little discrimination power in this variable in any case.
        }
\end{figure}

\begin{figure}
        \centering
        \subfloat[]{\includegraphics[width=0.49\textwidth]{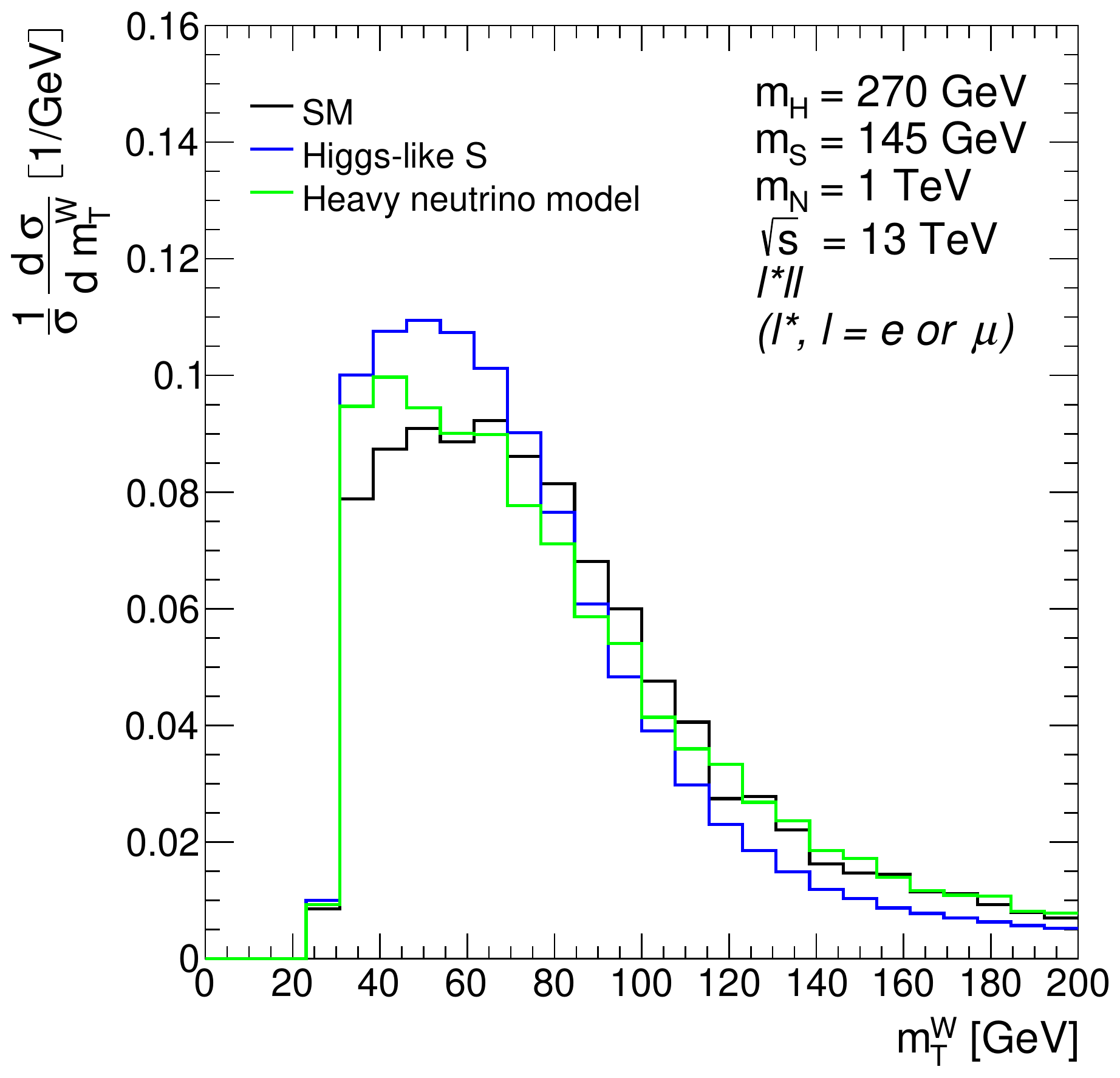}}
        \subfloat[]{\includegraphics[width=0.49\textwidth]{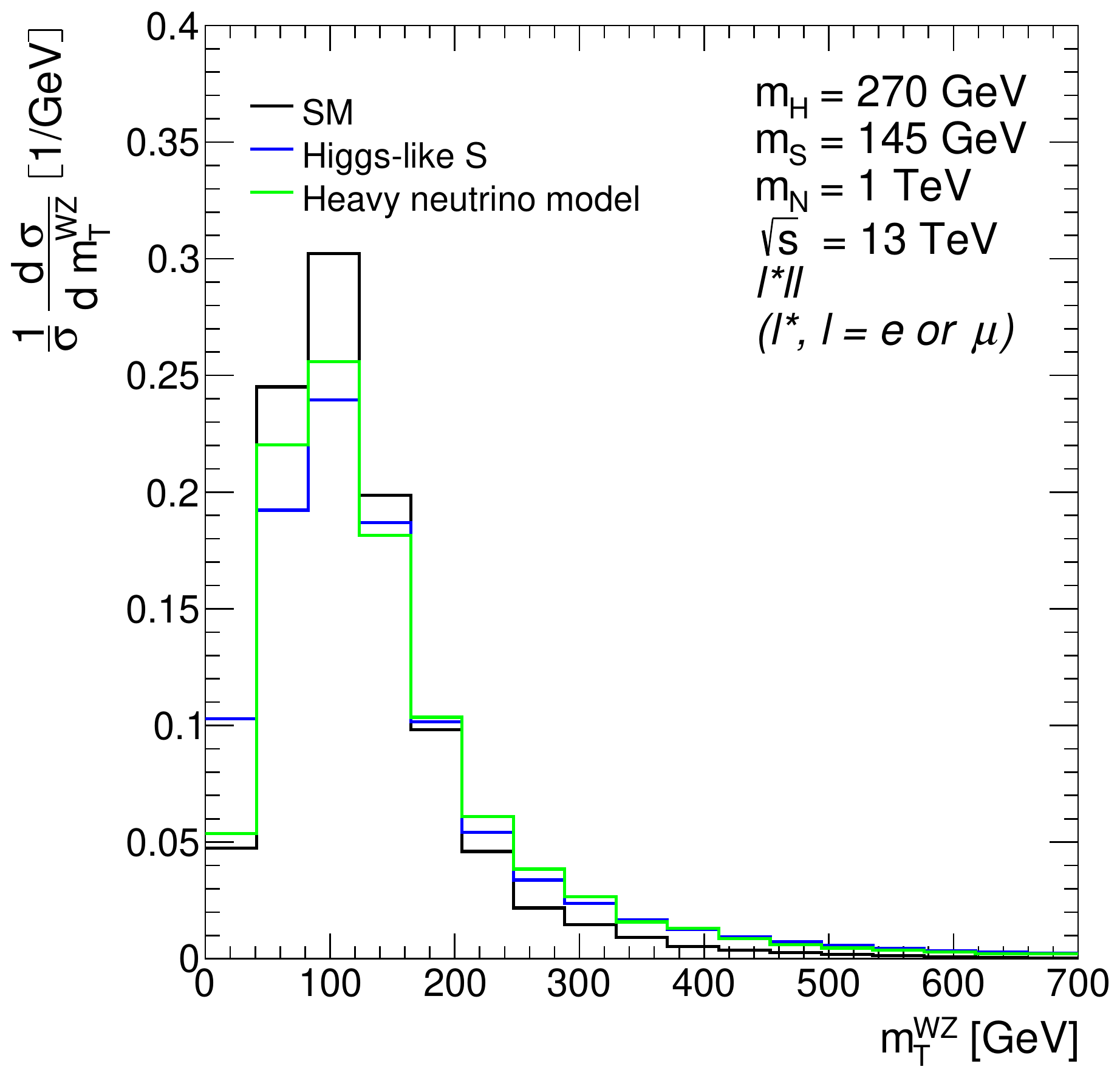}} \\
        \subfloat[]{\includegraphics[width=0.49\textwidth]{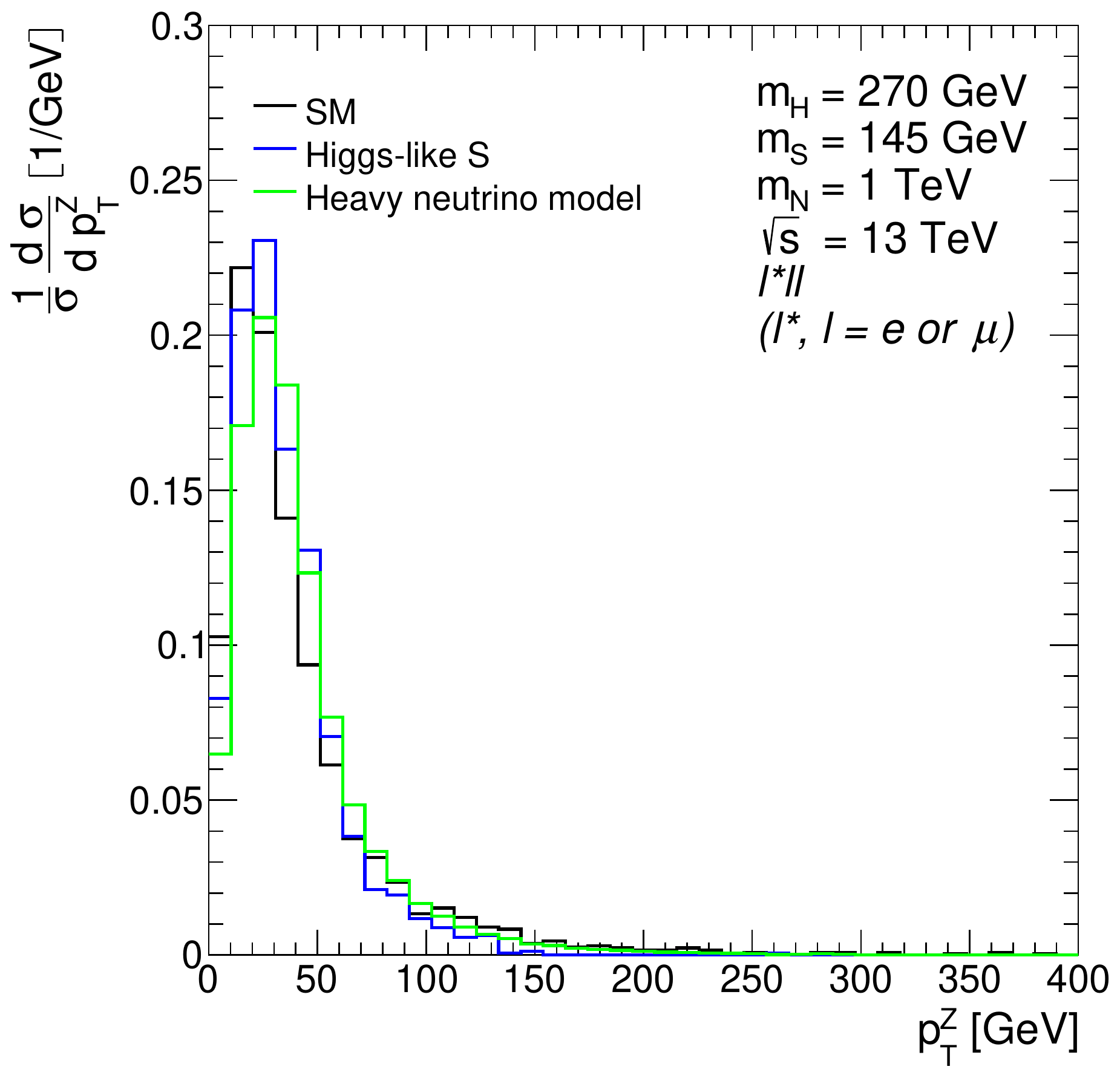}}
        \subfloat[]{\includegraphics[width=0.49\textwidth]{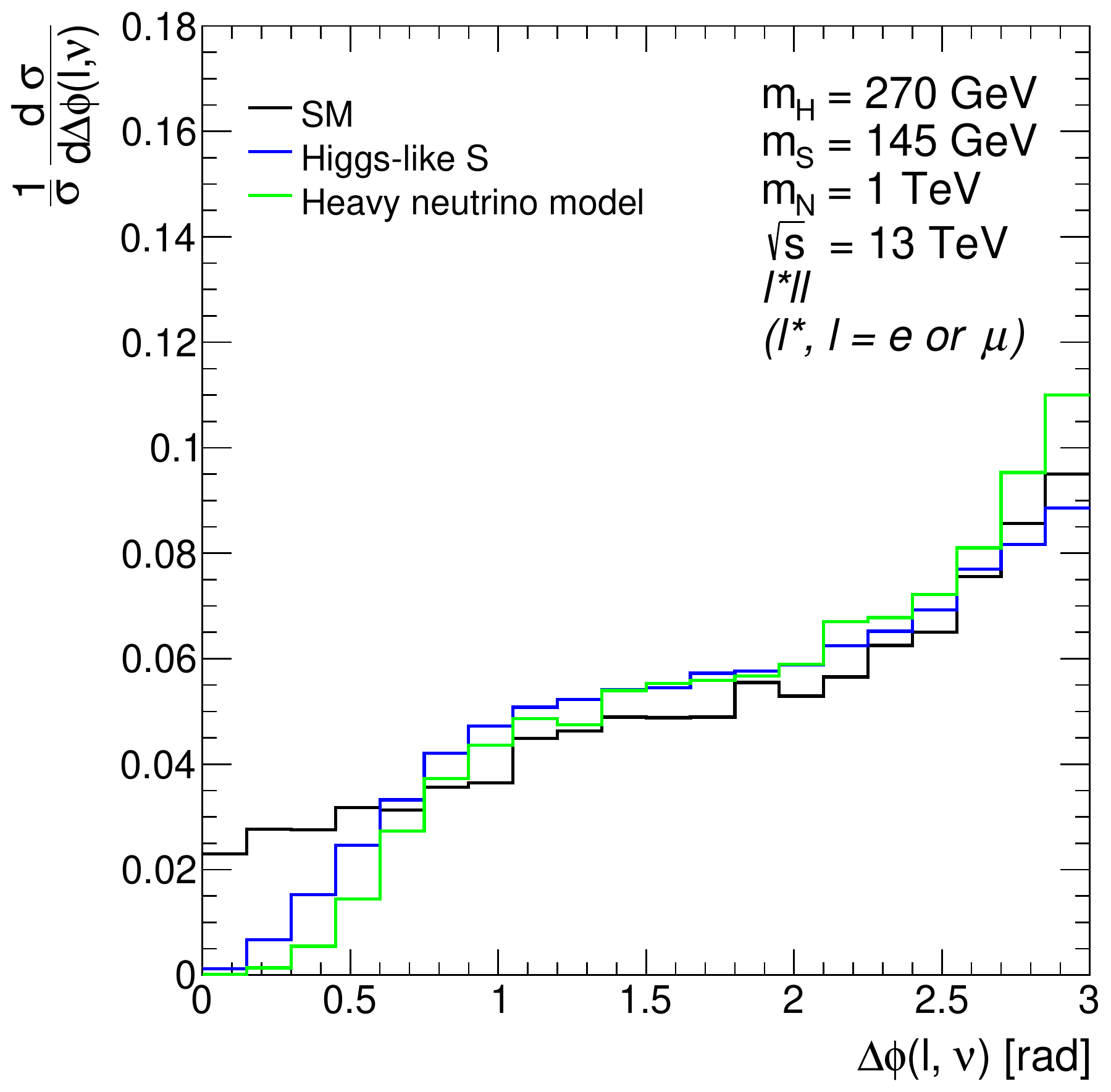}}
        \caption{\label{fig:tl} Normalised distributions of (a) the transverse mass of the reconstructed $W$ boson, (b) the transverse mass of reconstructed $WZ$ system, (c)the $p_\text{T}$ of $Z$ boson and (d) $\Delta\phi$ between leading lepton and missing energy for the event pre-selection in the tri-lepton channel, where the process is $gg  \to H \to Sh$.
        The SM prediction comes from the process $gg\to h$.
        }
\end{figure}

\begin{figure}
        \centering
        \subfloat[]{\includegraphics[width=0.49\textwidth]{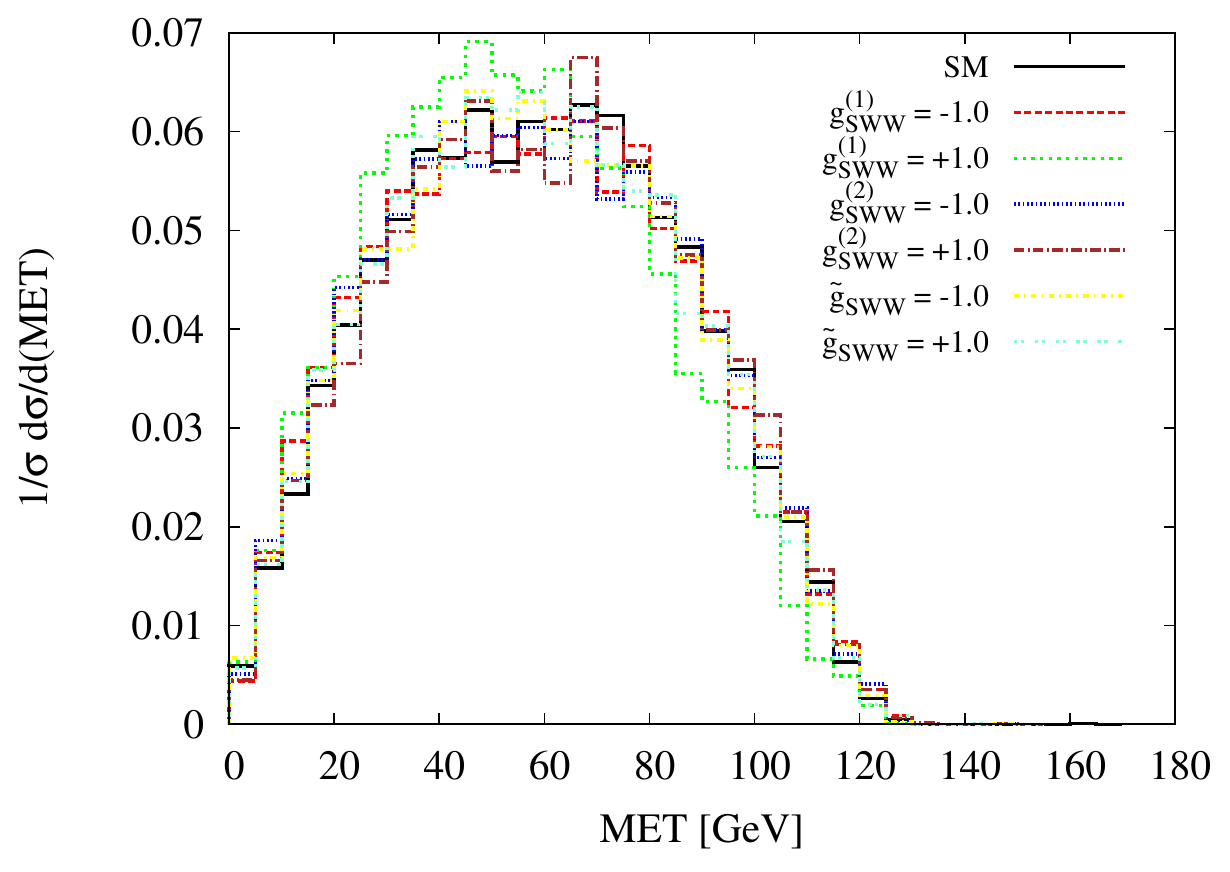}}
        \subfloat[]{\includegraphics[width=0.49\textwidth]{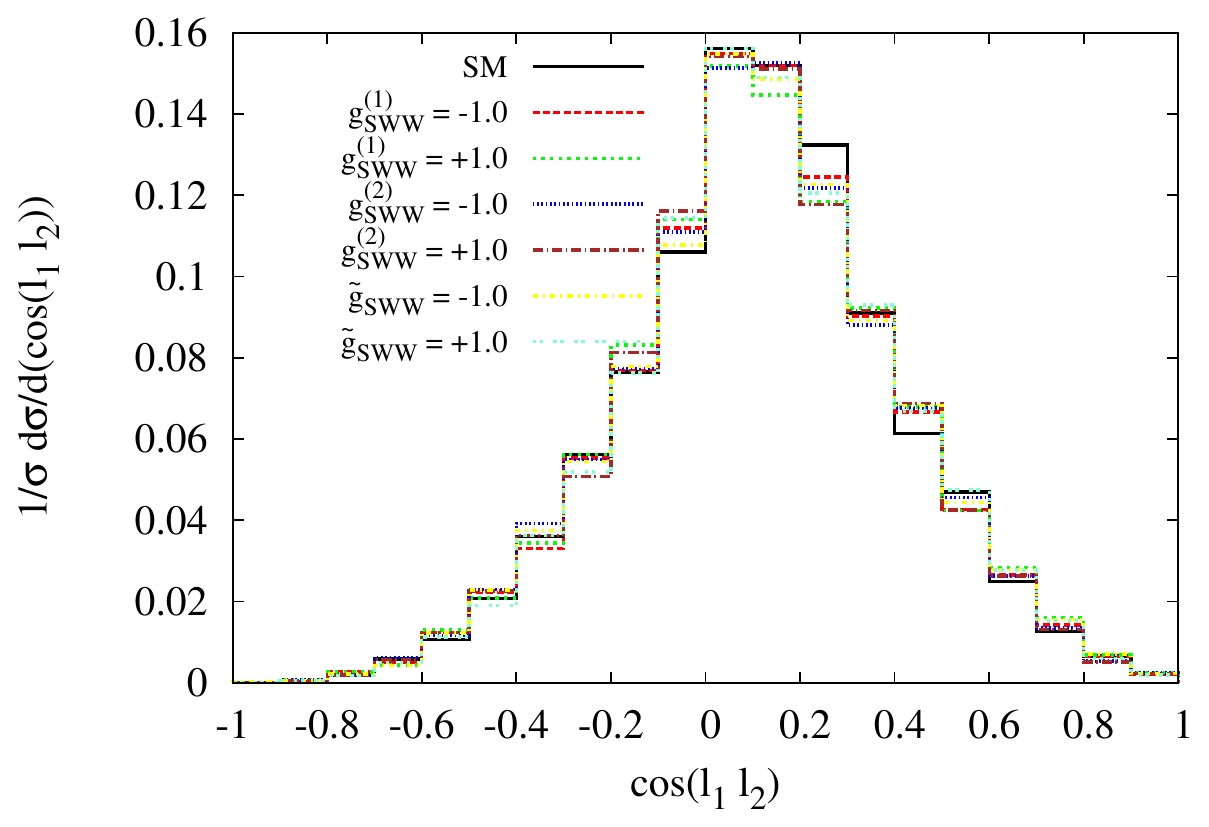}} \\
        \subfloat[]{\includegraphics[width=0.49\textwidth]{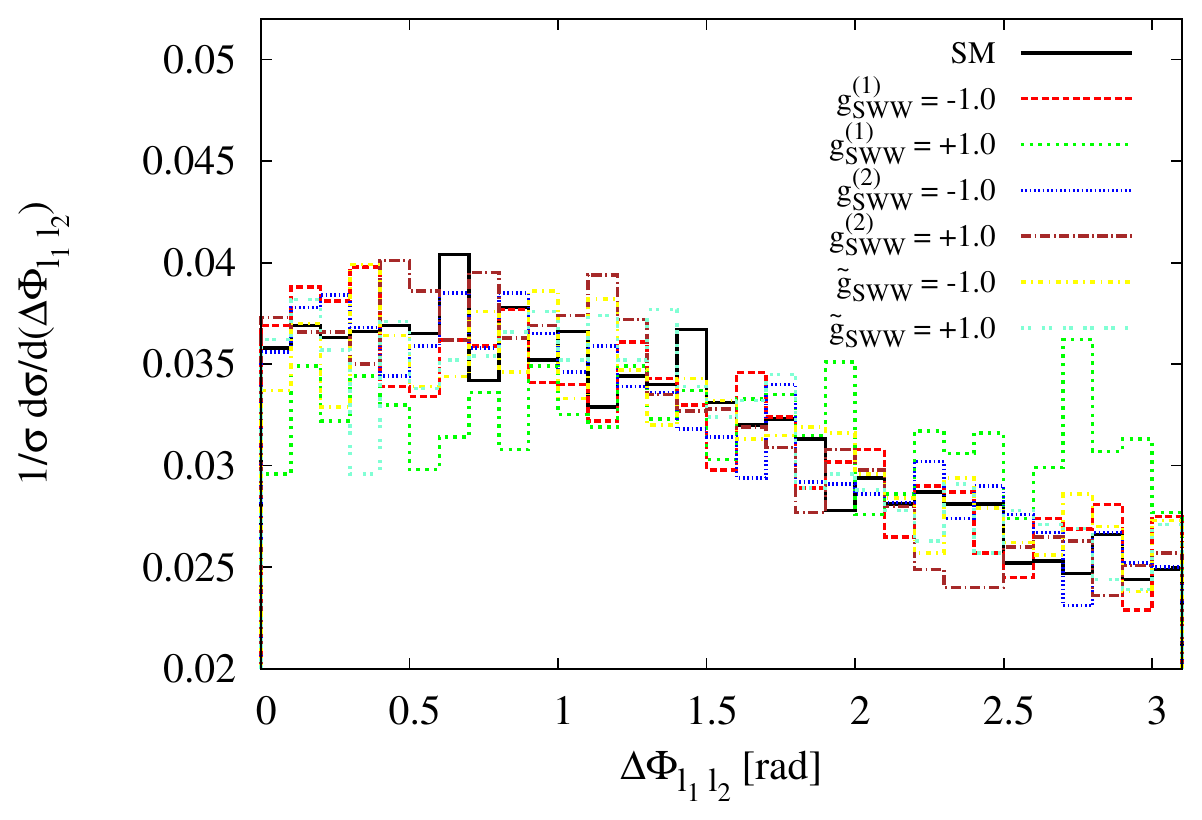}}
        \subfloat[]{\includegraphics[width=0.49\textwidth]{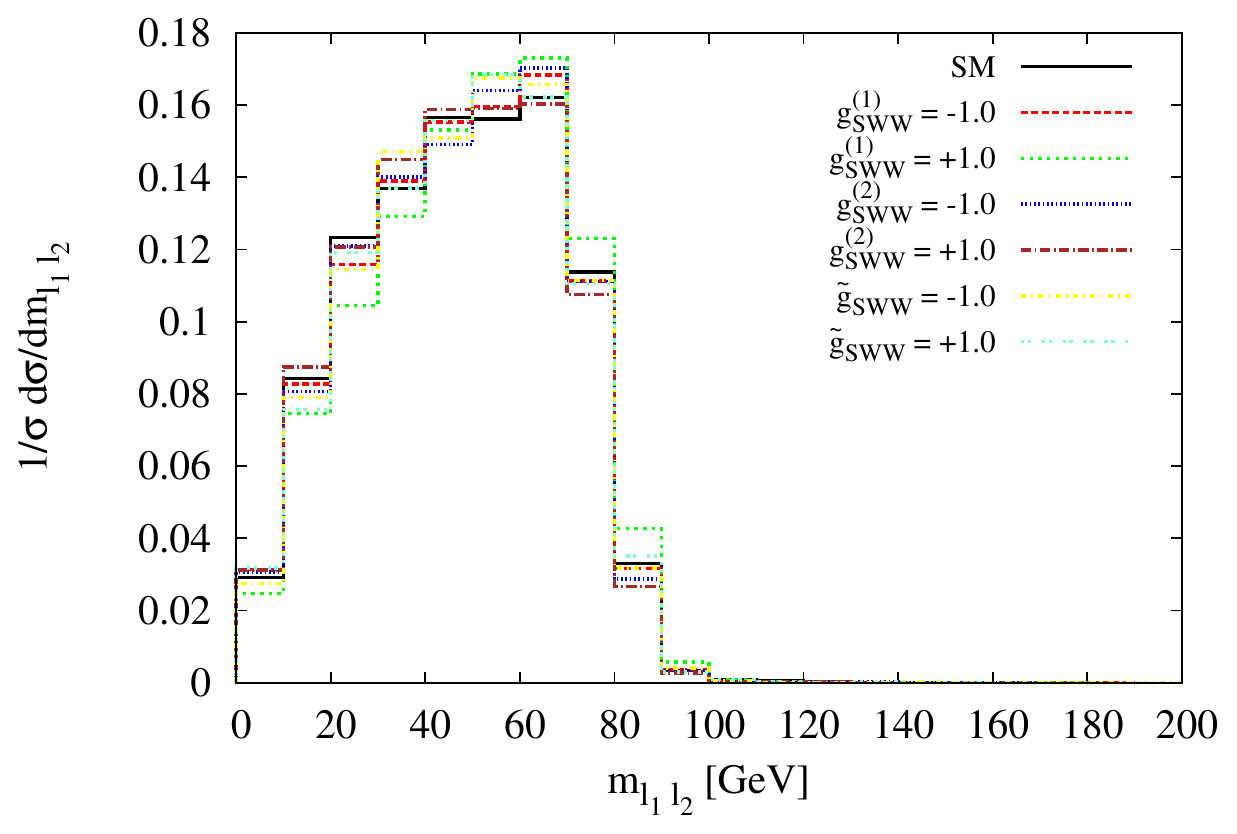}} \\
        \subfloat[]{\includegraphics[width=0.49\textwidth]{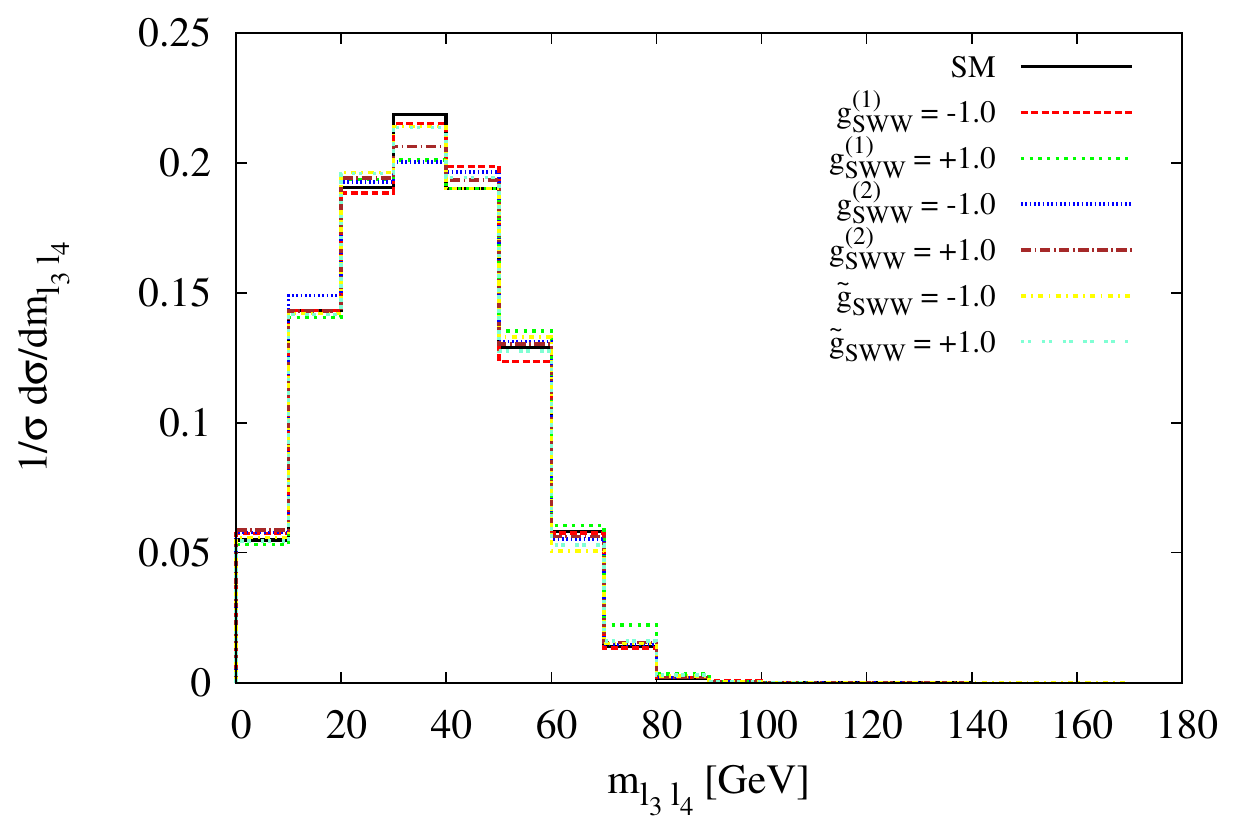}}
        \subfloat[]{\includegraphics[width=0.49\textwidth]{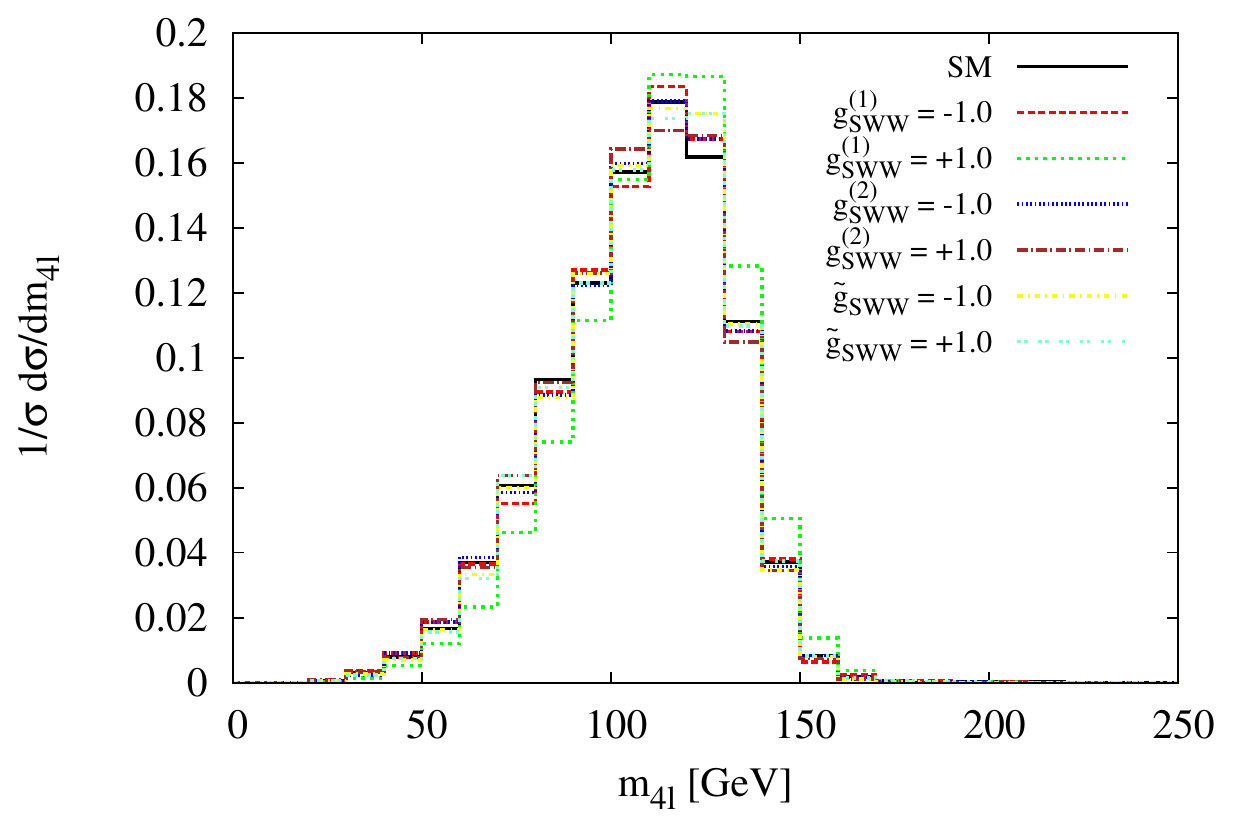}}
        \caption{\label{fig:fl} Discriminating kinematic distributions for ad-hoc values of new physics couplings relating to the CP-studies from \autoref{cplag}.
        Here the process that is considered is $p p \to g g \to H \to S h \to 4W$.
        The distributions include (a) the missing transverse energy, (b) the cosine of the angle between the two leading leptons, (c) the difference in azimuthal angle between the two leading leptons, and (d), (e) and (f), the invariant mass distributions for different possible groupings of the leptons.
        Due to a lack of statistics, the discrimination power of the missing transverse energy and $\Delta\phi$ distributions can only be estimated, although trends in the differences of the curves can be identified can be identified.}
\end{figure}

Various kinematic distributions arising from the direct production of $H$ through $gg$F are presented here as a reference.
All plots show the process $gg\to H\to Sh$ at a centre of mass energy of 13~TeV.
Both the Higgs-like $S$ model and the heavy neutrino model are compared.
\autoref{fig:dl1}, \autoref{fig:dl2} and \autoref{fig:dl3} show distributions for di-leptonic signatures, while \autoref{fig:tl} shows distributions for tri-lepton signatures.
Jet and $b$-tagged jet multiplicities are shown in \autoref{fig:dl1}.
The $p_\text{T}$ of the leptons and di-lepton pairs are presented in \autoref{fig:dl2}, while for the di-leptonic system, $\Delta\phi$ and invariant masses are shown in \autoref{fig:dl3}.
For the $W$ mass reconstruction, the transverse mass ($m_\text{T}^W$) observable is defined as
\begin{equation}
m_\text{T}^W = \sqrt{2\, E_T^{\rm miss}\, p_\text{T}^{\ell} \left[ 1- \cos\Delta\phi_{\ell\nu}\right]},
\end{equation}
where $E_T^{\rm miss}$ is the total transverse missing energy in the event (defined as the magnitude of the vector sum of all neutrinos) and $\ell$ and $\nu$ are the $W$ boson decay products.
Similarly, the transverse mass observable for the $WZ$-system ($m_\text{T}^{WZ}$) is defined as
\begin{align}
\left(m_\text{T}^{WZ}\right)^2 =&\, \left(\sum_{\ell=1}^{3}p_\text{T}^\ell + E_T^{\rm miss} \right)^2  -  \left(\sum_{\ell=1}^{3}p_x^\ell + E_x^{\rm miss} \right)^2  \nonumber \\
&\,-  \left(\sum_{\ell=1}^{3}p_y^\ell + E_y^{\rm miss} \right)^2,
\end{align}
where $\sum_{\ell=1}^{3}p_\text{T}^\ell $ is the sum of the transverse momentum of the three leptons etc.
These observables are shown in \autoref{fig:tl}, including the $p_\text{T}$ and reconstructed $Z$ boson mass with $\Delta\phi$ between leading lepton and missing energy vector.
For SS lepton pairs, the SM prediction is omitted due to the fact that the SM production cross section of SS leptons pairs is small.
In \autoref{fig:fl}, the effects of varying the CP structure of the $SWW$ coupling are shown for different interesting observables.
\section{Comparison with data}
\label{app:mll}
\begin{figure}
        \centering
        \subfloat[]{\includegraphics[width=0.49\textwidth]{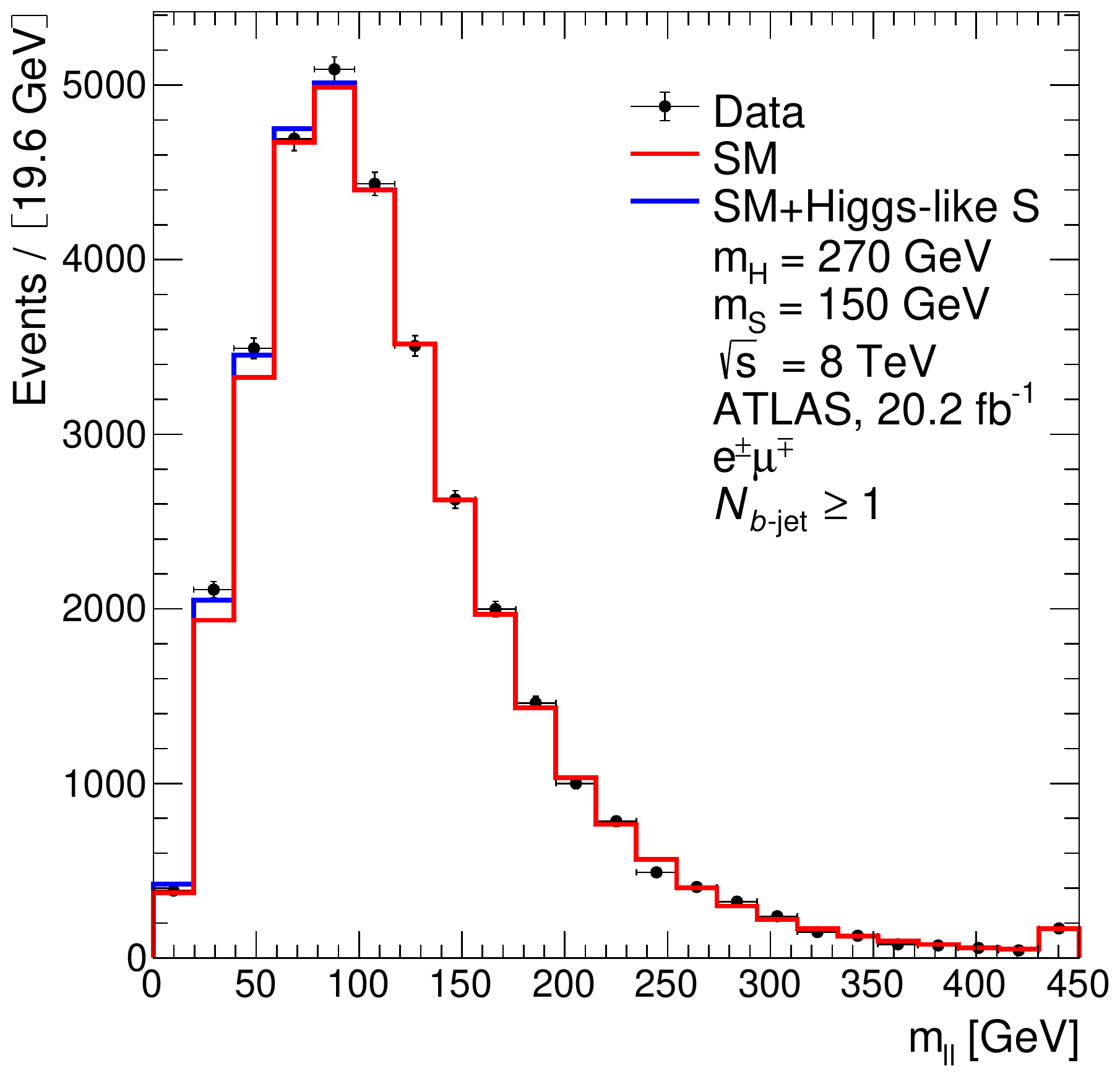}}
        \subfloat[]{\includegraphics[width=0.49\textwidth]{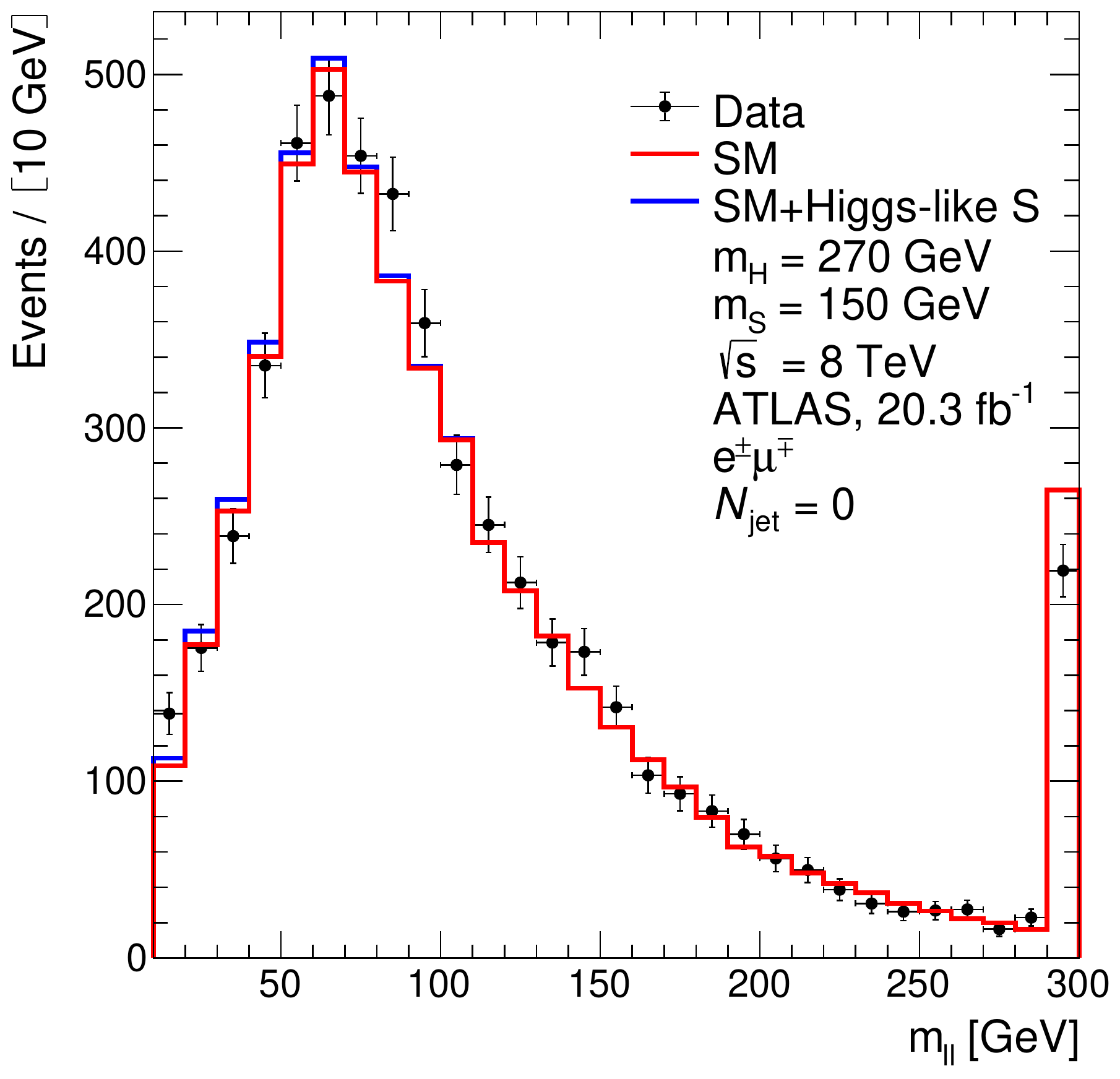}} \\
        \subfloat[]{\includegraphics[width=0.49\textwidth]{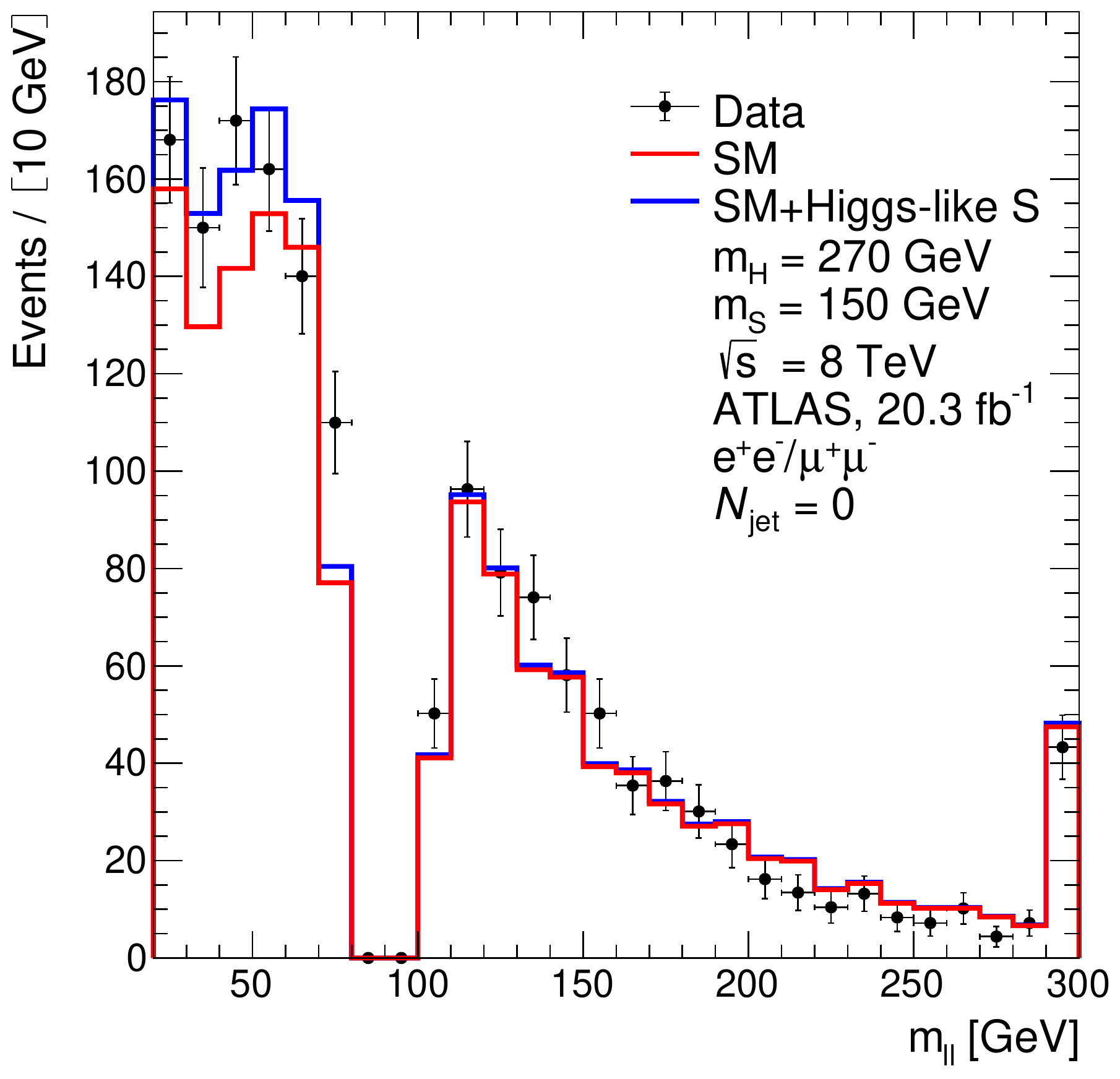}}
      \subfloat[]{\includegraphics[width=0.49\textwidth]{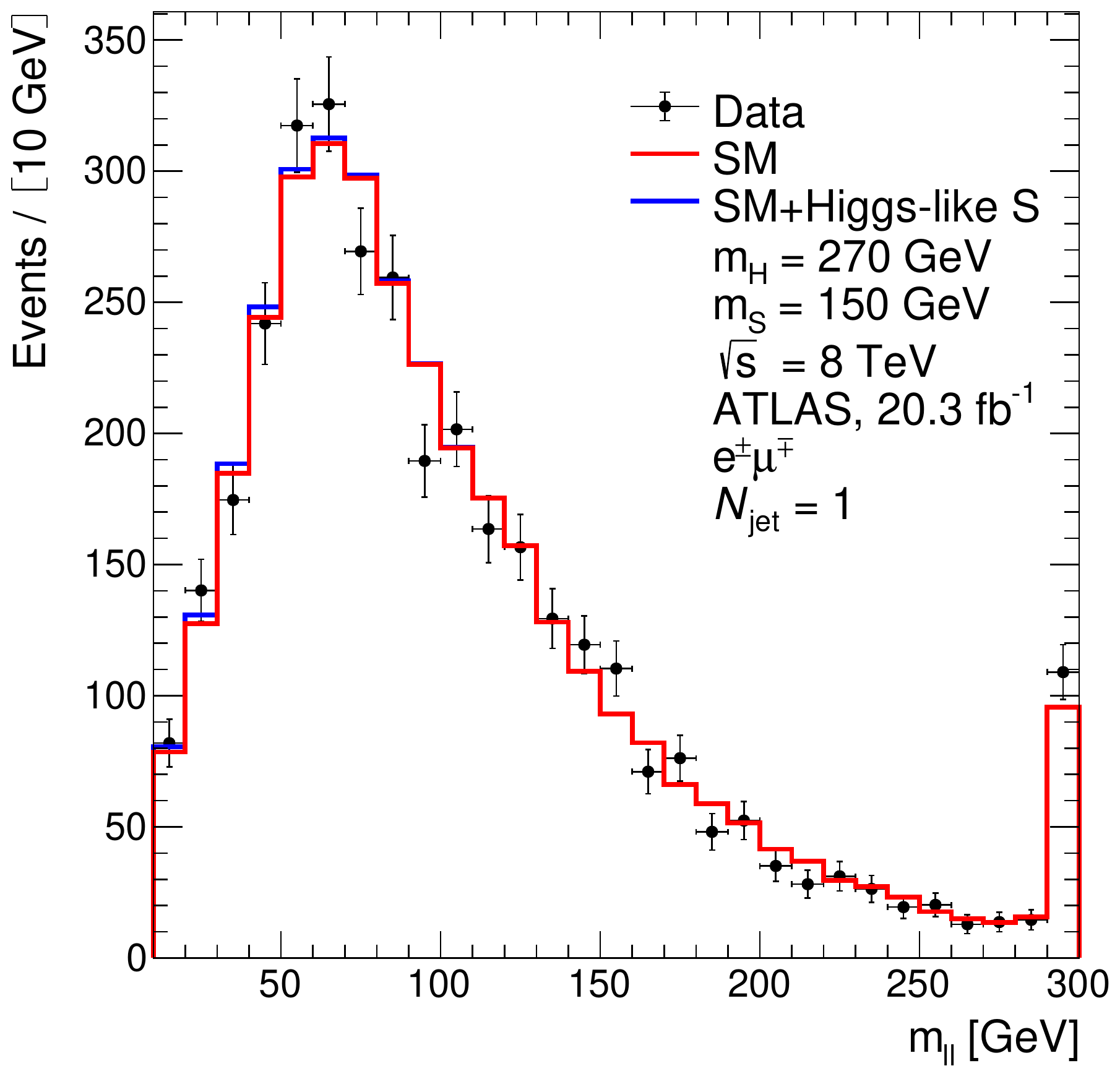}}
        \caption{\label{Fig:12} Distributions of the ATLAS data and SM background comparing to the BSM signal for the di-lepton invariant mass.
        In (a) events are required to have two opposite-charge leptons ($e$, $\mu$) with at least one $b$-tagged jet~\cite{Aaboud:2017ujq}.
        In (b) events are required to have two opposite-charge leptons with different-flavour $(e\mu)$ and (c) with same-flavour $(ee/\mu\mu)$, where both analyses require zero jets~\cite{Aad:2016wpd}.
        In the case of (d) events are required to have two opposite-charge leptons ($e$, $\mu$) with  exactly one jet.
        The data used here is from $pp$ collision collected by the ATLAS experiment at $\sqrt{s} = 8$~TeV with an integrated luminosity of 20.2~fb$^{-1}$ for (a), (b) and (d), and 20.3 fb$^{-1}$ for (c).}
\end{figure}

\begin{figure}
        \centering
        \subfloat[]{\includegraphics[width=0.49\textwidth]{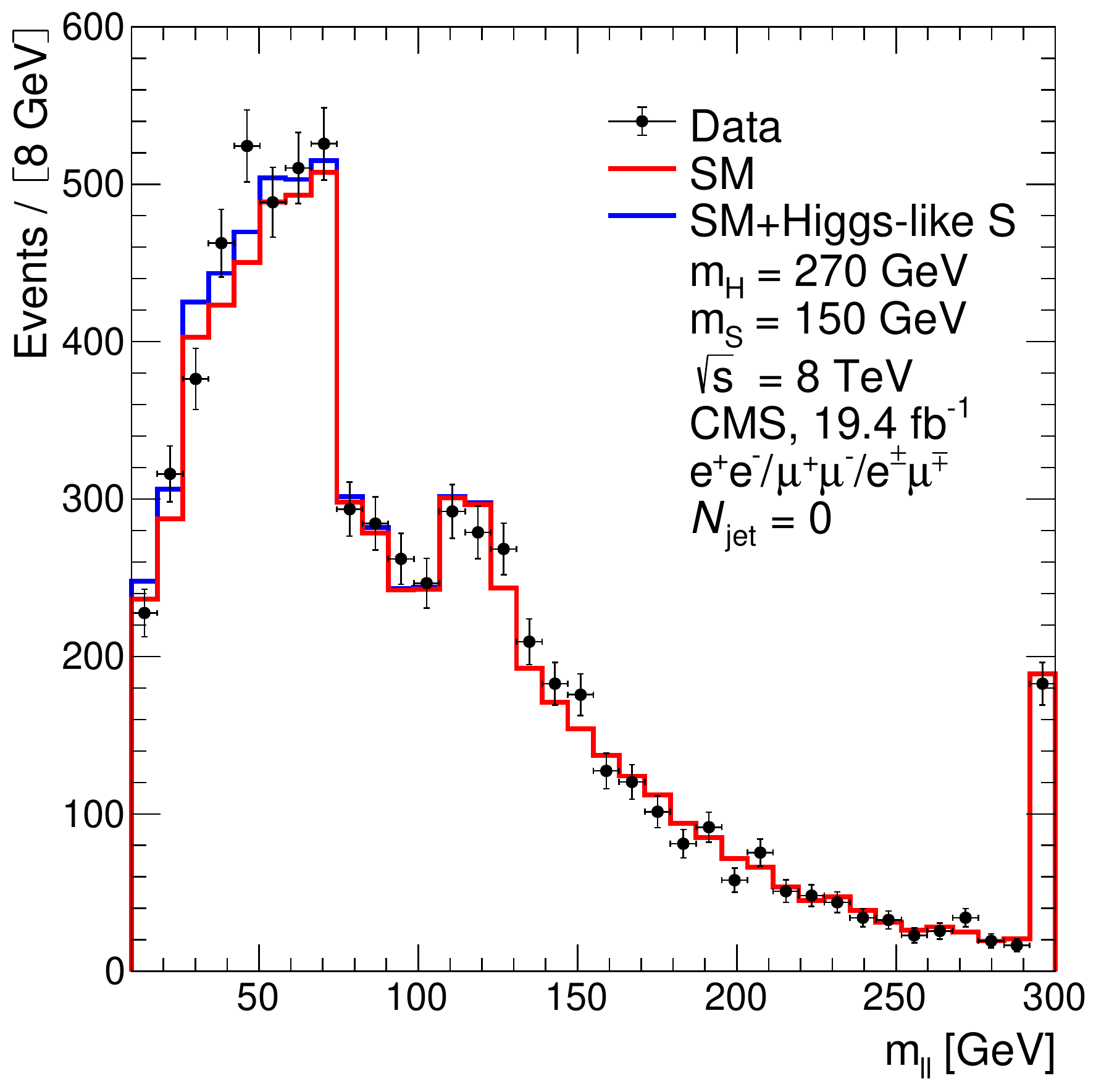}}
        \subfloat[]{\includegraphics[width=0.49\textwidth]{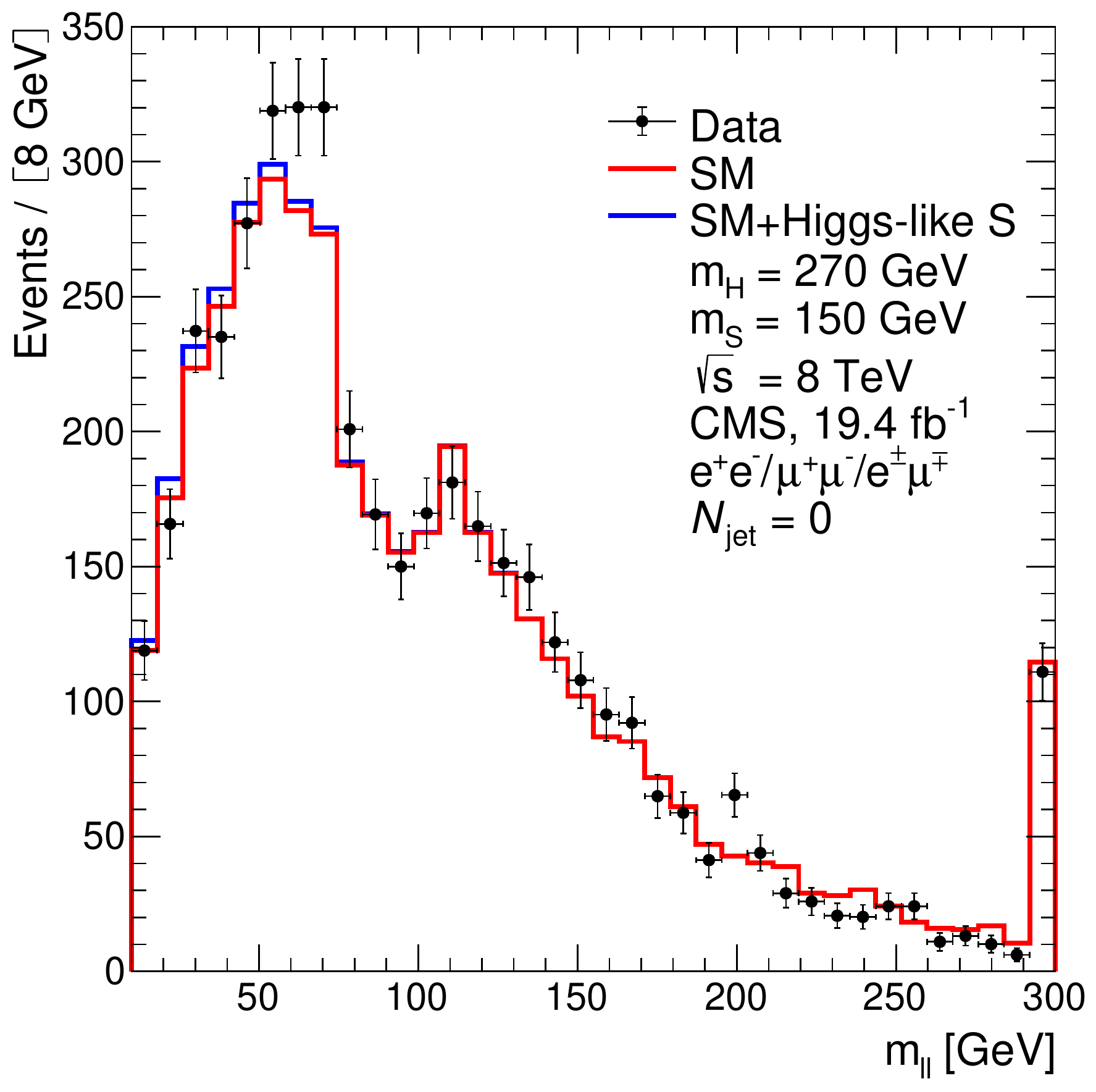}} \\
        \subfloat[]{\includegraphics[width=0.49\textwidth]{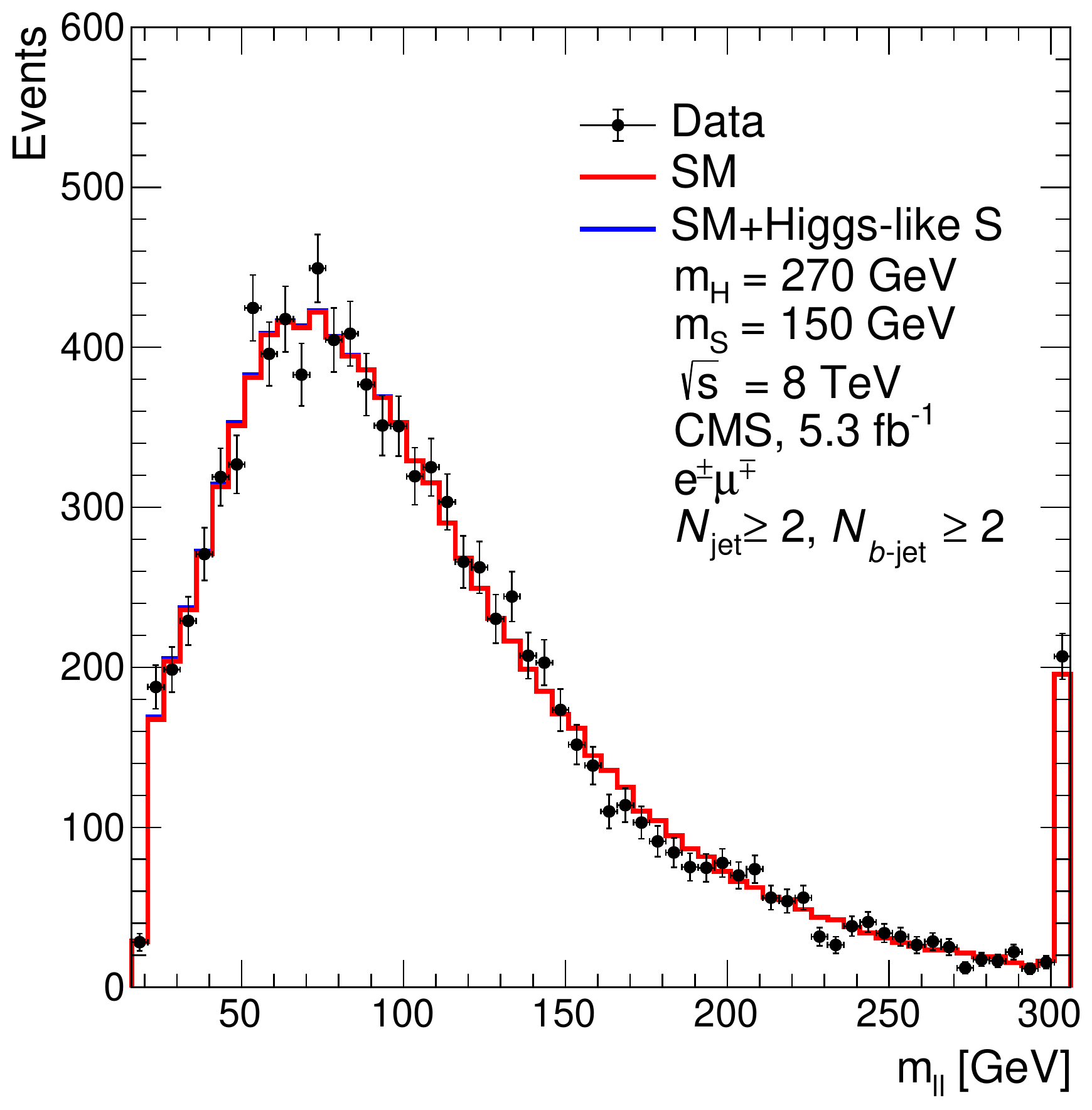}}
        \caption{\label{Fig:13} Distributions of the CMS data and SM background comparing to the BSM signal for the di-lepton invariant mass $(m_{ll})$.
        Events are required to have two opposite-charge leptons with (a) zero jets, (b) one jet~\cite{Khachatryan:2015sga} and (c) $e\mu$ channel~\cite{Chatrchyan:2013faa} with at least two jets and one $b$-tagged jet.
        The data used here are from the $W^+W^-$ and $t\bar{t}$ production cross section measurements in $pp$ collision at $\sqrt{s} = 8$~TeV with an integrated luminosity of 19.4 fb$^{-1}$ and 5.3~fb$^{-1}$ for top and bottom plots, respectively.}
\end{figure}

\begin{figure}
        \centering
        \vspace{-20pt}
        \subfloat[]{\includegraphics[width=0.42\textwidth]{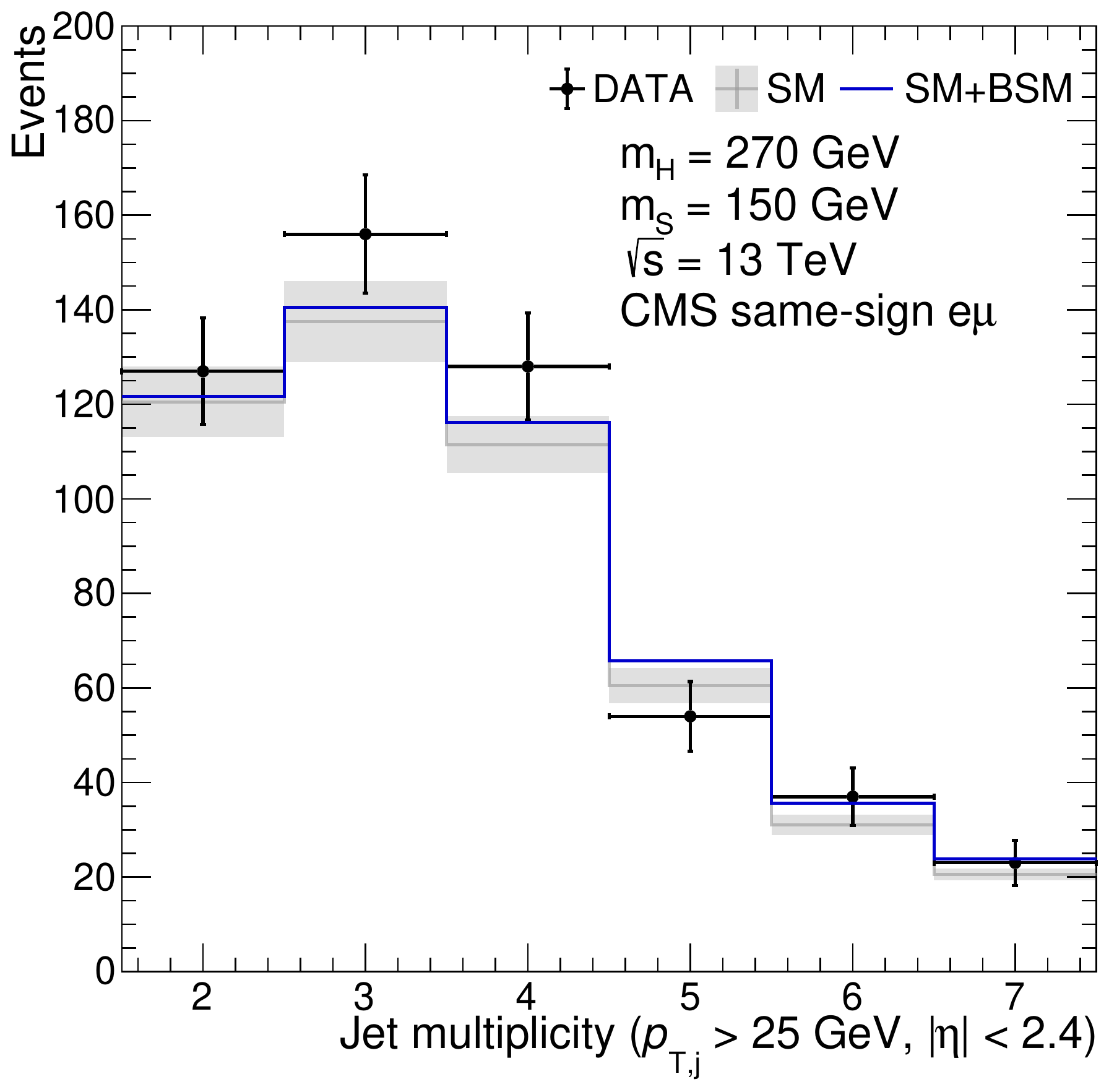}}
        \subfloat[]{\includegraphics[width=0.42\textwidth]{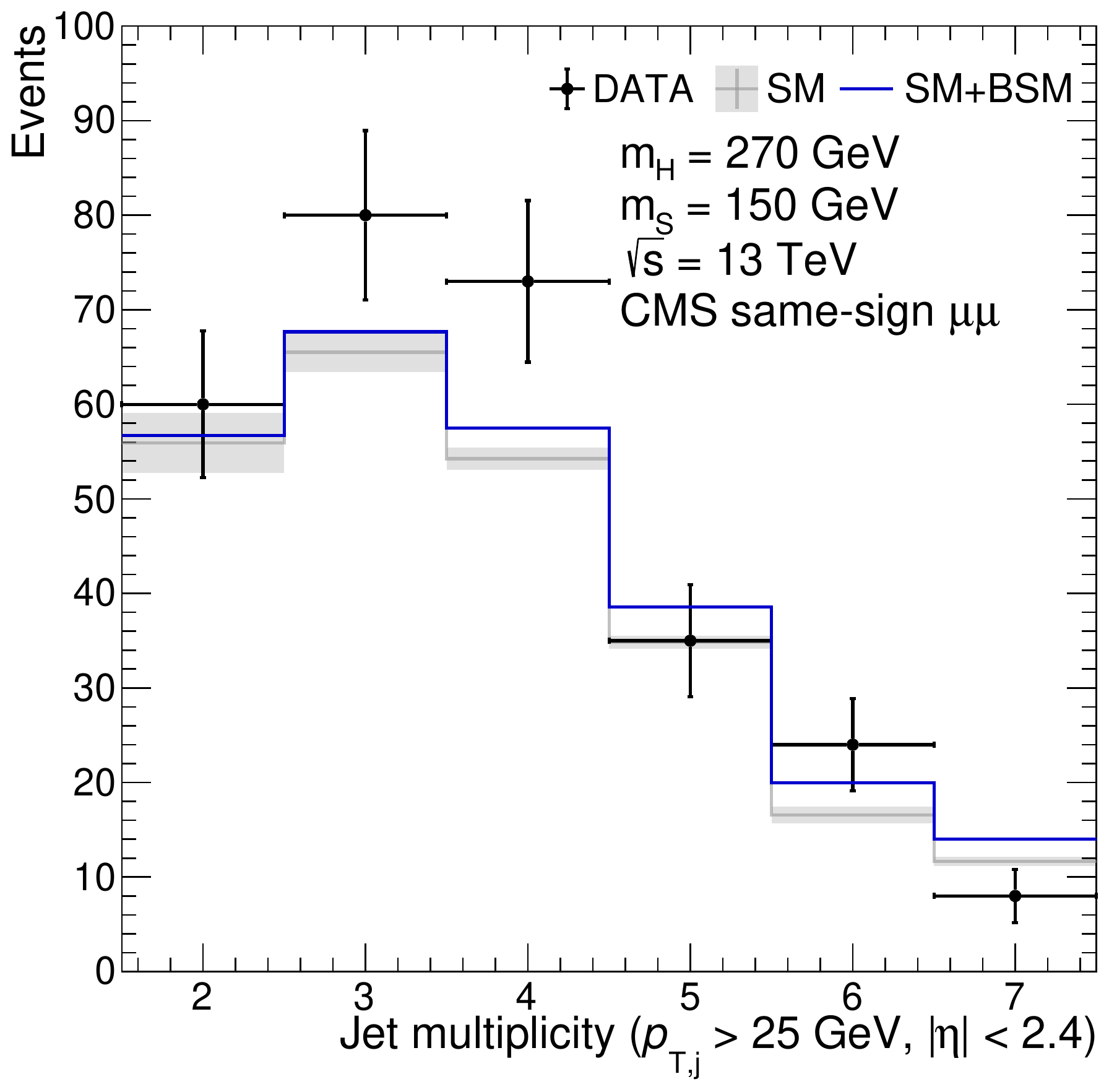}}\vspace{-10pt} \\
        \subfloat[]{\includegraphics[width=0.42\textwidth]{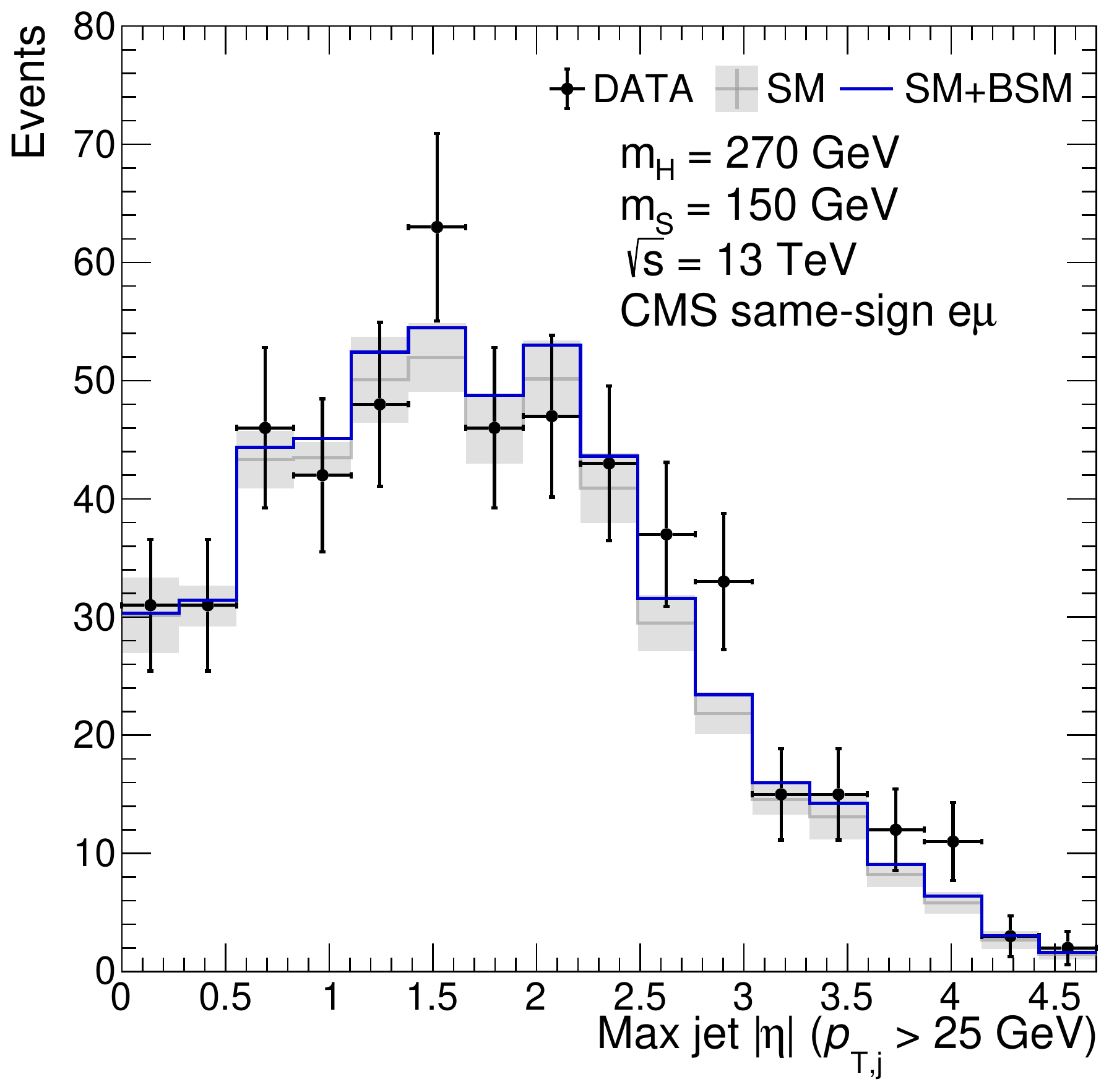}}
        \subfloat[]{\includegraphics[width=0.42\textwidth]{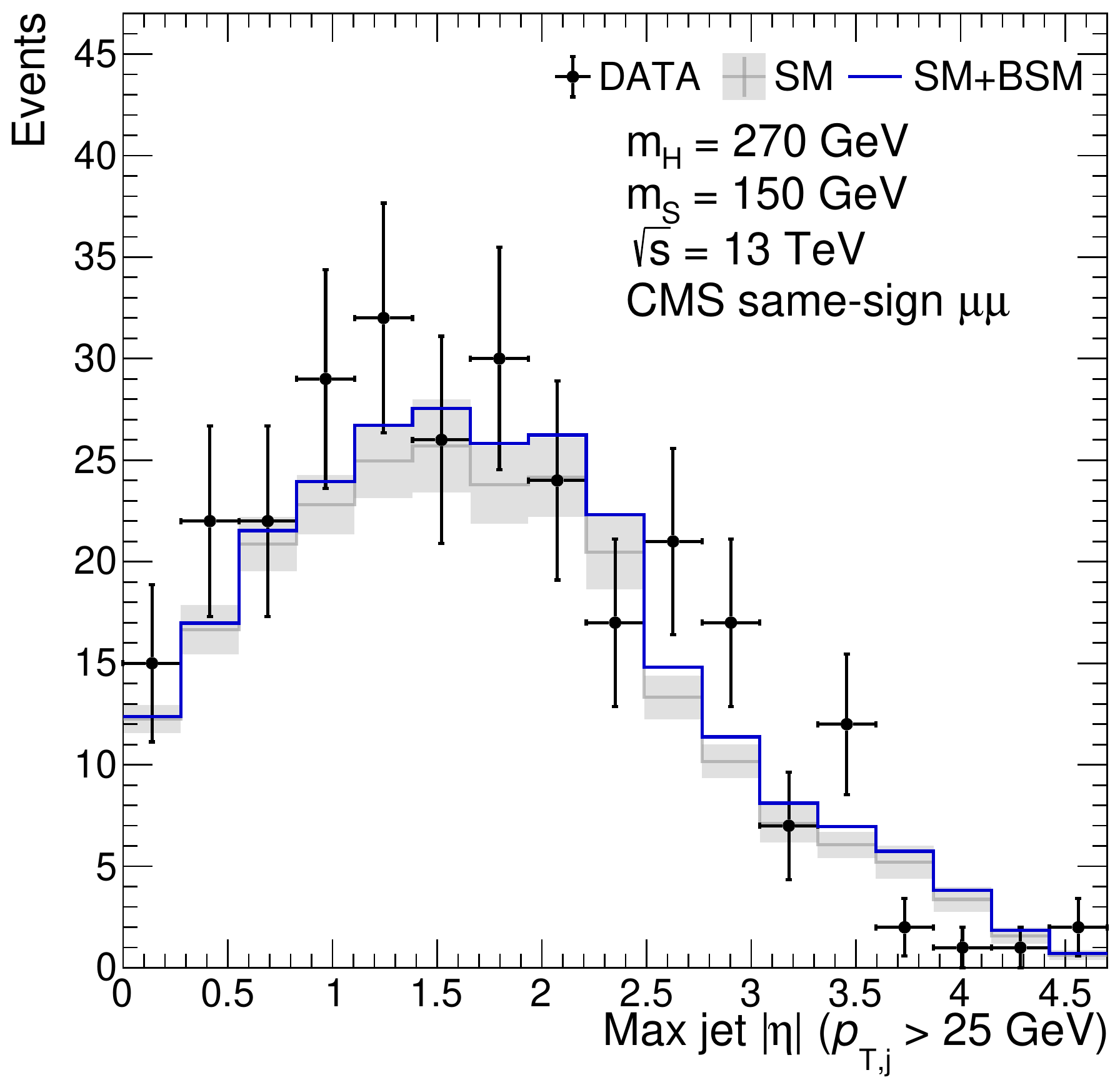}}\vspace{-10pt} \\
        \subfloat[]{\includegraphics[width=0.42\textwidth]{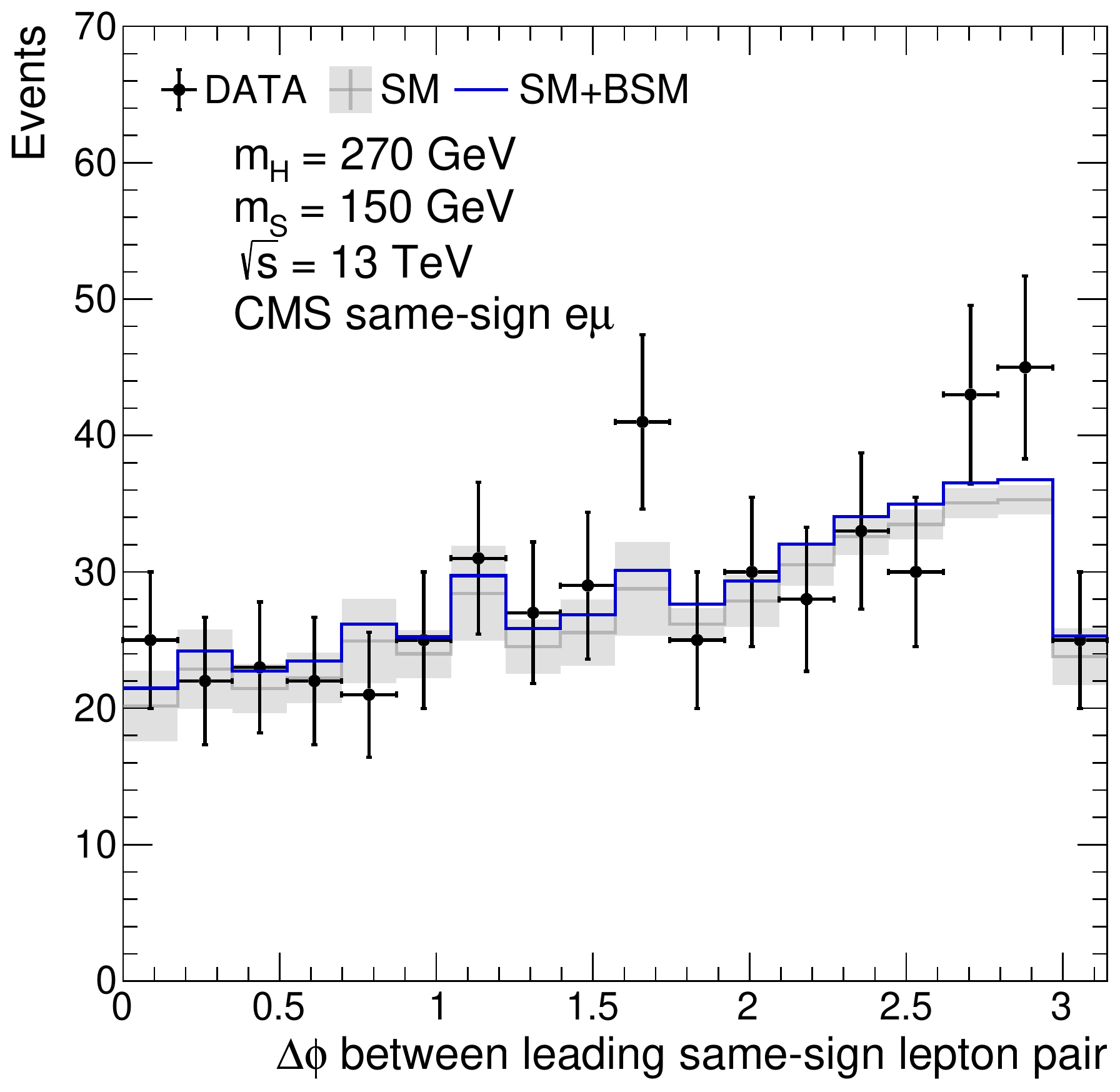}}
        \subfloat[]{\includegraphics[width=0.42\textwidth]{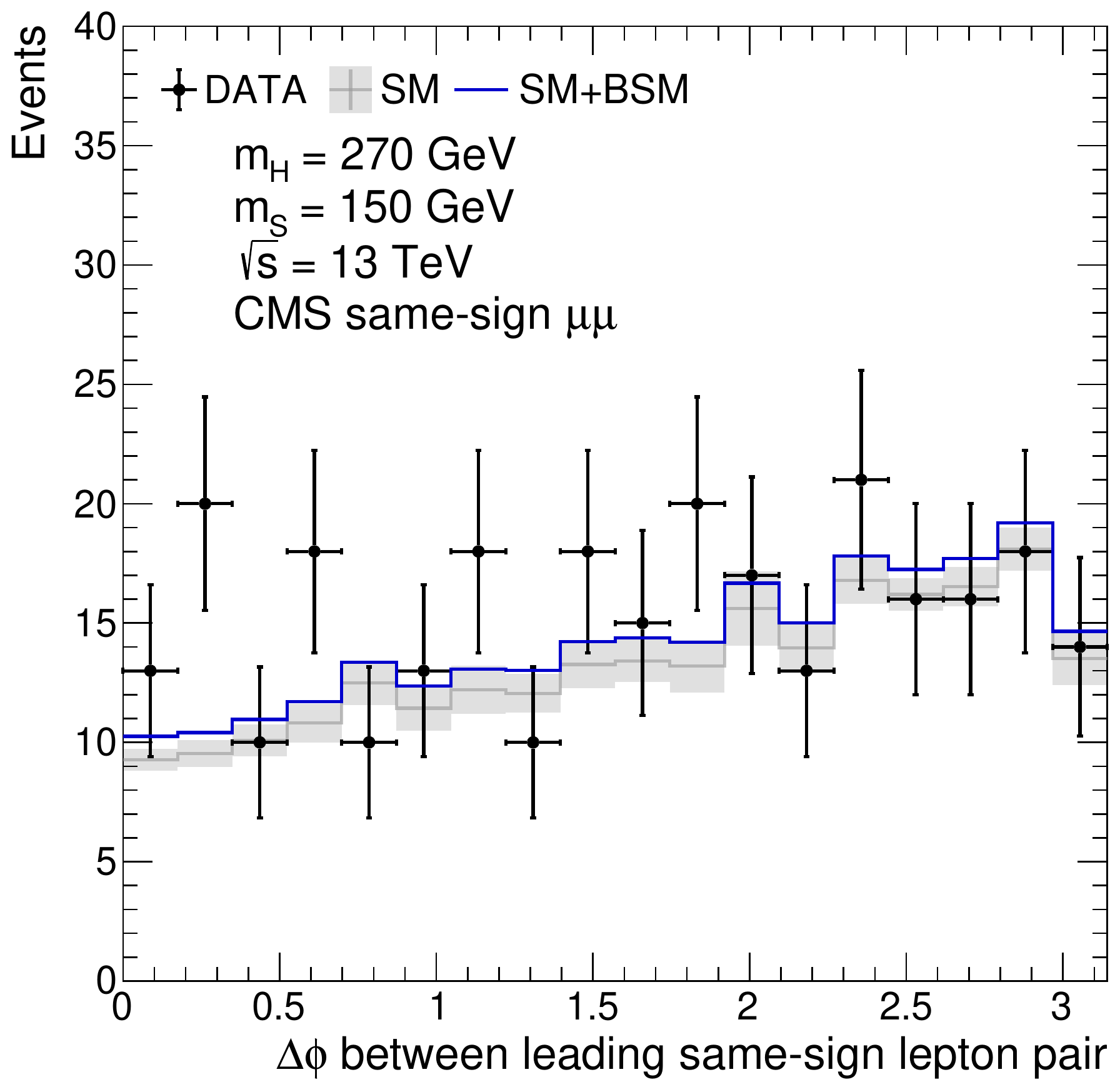}}
        \caption{\label{Fig:tth_dilepton} The distributions of jet multiplicity, maximum jet pseudo-rapidity and azimuthal separation between the leading SS
        lepton pair from the CMS Run~2 single top search.
        These have been separated into the $e\mu$ channel (left) and the $\mu\mu$ channel (right).
        The BSM predictions are scaled by the best fit values of $\beta_g^2$ as described in \autoref{sec:tth}.
        The mass point considered for the BSM prediction is $m_H=270$~GeV and $m_S=150$~GeV.}
\end{figure}

\begin{figure}
        \centering
        \subfloat[]{\includegraphics[width=0.49\textwidth]{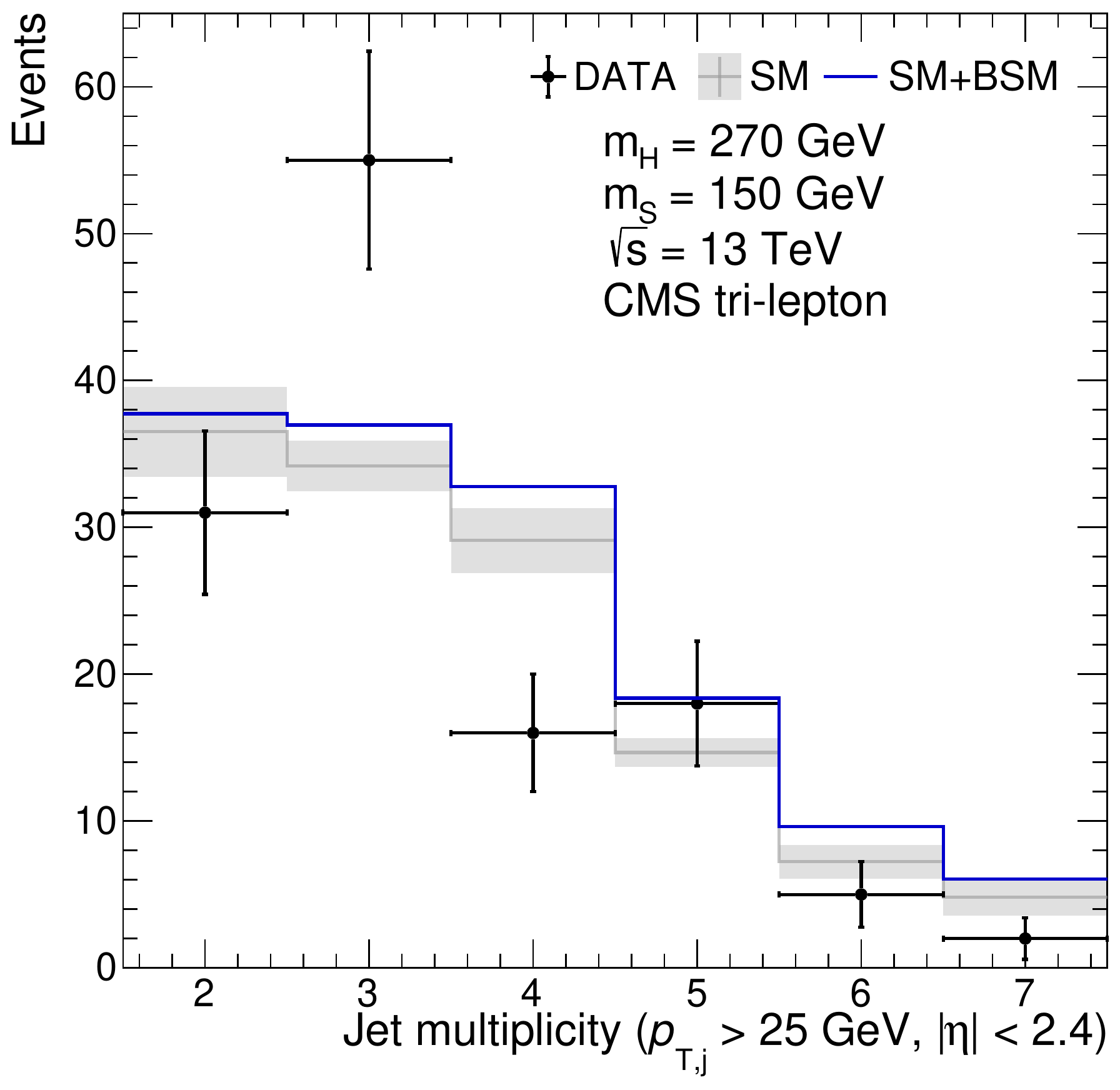}}
        \subfloat[]{\includegraphics[width=0.49\textwidth]{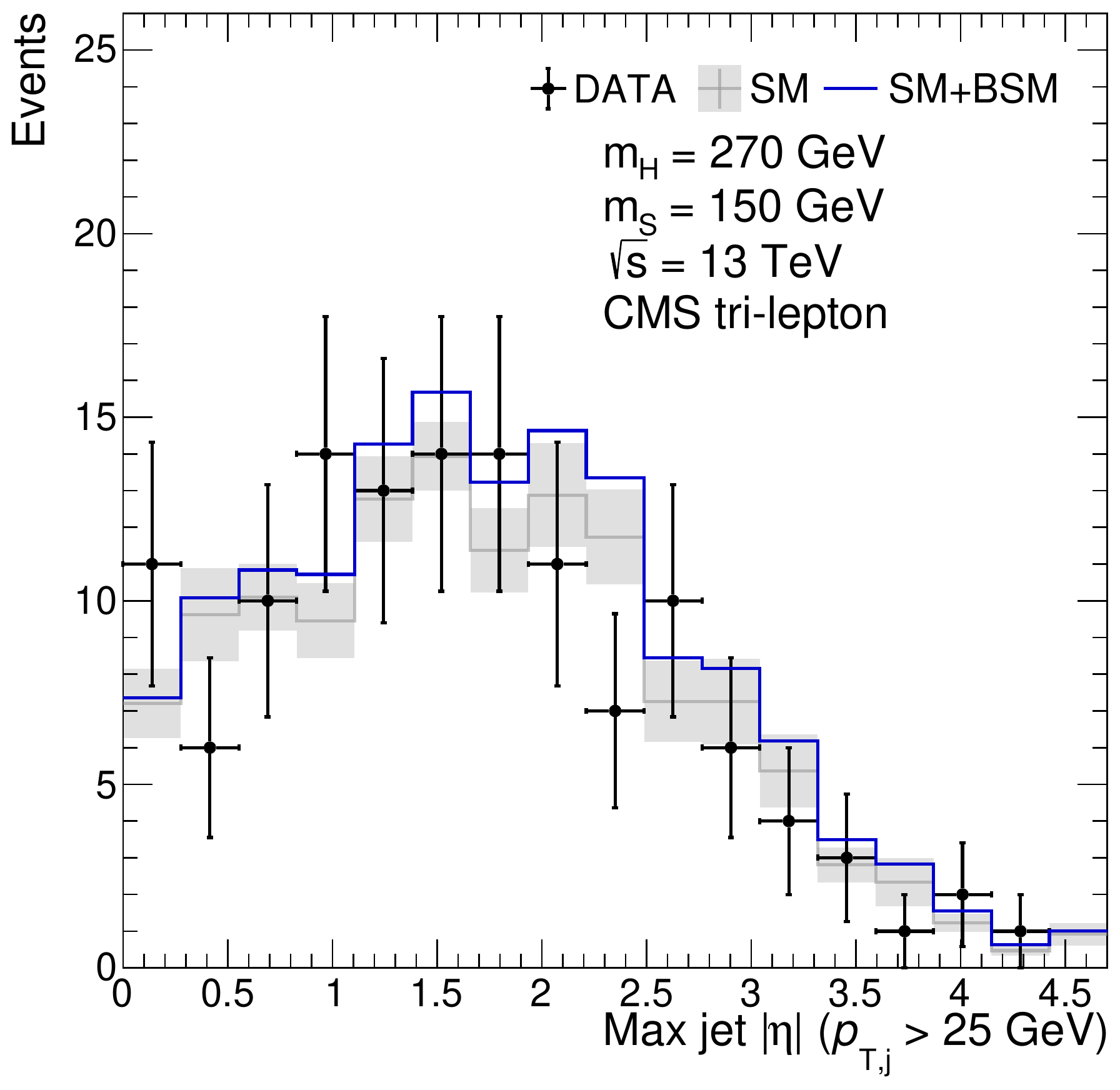}} \\
        \subfloat[]{\includegraphics[width=0.49\textwidth]{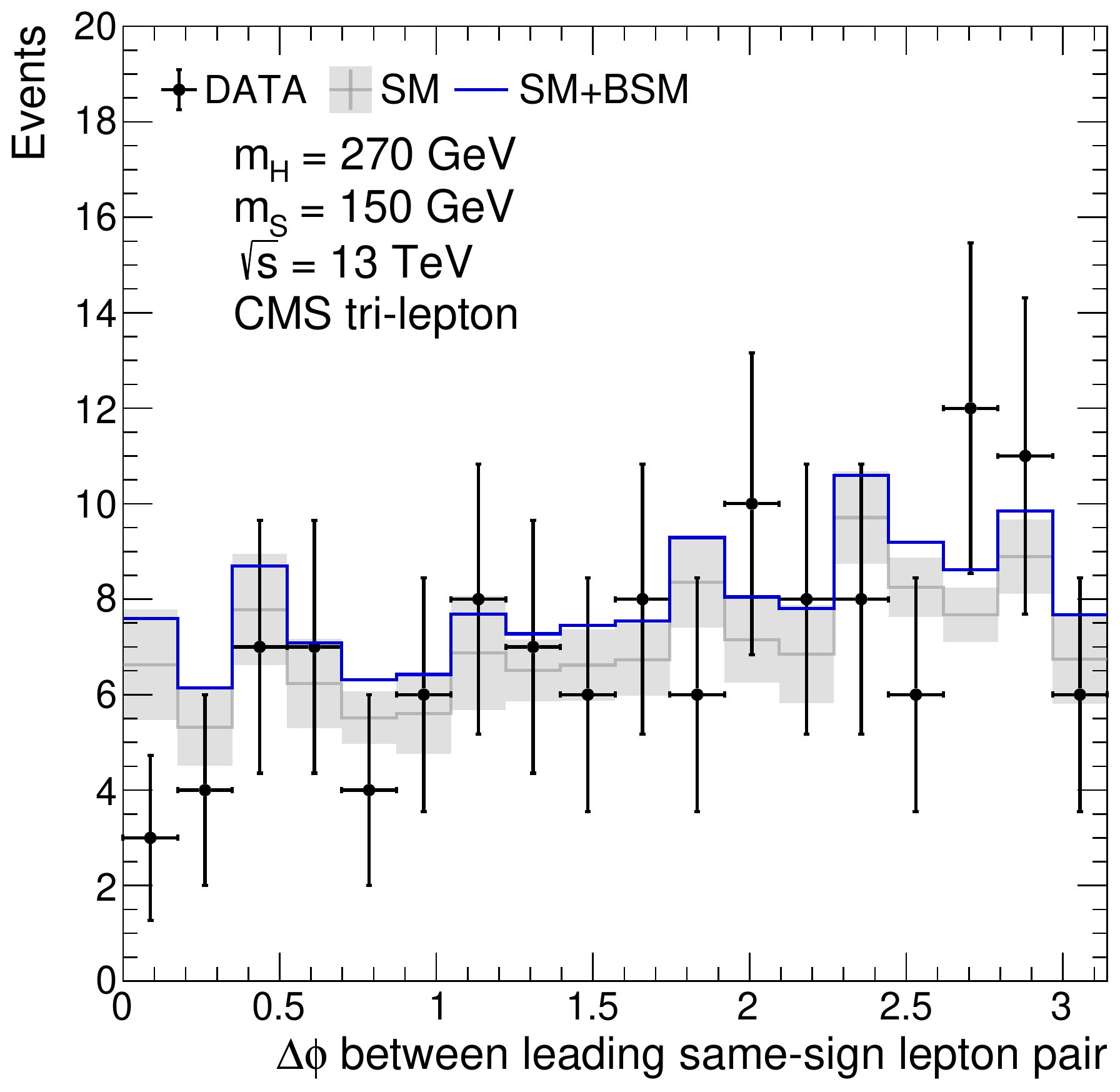}}
        \caption{\label{Fig:tth_trilepton} The distributions of jet multiplicity, maximum jet pseudo-rapidity and  azimuthal separation between the leading SS lepton pair from the CMS Run~2 single top search.
        The BSM predictions are scaled by the best fit values of $\beta_g^2$ as described in \autoref{sec:tth}.
        The mass point considered for the BSM prediction is $m_H=270$~GeV and $m_S=150$~GeV.}
\end{figure}

The calculation of the numbers in \autoref{tab:dlfits} (\autoref{sec:2lcomp}) made use of the di-lepton invariant mass distributions from the ATLAS and CMS collaborations.
Those differential distributions are shown in \autoref{Fig:12} and \autoref{Fig:13} with the best fit data points from ATLAS and CMS, respectively.
In addition to this, fits are made to the distributions in the CMS Run 2 search for a single top in association with a Higgs boson in \autoref{sec:tth}.
The distributions showing the results of these fits can be seen in \autoref{Fig:tth_dilepton} for the di-leptonic categories and \autoref{Fig:tth_trilepton} for the tri-lepton category.
%
%

\pagebreak
\bibliographystyle{spphys}
{\small\bibliography{references.bib}}

\begin{thebibliography}{10}
\providecommand{\url}[1]{{#1}}
\providecommand{\urlprefix}{URL }
\expandafter\ifx\csname urlstyle\endcsname\relax
  \providecommand{\doi}[1]{DOI \discretionary{}{}{}#1}\else
  \providecommand{\doi}{DOI \discretionary{}{}{}\begingroup
  \urlstyle{rm}\Url}\fi

\bibitem{vonBuddenbrock:2015ema}
S.~von Buddenbrock, N.~Chakrabarty, A.S. Cornell, D.~Kar, M.~Kumar, T.~Mandal,
  B.~Mellado, B.~Mukhopadhyaya, R.G. Reed,   (2015).
\newblock {arXiv:1506.00612 [hep-ph]}

\bibitem{Kumar:2016vut}
M.~Kumar, S.~von Buddenbrock, N.~Chakrabarty, A.S. Cornell, D.~Kar, T.~Mandal,
  B.~Mellado, B.~Mukhopadhyaya, R.G. Reed, J. Phys. Conf. Ser. \textbf{802}(1),
  012007 (2017).
\newblock \doi{10.1088/1742-6596/802/1/012007}

\bibitem{vonBuddenbrock:2016rmr}
S.~von Buddenbrock, N.~Chakrabarty, A.S. Cornell, D.~Kar, M.~Kumar, T.~Mandal,
  B.~Mellado, B.~Mukhopadhyaya, R.G. Reed, X.~Ruan, Eur. Phys. J.
  \textbf{C76}(10), 580 (2016).
\newblock \doi{10.1140/epjc/s10052-016-4435-8}

\bibitem{Aad:2012tfa}
{ATLAS Collaboration}, Phys. Lett. \textbf{B716}, 1 (2012).
\newblock \doi{10.1016/j.physletb.2012.08.020}

\bibitem{Chatrchyan:2012xdj}
{CMS Collaboration}, Phys. Lett. \textbf{B716}, 30 (2012).
\newblock \doi{10.1016/j.physletb.2012.08.021}

\bibitem{Aad:2013xqa}
G.~Aad, et~al., Phys. Lett. \textbf{B726}, 120 (2013).
\newblock \doi{10.1016/j.physletb.2013.08.026}

\bibitem{Chatrchyan:2012jja}
S.~Chatrchyan, et~al., Phys. Rev. Lett. \textbf{110}(8), 081803 (2013).
\newblock \doi{10.1103/PhysRevLett.110.081803}

\bibitem{Aad:2015zhl}
{ATLAS and CMS Collaborations}, Phys. Rev. Lett. \textbf{114}, 191803 (2015).
\newblock \doi{10.1103/PhysRevLett.114.191803}

\bibitem{Khachatryan:2016vau}
G.~Aad, et~al., JHEP \textbf{08}, 045 (2016).
\newblock \doi{10.1007/JHEP08(2016)045}

\bibitem{Atre:2009rg}
A.~Atre, T.~Han, S.~Pascoli, B.~Zhang, JHEP \textbf{05}, 030 (2009).
\newblock \doi{10.1088/1126-6708/2009/05/030}

\bibitem{Degrande:2016aje}
C.~Degrande, O.~Mattelaer, R.~Ruiz, J.~Turner, Phys. Rev. \textbf{D94}(5),
  053002 (2016).
\newblock \doi{10.1103/PhysRevD.94.053002}

\bibitem{Alva:2014gxa}
D.~Alva, T.~Han, R.~Ruiz, JHEP \textbf{02}, 072 (2015).
\newblock \doi{10.1007/JHEP02(2015)072}

\bibitem{Golling:2016gvc}
T.~Golling, et~al., CERN Yellow Report (3), 441 (2017).
\newblock \doi{10.23731/CYRM-2017-003.441}

\bibitem{Kumar:2016wzt}
M.~Kumar, S.~von Buddenbrock, N.~Chakrabarty, A.S. Cornell, D.~Kar, T.~Mandal,
  B.~Mellado, B.~Mukhopadhyaya, R.G. Reed, X.~Ruan, in \emph{{61st Annual
  Conference of the South African Institute of Physics (SAIP2016) Johannesburg,
  South Africa, July 4-8, 2016}} (2016)

\bibitem{Mthembu:2017vix}
S.~Mthembu, M.~Kumar,   (2017).
\newblock {arXiv:1702.08772 [hep-ph]}

\bibitem{vonBuddenbrock:2017bqf}
S.~von Buddenbrock, J. Phys. Conf. Ser. \textbf{878}(1), 012030 (2017).
\newblock \doi{10.1088/1742-6596/878/1/012030}

\bibitem{Fang:2017tmh}
Y.~Fang, M.~Kumar, B.~Mellado, Y.~Zhang, M.~Zhu, Int. J. Mod. Phys.
  \textbf{A32}(34), 1746010 (2017).
\newblock \doi{10.1142/S0217751X17460101}

\bibitem{vonBuddenbrock:2017jqp}
S.~von Buddenbrock, A.S. Cornell, M.~Kumar, B.~Mellado, J. Phys. Conf. Ser.
  \textbf{889}(1), 012020 (2017).
\newblock \doi{10.1088/1742-6596/889/1/012020}

\bibitem{Mellado_HDAYS2017}
B.~Mellado.
\newblock {The production of additional bosons and the impact on the Large
  Hadron Collider, Talk given at HDAYS2017, Santander, Spain, } (2017).
\newblock \urlprefix\url{http://hdays.csic.es/HDays17/}

\bibitem{Antusch:2014woa}
S.~Antusch, O.~Fischer, JHEP \textbf{10}, 094 (2014).
\newblock \doi{10.1007/JHEP10(2014)094}

\bibitem{deGouvea:2015euy}
A.~de~Gouvêa, A.~Kobach, Phys. Rev. \textbf{D93}(3), 033005 (2016).
\newblock \doi{10.1103/PhysRevD.93.033005}

\bibitem{Das:2015toa}
A.~Das, N.~Okada, Phys. Rev. \textbf{D93}(3), 033003 (2016).
\newblock \doi{10.1103/PhysRevD.93.033003}

\bibitem{Das:2016hof}
A.~Das, P.~Konar, S.~Majhi, JHEP \textbf{06}, 019 (2016).
\newblock \doi{10.1007/JHEP06(2016)019}

\bibitem{Alloul:2013naa}
A.~Alloul, B.~Fuks, V.~Sanz, JHEP \textbf{04}, 110 (2014).
\newblock \doi{10.1007/JHEP04(2014)110}

\bibitem{Alwall:2014hca}
J.~Alwall, R.~Frederix, S.~Frixione, V.~Hirschi, F.~Maltoni, O.~Mattelaer, H.S.
  Shao, T.~Stelzer, P.~Torrielli, M.~Zaro, JHEP \textbf{07}, 079 (2014).
\newblock \doi{10.1007/JHEP07(2014)079}

\bibitem{Sjostrand:2014zea}
T.~Sjostrand, S.~Ask, J.R. Christiansen, R.~Corke, N.~Desai, P.~Ilten,
  S.~Mrenna, S.~Prestel, C.O. Rasmussen, P.Z. Skands, Comput. Phys. Commun.
  \textbf{191}, 159 (2015).
\newblock \doi{10.1016/j.cpc.2015.01.024}

\bibitem{deFavereau:2013fsa}
J.~de~Favereau, C.~Delaere, P.~Demin, A.~Giammanco, V.~Lemaître, A.~Mertens,
  M.~Selvaggi, JHEP \textbf{02}, 057 (2014).
\newblock \doi{10.1007/JHEP02(2014)057}

\bibitem{Cacciari:2011ma}
M.~Cacciari, G.P. Salam, G.~Soyez, Eur. Phys. J. \textbf{C72}, 1896 (2012).
\newblock \doi{10.1140/epjc/s10052-012-1896-2}

\bibitem{Cacciari:2008gp}
M.~Cacciari, G.P. Salam, G.~Soyez, JHEP \textbf{04}, 063 (2008).
\newblock \doi{10.1088/1126-6708/2008/04/063}

\bibitem{Aaboud:2017ujq}
M.~Aaboud, et~al., Eur. Phys. J. \textbf{C77}(11), 804 (2017).
\newblock \doi{10.1140/epjc/s10052-017-5349-9}

\bibitem{Aad:2016wpd}
{ATLAS Collaboration}, JHEP \textbf{09}, 029 (2016).
\newblock \doi{10.1007/JHEP09(2016)029}

\bibitem{Aaboud:2016mrt}
{ATLAS Collaboration}, Phys. Lett. \textbf{B763}, 114 (2016).
\newblock \doi{10.1016/j.physletb.2016.10.014}

\bibitem{Khachatryan:2015sga}
{CMS Collaboration}, Eur. Phys. J. \textbf{C76}(7), 401 (2016).
\newblock \doi{10.1140/epjc/s10052-016-4219-1}

\bibitem{Chatrchyan:2013faa}
{CMS Collaboration}, JHEP \textbf{02}, 024 (2014).
\newblock \doi{10.1007/JHEP02(2014)024, 10.1007/JHEP02(2014)102}.
\newblock [Erratum: JHEP02,102(2014)]

\bibitem{Aad:2014mfk}
G.~Aad, et~al., Phys. Rev. Lett. \textbf{114}(14), 142001 (2015).
\newblock \doi{10.1103/PhysRevLett.114.142001}

\bibitem{Khachatryan:2016xws}
V.~Khachatryan, et~al., Phys. Rev. \textbf{D93}(5), 052007 (2016).
\newblock \doi{10.1103/PhysRevD.93.052007}

\bibitem{ATLAS:2018gcr}
{ATLAS collaboration}, ATLAS-CONF-2018-004  (2018)

\bibitem{Biswal:2012mp}
S.S. Biswal, R.M. Godbole, B.~Mellado, S.~Raychaudhuri, Phys. Rev. Lett.
  \textbf{109}, 261801 (2012).
\newblock \doi{10.1103/PhysRevLett.109.261801}

\bibitem{Kumar:2015kca}
M.~Kumar, X.~Ruan, R.~Islam, A.S. Cornell, M.~Klein, U.~Klein, B.~Mellado,
  Phys. Lett. \textbf{B764}, 247 (2017).
\newblock \doi{10.1016/j.physletb.2016.11.039}

\bibitem{Aaboud:2017jvq}
M.~Aaboud, et~al., Phys. Rev. \textbf{D97}(7), 072003 (2018).
\newblock \doi{10.1103/PhysRevD.97.072003}

\bibitem{Sirunyan:2018shy}
A.M. Sirunyan, et~al.,   (2018)

\bibitem{Farina:2012xp}
M.~Farina, C.~Grojean, F.~Maltoni, E.~Salvioni, A.~Thamm, JHEP \textbf{05}, 022
  (2013).
\newblock \doi{10.1007/JHEP05(2013)022}

\bibitem{deFlorian:2016spz}
D.~de~Florian, et~al.,   (2016).
\newblock \doi{10.23731/CYRM-2017-002}

\bibitem{CMS:2017uzk}
{CMS Collaboration},   (2017).
\newblock CMS-PAS-HIG-17-005

\bibitem{Sirunyan:2018lcp}
A.M. Sirunyan, et~al., JHEP  (2018).
\newblock \doi{10.3204/PUBDB-2018-02197}

\bibitem{ATLAS:2018rgl}
{ATLAS collaboration},   (2018).
\newblock ATLAS-CONF-2018-027

\end{thebibliography}

\end{document}